\documentclass[a4paper,11pt]{article}
\pdfoutput=1
\usepackage{jheppub}
\usepackage{tablefootnote}
\usepackage[T1]{fontenc}
\usepackage{adjustbox}
\usepackage{multicol}
\usepackage{multirow}
\usepackage{bbm}
\usepackage{xcolor}
\newcommand{\yong}[1]{{\color{black}#1}}
\newcommand{\hl}[1]{{\color{black}#1}}
\newcommand{\eqal}[1]{\begin{align}#1\end{align}}
\newcommand{\juno}[1]{{\color{black}#1}}

\title{\boldmath Exploring SMEFT Induced Non-Standard Interactions from COHERENT to Neutrino Oscillations}

\author[a]{Yong Du,}
\author[a]{Hao-Lin Li,}
\author[c]{Jian Tang,}
\author[c]{Sampsa Vihonen,}
\author[a,d,e,f,g]{Jiang-Hao Yu}

\affiliation[a]{CAS Key Laboratory of Theoretical Physics, Institute of Theoretical Physics, Chinese Academy of Sciences, Beijing 100190, P. R. China}
\affiliation[c]{School of Physics, Sun Yat-sen University, Guangzhou 510275, China}
\affiliation[d]{School of Physical Sciences, University of Chinese Academy of Sciences, Beijing 100049, P.R. China}
\affiliation[e]{Center for High Energy Physics, Peking University, Beijing 100871, China}
\affiliation[f]{School of Fundamental Physics and Mathematical Sciences, Hangzhou Institute for Advanced Study, UCAS, Hangzhou 310024, China}
\affiliation[g]{International Centre for Theoretical Physics Asia-Pacific, Beijing/Hangzhou, China}

\emailAdd{yongdu@itp.ac.cn}
\emailAdd{lihaolin@itp.ac.cn}
\emailAdd{tangjian5@mail.sysu.edu.cn}
\emailAdd{sampsa@mail.sysu.edu.cn}
\emailAdd{jhyu@itp.ac.cn}

\abstract{
We investigate the prospects of next-generation neutrino oscillation experiments DUNE, T2HK and JUNO including TAO within Standard Model Effective Field Theory (SMEFT). We also re-interpret COHERENT data in this framework. Considering both charged and neutral current neutrino Non-Standard Interactions (NSIs), we analyse dimension-6 SMEFT operators and derive lower bounds to UV scale $\Lambda$. The most powerful probe is obtained on ${\cal O}_{{ledq}_{1211}}$ with $\Lambda \gtrsim$ 450\,TeV due to the electron neutrino sample in T2HK near detector. We find DUNE and JUNO to be complementary to T2HK in exploring different subsets of SMEFT operators at about 25\,TeV. We conclude that near detectors play a significant role in each experiment. We also find COHERENT with CsI and LAr targets to be sensitive to new physics up to $\sim$900\,GeV.}

\begin{document}
\maketitle
\flushbottom

\section{Introduction}\label{sec:intro}
The Standard Model of particle physics (SM) has been one of the most successful mathematical formulations to describe the physics of elementary particles and their interactions. It has been tested in numerous experiments and validated to very high energies. There are yet several issues which SM is not able to explain. It is not possible to tell, for example, whether neutrino masses have their origin in the Higgs mechanism or how gravity arises between elementary particles in SM. In the absence of new particle discoveries, there is a strong motivation to adopt a bottom-up approach to study physics beyond the Standard Model while utilising the precision data from diverse experimental setups. The recently developed Standard Model Effective Field Theory (SMEFT) framework ~\cite{Weinberg:1979sa,Buchmuller:1985jz,Grzadkowski:2010es,Lehman:2014jma,Li:2020gnx,Murphy:2020rsh,Li:2020xlh,Liao:2020jmn,Liao:2016hru} is very suitable for this purpose as it provides the methodology to study physics beyond the SM in a model-independent universal framework. In the present work, we apply SMEFT framework in neutrino experiments and assess the physics programs of the next-generation experiments of accelerator and reactor-based neutrino sources.

Next-generation neutrino oscillation experiments are multi-national research infrastructure projects aiming to resolve the remaining unknowns in the theoretical framework that describes the oscillations of three neutrinos. The next-generation long-baseline experiments, Tokai-to-HyperKamiokande (T2HK) experiment in Japan~\cite{Abe:2015zbg} and Deep Underground Neutrino Experiment (DUNE) in the United States~\cite{Acciarri:2016crz} are currently being constructed to search for \emph{CP} violation and establish the ordering of neutrino masses $m_1$, $m_2$ and $m_3$. Other experiment goals include determining the octant of the atmospheric mixing angle $\theta_{23}$. In the meantime, Jiangmen Underground Neutrino Observatory (JUNO) and Taishan Antineutrino Observatory (TAO) are prepared to take data in China~\cite{An:2015jdp,Abusleme:2020bzt}. Whereas T2HK and DUNE rely on accelerator-based muon neutrino sources for neutrino production in J-PARC and Fermilab, respectively, JUNO and TAO shall measure the electron antineutrino flux coming from the Yangjiang and Taishan nuclear power plants. Once operational, these experiments will represent the state of art with their respective technologies. There are also a number of neutrino observatories such as KM3NeT~\cite{Katz:2006wv} and IceCube~\cite{Ahrens:2003ix}, which shall complement the terrestrial experiments with observations of neutrinos of solar, atmospheric and astrophysical origin.

Though the main physics questions in the next-generation neutrino experiments are focused on the physics described by the SM fields and interactions, the aforementioned oscillation experiments shall also begin an era of SM precision tests. In the past, deviations from the SM interactions of neutrinos have usually been described by non-standard interactions (NSIs)~\cite{Ohlsson:2012kf,Dev:2019anc} with several studies delving into EFT~\cite{Falkowski:2019kfn,Falkowski:2019xoe,Terol-Calvo:2019vck,Escrihuela:2021mud,Bischer:2019ttk,Bischer:2018zcz,Li:2020lba,Li:2020wxi,Li:2019fhz}. The former has been studied in great extent in both accelerator and reactor neutrino experiments~\cite{Huitu:2016bmb,Blennow:2016etl,Farzan:2017xzy,Han:2019zkz,Liao:2019qbb,Feng:2019mno,Bakhti:2020fde,KumarAgarwalla:2021twp}, where NSIs have been shown to be a hindrance to reaching \emph{CP} violation, mass hierarchy and octant targets~\cite{Das:2017fcz,Capozzi:2019iqn,Esteban:2020itz,Esteban:2019lfo} but also an opening to perform indirect searches for new particles~\cite{Huitu:2017cpc,Brdar:2017kbt}. Neutrino NSIs can also be probed in scattering processes, such as the coherent elastic neutrino-nucleus scattering (CE$\nu$NS) recently observed in COHERENT and CONUS experiments~\cite{Akimov:2018vzs,Bonet:2020awv}. When the prospects to study neutrino NSIs are re-analysed in the EFT framework, the physics potential of neutrino experiments can be connected to experiments representing the energy and cosmic frontiers. An example of the latter can be found in Refs.~\cite{Du:2021idh,Du:2021nyb} which describe a methodology to connect neutral current neutrino NSIs to number of relativistic species $N_{\rm eff}$.

In the present work, we investigate the prospects of next-generation neutrino oscillation experiments in the search for neutrino NSI. We simulate the neutrino oscillations in T2HK, DUNE and JUNO experiments and project their expected sensitivities in SMEFT. Our main result is a systematic scan of 1635 dimension-6 SMEFT operators, where the UV-scale $\Lambda$ associated with the individual operators are constrained from below assuming non-observation of new particles. This work is continuation to our previous article~\cite{Du:2020dwr}, where the recent data from on-going long-baseline and reactor neutrino experiments were analysed in the SMEFT framework. Our present work develops this approach by adding neutral current NSI effects into consideration, in addition to charged current NSIs examined in Ref.~\cite{Du:2020dwr}. We furthermore extend this analysis to include the neutral current NSIs that can be searched in CE$\nu$NS experiments. This methodology is tested with the recent COHERENT data using the data releases reported from CsI and Ar targets~\cite{Akimov:2018vzs,Akimov:2020czh}. As we shall see in this work, in the absence of new particle discoveries the most stringent constraints to neutral current NSIs can be expected from the far detector data of T2HK and DUNE, whilst the charged current NSIs can be tested to unprecedented precision in JUNO and TAO as well as in the near detector data of T2HK and DUNE. We note that the most stringent constraint to a single dimension-6 SMEFT operator is expected from the T2HK near detector, which is expected to probe ${\cal O}_{{ledq}_{1211}}$ up to $\Lambda \sim$ 450\,TeV with its corresponding Wilson coefficient set to unity. Our results emphasize the importance of achieving complementarity through probing neutrino NSIs with multiple production and detection techniques and with various baseline lengths.

This article is organized as follows: We discuss the formalism of neutrino NSI in section\,\ref{sec:nsi_intro} and describe the matching process between NSI and SMEFT frameworks in section\,\ref{sec:smeft-to-nsi}. Details of the simulation techniques and the considered experiments are presented in section\,\ref{sec:experiments}. The results of this study are shown in section\,\ref{sec:results}, with the full list of lower bounds for $\Lambda$ presented in Appendix\,\ref{app:A}. \footnote{The plots presented in Appendix\,\ref{app:A}, as well as some additional plots, are also available at \url{https://github.com/Yong-Du/SMEFT_NSIs}.} The analysis of the COHERENT data is provided in section\,\ref{sec:coherent}. We conclude in section\,\ref{sec:concl}.

\section{\label{sec:nsi_intro}Neutrino oscillation and Non-Standard Interactions}
In Ref.~\cite{Du:2020dwr}, we have studied the effects of the charged current (CC) neutrino NSIs in the oscillation experiments, where the source and detection NSI parameters $\epsilon^s$ and $\epsilon^d$ at quantum mechanical (QM) level encode the modification of the flavor eigenstates in the two processes defined in the following equations:
\begin{eqnarray}\label{eq:NSIQM}
|\nu_\alpha^s\rangle = \frac{(1+\epsilon^s)_{\alpha\gamma}}{N^s_\alpha}|\nu_\gamma\rangle,\ \langle\nu_\beta^d| = \langle\nu_\gamma| \frac{(1+\epsilon^d)_{\gamma\beta}}{N^d_\beta},
\end{eqnarray}
$N_{\alpha}^s=\sqrt{[(1+\epsilon^s)(1+\epsilon^{s\dagger})]_{\alpha\alpha}}$ and $N_{\beta}^d=\sqrt{[(1+\epsilon^{d{\dagger}})(1+\epsilon^d)]_{\beta\beta}}$ in the above equations are normalization factors, and the Greek letters represent lepton flavors that can be taken from $e,\mu,\tau$. 
On the other hand the general parameterization of the neutrino CC NSIs in the LEFT Lagrangian is given by:
\begin{align}
\mathcal{L}_{\rm CC} \supset &-2 \sqrt{2}G_F V^{\rm SM}_{u d}\left\{\left[\mathbf{1}+\epsilon_{L}\right]^{ij}_{\alpha \beta}\left(\bar{u}_i \gamma^{\mu} P_{L} d_j\right)\left(\bar{\ell}_{\alpha} \gamma_{\mu} P_{L} \nu_{\beta}\right)+\left[\epsilon_{R}\right]^{ij}_{\alpha \beta}\left(\bar{u}_i \gamma^{\mu} P_{R} d_j\right)\left(\bar{\ell}_{\alpha} \gamma_{\mu} P_{L} \nu_{\beta}\right)\right.\nonumber \\
& + \frac{1}{2}\left[\epsilon_{S}\right]^{ij}_{\alpha \beta}(\bar{u}_i d_j)\left(\bar{\ell}_{\alpha} P_{L} \nu_{\beta}\right)-\frac{1}{2}\left[\epsilon_{P}\right]^{ij}_{\alpha \beta}\left(\bar{u}_i \gamma_{5} d_j\right)\left(\bar{\ell}_{\alpha} P_{L} \nu_{\beta}\right)\nonumber \\
&+\left.\frac{1}{4}\left[\epsilon_{T}\right]^{ij}_{\alpha \beta}\left(\bar{u}_i \sigma^{\mu \nu} P_{L} d_j\right)\left(\bar{\ell}_{\alpha} \sigma_{\mu \nu} P_{L} \nu_{\beta}\right)+\mathrm{h.c.} \right\}.\label{eq:CC}
\end{align}
\hl{In the above equations, $V^{\rm SM}_{ud}=0.97439$ is the $ud$ components of the CKM matrix elements obtained from Flavour Lattice Averaging Group (FLAG)~\cite{Aoki:2016frl}, which is different from the $V_{\rm CKM}$ we defined in section.~\ref{sec:smeft-to-nsi} in the presence of SMEFT Wilson coefficients, and $G_F$ is the Fermi-Constant measured from the muon decay, these are just normalization conventions for $\epsilon$'s present in the above eq.~\eqref{eq:CC} }
The matching between QM source and detection NSI parameters and NSI parameters in the LEFT Lagrangians are studied in Ref.~\cite{Falkowski:2019kfn,Du:2020dwr}, and we list them as a dictionary in table.~\ref{table:matching}, where $g_S=1.022(10)$, $g_A=1.251(33)$ and $g_T=0.987(55)$ are the scalar, axial-vector and tensor charges of the nucleon~\cite{Chang:2018uxx,Gupta:2018qil,Aoki:2019cca,Gonzalez-Alonso:2013ura}, $\Delta\equiv m_n-m_p\simeq 1.3 \rm ~MeV$, $E_\nu$ is the neutrino energy, and $f_{T}(E_\nu)$ is the nucleon form factor resulting from tensor-type NSIs in eq.\,\eqref{eq:CC}, for which we use the same parameterization as that in Ref.\,\cite{Falkowski:2019xoe}.
In the table we also take into account the matching of source NSIs from the muon decay that are derived in Ref. \cite{Falkowski:2019xoe,Du:2020dwr}, where the couplings $g_{22}$ and $h_{21}$ are defined as the anomalous couplings for the pure leptonic operators:
\begin{equation}
    {\cal L}_{e\mu} = \frac{G_F}{\sqrt{2}}\{g_{22}(\bar{e}_L\gamma^\mu \nu_{e L})(\bar{\nu}_{\mu L}\gamma_\mu \mu_L)-2h_{21}(\bar{\nu}_{eL}e_R)(\bar{\mu}_R\nu_{\mu L})+h.c.\}.
\end{equation}
\begin{table}[ht]
\caption{Matching between QM and QFT NSI parameters} 
\centering 
\begin{tabular}{ll} 
\hline\hline 
QM NSIs & Relations to QFT NSIs  \\ [0.5ex] 
\hline 
$\epsilon_{e \beta}^{s}$ ($\beta$ decay) & $\left[\epsilon_{L}-\epsilon_{R}-\frac{g_{T}}{g_{A}} \frac{m_{e}}{f_{T}\left(E_{\nu}\right)} \epsilon_{T}\right]_{e \beta}^{*}$ \\ 
$\epsilon_{\beta e}^{d}$ (inverse $\beta$ decay)& $\left[\epsilon_{L}+\frac{1-3 g_{A}^{2}}{1+3 g_{A}^{2}} \epsilon_{R}-\frac{m_{e}}{E_{\nu}-\Delta}\left(\frac{g_{S}}{1+3 g_{A}^{2}} \epsilon_{S}-\frac{3 g_{A} g_{T}}{1+3 g_{A}^{2}} \epsilon_{T}\right)\right]_{e \beta}$ \\
$\epsilon_{\mu \beta}^{s}$ (pion decay)& $\left[\epsilon_{L}-\epsilon_{R}-\frac{m_{\pi}^{2}}{m_{\mu}\left(m_{u}+m_{d}\right)} \epsilon_{P}\right]_{\mu \beta}^{*}$ \\
$\epsilon_{\mu\beta}^s$(muon decay) & $\left[ g_{22} + \frac{3m_em_\mu(m_\mu-2E_\nu)}{16m_\mu E_\nu^2+6m_\mu(m_\mu^2+m_e^2)-4E_\nu(5m_\mu^2+m_e^2)}h_{21} \right]_{\mu\beta}^*$ \\
$\epsilon_{e\beta}^s$ (muon decay)& $\left[ g_{22} + \frac{m_e}{4(m_\mu-2E_{\bar{\nu}})}h_{21} \right]_{e\beta}^*$\\ [1ex] 
\hline 
\end{tabular}
\label{table:matching} 
\end{table}

\yong{We comment on the neutrino-energy dependence in the matching formulas above. For SM-like interactions, meaning those of $V-A$ type ones, it has been shown gracefully in Ref.\,\cite{Falkowski:2019kfn} that the matching between the QM and the QFT formalisms is free of $E_\nu$ and valid up to all orders in $\epsilon$'s in the Lagrangian of the LEFT. However, beyond $V-A$, the matching generically becomes neutrino-energy dependent as are the cases for $\beta$, inverse $\beta$ and muon decay. Since the neutrino production through muon decays is never considered in our analysis as its effect has been absorbed in the definition of $G_F$ as we will discuss shortly, we focus here on $\beta$ and inverse $\beta$ decay. For these two processes, the dependence on the neutrino energy is a well-known result, known as the Fierz interference, that seeds in the chiral flip of the electrons and the interference between the vector- (axial-vector-) and scalar-type (tensor-type) interactions in the LEFT for Fermion (Gamow-Teller) transitions. For $\beta$ decay, the energy dependence arises from the tensor-type interactions in the LEFT and is parameterized as $f_T(E_\nu)$. This form factor is approximated following the procedure described in Ref.\,\cite{Falkowski:2019xoe} and takes a value of $\sim2.0$ in practice. On the other hand, for inverse $\beta$ decay, the energy dependence shows up in the chiral suppression factor $m_e/(E_\nu-\Delta)\simeq m_e/E_e$. For our purpose, we take $E_\nu=3$\,MeV that is consistent with its peak value for reactor type neutrino oscillation experiments. Varying $E_\nu$ around its peak value, we find negligible impact on our conclusion presented in this work.}

We also take into account the NSI-induced matter effects during propagation, which at the QM level are described by the parameters $\epsilon^m_{\alpha\beta}$ in the following Hamiltonian:
\begin{equation}
\label{eq:HNSI}
H = \frac{1}{2E_{\nu}}\left[U
\left(
\begin{array}{ccc}
0 \,\,\, & 0 & 0 \\
0 \,\,\, & \Delta m_{21}^2 & 0\\
0 \,\,\, & 0 & \Delta m_{31}^2 
\end{array}
\right) U^{\dagger}
+ A
\left(
\begin{array}{ccc}
1+\varepsilon_{ee}^m & \epsilon_{e\mu}^m & \,\,\, \epsilon_{e\tau}^m \\
\epsilon_{e\mu}^{m*} & \epsilon_{\mu\mu}^m & \,\,\, \epsilon_{\mu\tau}^m \\
\epsilon_{e\tau}^{m*} & \epsilon_{\mu\tau}^{m*} & \,\,\, \epsilon_{\tau\tau}^m
\end{array}
\right)
\right],
\end{equation}
where $U$ is the Pontecorvo-Maki-Nakagawa-Sakata matrix and $A = \sqrt{2} G_F N_e$, with $G_F$ denoting the Fermi constant and $N_e$ the electron number density in the medium, defines the standard matter potential~\cite{Wolfenstein:1977ue,Mikheev:1986gs}. Similar to eq.~\eqref{eq:CC}, we have the LEFT Lagrangian describing the NC NSI~\cite{Jenkins:2017jig}:
\begin{eqnarray}\label{eq:NC}
\mathcal{L}_{\rm NC} \supset 2\sqrt{2}G_F\left[\epsilon^{fL}_{\alpha\beta}(\bar{\nu}_\alpha \gamma^\mu P_L\nu_\beta)(\bar{f}\gamma_\mu P_L f) +\epsilon^{fR}_{\alpha\beta}(\bar{\nu}_\alpha \gamma^\mu P_L\nu_\beta)(\bar{f}\gamma_\mu P_R f)\right]+h.c.\ ,
\end{eqnarray}
where $f=u,d,e$ for the terrestrial experiments we considere here and Greek letters again represent lepton flavor $e,\mu,\tau$. In principle, other types of the Lorentz structures exist, while only the vector current can generate the neutrino matter effect potential. The relations between QM parameters $\epsilon^m$ and $\epsilon^{fL(R)}$ in eq.~\eqref{eq:NC} are:
\begin{eqnarray}
\epsilon^m_{\alpha\beta} = \sum_f\left\langle\frac{N_f(x)}{N_e(x)}\right\rangle (\epsilon^{fL}_{\alpha\beta} +\epsilon^{fR}_{\alpha\beta}),
\end{eqnarray}
where $N_f(x)$ and $N_e(x)$ are average numbers of electron and fermion $f$ in the matter at the position $x$.

NSIs in neutrino propagation are generally significant in experiments where the neutrinos are exposed to relatively large electron number density $N_e(x)$. The modeling of the matter density profile therefore becomes important in estimating NSI effects in long baseline neutrino oscillation experiments. In most studies, including this work, the matter density profile is computed from the Preliminary Reference Earth Model (PREM)~\cite{Dziewonski:1981xy}. There have been recently several studies in the literature investigating the impact of applying PREM with respect to using more detailed matter density profiles, see e.g. Refs.~\cite{Chatterjee:2018dyd,Kelly:2018kmb,King:2020ydu}. We consider PREM to be sufficient for the scope of the present study.

\section{\label{sec:smeft-to-nsi}Connecting SMEFT to Neutrino NSIs}
Since no new particle have been observed after the discovery of the Higgs, the assumption that new physics living at a higher scale is well-motivated. The SMEFT provides a framework to study the effects of new physics of which the scale is much larger than the electroweak scale.  In our work, we focus on the scenario where neutrino NSIs are indeed originated from some heavy new physics particles such that after integrating them out, their effects can be fully captured by the SMEFT with a new physics scale $\Lambda$.
The SMEFT can be organized by the non-renormalizable operators of different dimensions $D$ that are invariants under the SM gauge groups:
\begin{equation}\label{eq:defsmeft}
    {\cal L}_\text{SMEFT} = \sum_D \sum_{i_D} \frac{c_{i_D}}{\Lambda^{D-4}} {\cal O}_i^{D},
\end{equation}
where $c_{i_D}$ represents the dimensionless Wilson coefficients of the operator $O_i^D$. As the operators is further suppressed with the $1/\Lambda^{D-4}$ factor for a higher dimension $D$, we will focus on the lowest order effects that come from dimension-6 operators.\footnote{We assume that the dimension-5 operator $LLHH$ gives the correct neutrino masses, and consequently the corresponding scale $\Lambda$ is much greater than the TeV scale around which we aim to probe the dimension-6 SMEFT operators in our work.} It is the constraints on these $c_{i_6}/\Lambda^2$'s that we will derive using the next generation neutrino experiments. 

In the previous section we demonstrate the relations between the neutrino NSIs in the LEFT and in the QM formalisms, and the latter one serves as an input in the neutrino oscillation experiment event generator General Long Baseline Experiment Simulator ({\tt GLoBES})~\cite{Huber:2004ka,Huber:2007ji}, which we will use for deriving the $\chi^2$s in the following related sections. Therefore, to complete the goal of deriving the constraints on the Wilson coefficients of the SMEFT, relations between the the NSI parameters in the LEFT and the SMEFT Wilson coefficients are needed. A complete tree-level matching between the SMEFT and the LEFT has been studied in Ref.~\cite{Jenkins:2017jig}, where we will summarize the relevant results for the neutral current NSIs below, and the charged current part has been studied in our previous work~\cite{Du:2020dwr}.

Through the tree-level matching, by setting the Higgs to its vev $v_T$ and integrating out heavy gauge bosons $W^{\pm}$ and $Z$, the $\epsilon^{fL(R)}_{\alpha\beta}$ have the following relations to the Wilson coefficients of the dimension-6 operators:
\begin{eqnarray}
\epsilon^{eL}_{\alpha\beta}&=&C_{\substack{ll \\ \alpha\beta 11}}+C_{\substack{ll \\ 11\alpha\beta }}-\frac{\bar{g}_{2}^{2}}{2 M_{W}^{2}}\left[W_{l}\right]_{\alpha1}\left[W_{l}\right]_{\beta1}^{*}-\frac{\bar{g}_{Z}^{2}}{M_{Z}^{2}}\left[Z_{\nu}\right]_{\alpha\beta}\left[Z_{e_{L}}\right]_{11}\nonumber\\
\epsilon^{u(d)L}_{\alpha\beta}&=&C_{\substack{lq \\ \alpha\beta 11}}^{(1)}+C_{\substack{lq \\ \alpha\beta 11}}^{(3)}-\frac{\bar{g}_{Z}^{2}}{M_{Z}^{2}}\left[Z_{\nu}\right]_{\alpha\beta}\left[Z_{u(d)_{L}}\right]_{11}\nonumber \\
\epsilon^{fR}_{\alpha\beta}&=&C_{\substack{lf \\ \alpha\beta 11}}-\frac{\bar{g}_{Z}^{2}}{M_{Z}^{2}}\left[Z_{\nu}\right]_{\alpha\beta}\left[Z_{f_{R}}\right]_{11},\ (f=e,u,d)\ ,\label{eq:matching}
\end{eqnarray}
where $[W_l]$'s are anomalous couplings between the corresponding fermion currents $\bar\nu_\alpha\gamma^\mu P_L e$ to $W$ bosons, and similarly,  $[Z_\nu]$ and $[Z_{f_{L(R)}}]$ are anomalous couplings between currents $\bar\nu_\alpha\gamma^\mu P_L \nu_\beta$ and $\bar f\gamma^\mu P_L(P_R)f$ and the $Z$ boson. The inclusion of seemingly charged current coupling with $W$ in $\epsilon^{eL}_{\alpha\beta}$ results from the convention for the choices of the LEFT operator basis, where $(\bar\nu_\alpha\gamma^\mu P_L e)(\bar\nu_\beta\gamma^\mu P_L e)$ can be converted to $(\bar\nu_\alpha\gamma^\mu P_L \nu_\beta)(\bar  e \gamma^\mu P_L e)$ using the Fierz identity. 
In terms of Wilson coefficients in the SMEFT, these anomalous couplings can be expressed in the following formulae:\footnote{Comparing to formulas in Ref.~\cite{Jenkins:2017jig}, we have subtracted the SM contributions which are proportional to $\delta_{pr}$, this is why we call them ``anomalous'' couplings.}
\begin{eqnarray}
&&[W_l]_{\alpha\beta}=v_T^2C^{(3)}_{\substack{Hl \\ \alpha\beta}}, \quad [Z_\nu]_{\alpha\beta}=\frac{v_T^2}{2}\left(C^{(3)}_{\substack{Hl \\ \alpha\beta}}-C^{(1)}_{\substack{Hl \\ \alpha\beta}}\right),\\
&&[Z_{e_L}]_{11}=-\frac{v_T^2}{2}\left(C^{(3)}_{\substack{Hl \\ 11}}+C^{(1)}_{\substack{Hl \\ 11}}\right), \quad [Z_{u_L}]_{11}=\frac{v_T^2}{2}\left(C^{(3)}_{\substack{Hq \\ 11}}-C^{(1)}_{\substack{Hq \\ 11}}\right),\\
&& [Z_{d_L}]_{11}=-\frac{v_T^2}{2}\left(C^{(3)}_{\substack{Hq \\ 11}}+C^{(1)}_{\substack{Hq \\ 11}}\right),\quad
[Z_{f_R}]=-\frac{v_T^2}{2}\left(C^{(1)}_{\substack{Hf}}\right).
\end{eqnarray}
The mass of gauge bosons in eq.~\eqref{eq:matching} has the relation to the renormalized gauge couplings $\bar{g}_{1,2}$:
\begin{eqnarray}
\begin{aligned}\label{eq:electroweak1}
M_{W}^{2} &=\frac{1}{4} \bar{g}_{2}^{2} v_{T}^{2}\ , \\
M_{Z}^{2} &=\frac{1}{4}\left(\bar{g}_{2}^{2}+\bar{g}_{1}^{2}\right) v_{T}^{2}\left(1+\frac{1}{2} C_{H D} v_{T}^{2}\right)+\frac{\epsilon}{2} \bar{g}_{1} \bar{g}_{2} v_{T}^{2}\ ,
\end{aligned}
\end{eqnarray}
with $\epsilon$, and $v_T$ defined by:
\begin{eqnarray}
\begin{aligned}\label{eq:electroweak2}
\epsilon &\equiv C_{H W B} v_{T}^{2}\ ,\\
\frac{2}{v_{T}^{2}}&=\frac{4 G_F}{\sqrt{2}}+C_{\substack{ll\\  2112 }}+C_{\substack{ll\\  1221 }}-2 C_{\substack{Hl\\  22 }}^{(3)}-2 C_{\substack{Hl\\  11 }}^{(3)}\ ,
\end{aligned}
\end{eqnarray}
where the Fermi Constant $G_F$ is measured from the muon decay. 
The remaining coupling $\bar{g}_{Z}$ in the eq.~\eqref{eq:matching} along with the electric coupling $\bar{e}$ are related to couplings $\bar{g}_{1,2}$ and Weinberg angle $\bar\theta$ in the following forms:
\begin{eqnarray}
\begin{aligned}
\bar{e} &=\bar{g}_{2} \sin \bar{\theta}-\frac{1}{2} \cos \bar{\theta} \bar{g}_{2} v_{T}^{2} C_{H W B} \\
\bar{g}_{Z} &=\frac{\bar{e}}{\sin \bar{\theta} \cos \bar{\theta}}\left[1+\frac{\bar{g}_{1}^{2}+\bar{g}_{2}^{2}}{2 \bar{g}_{1} \bar{g}_{2}} v_{T}^{2} C_{H W B}\right]
\end{aligned}\label{eq:electroweak3}
\end{eqnarray}
In this sense we treat $G_F$, $M_{W,Z}$ and $\bar e$ as input parameters for our analysis, using eq.~\eqref{eq:electroweak1}-\eqref{eq:electroweak3} to solve for $\bar{g}_{1,2}$ and $\bar{g}_{Z}$, and then put them back to ~\eqref{eq:matching} to fully express the LEFT neutrino NSIs in terms of SMEFT Wilson coefficients. 

\hl{Similar to the $\bar{g}_{1,2}$ and $\bar{g}_Z$, another subtle problem one needs to consider is the value of the quarks Yukawa matrices ${\cal Y}_{f=e,u,d}$, which related to the calibration of the CKM matrix. Let us first specify the convention of gauge eigenstates for fermion fields and the definition of CKM matrix in the presence of SMEFT Wilson coefficients.  The mass matrices for u-type and d-type quarks and charged leptons defined by $\bar{f}_{L,i}M_{f,ij}f_{R,j}+h.c.$ is modified to the following form in the presence of the Wilson coefficients $C_{fH}$ ($f=e,u,d$):
\begin{eqnarray}\label{eq:masstovev}
M_{e}=\frac{v}{\sqrt{2}}\left({\cal Y}_{e}-C_{ eH} \frac{v^{2}}{2}\right), &
M_{u}=\frac{v}{\sqrt{2}}\left({\cal Y}_{u}-C_{ uH} \frac{v^{2}}{2}\right), & M_{d}=\frac{v}{\sqrt{2}}\left({\cal Y}_{d}-C_{dH} \frac{v^{2}}{2}\right),
\end{eqnarray}
which can be diagonalized by the singular value decomposition:
\begin{eqnarray}
M_{e}=U_{e_{L}}\operatorname{diag}\left(m_{e}, m_{\mu}, m_{\tau}\right)U_{e_{R}}^\dagger, \\
M_{u}=U_{u_{L}}\operatorname{diag}\left(m_{u}, m_{c}, m_{t}\right)U_{u_{R}}^\dagger, \\
M_{d}=U_{d_{L}}\operatorname{diag}\left(m_{d}, m_{s}, m_{b}\right)U_{d_{R}}^\dagger.
\end{eqnarray}
One can always rotate the original the gauge eigenstates $Q,L,e_R,u_R,d_R$ with $U_{d_L}$, $U_{e_L}$, $U_{e_R}$, $U_{u_R}$, $U_{d_R}$ in advance such that $M_{e,d}$ are diagonal in the new basis, and $d_L,e_L,e_R,u_R,d_R$ are already in their mass eigenstates, and in the new basis the CKM matrix defined as in the SM by
\begin{equation}\label{eq:defCKM}
V_{\rm CKM}=U_{u_L}^\dagger U_{d_L}=U_{u_L}^\dagger\ \mathbbm{1},
\end{equation}
is still a unitary matrix.
We emphasis that, all the SMEFT Wilson coefficients in this work are defined in this basis where $M_{e,d}$ are already diagonal or referred to the \textit{Waswar Down} basis in the literature. 

From eq.~\eqref{eq:masstovev} to eq.~\eqref{eq:defCKM}, one can find that the determination of the SM Yukawa couplings ${\cal Y}_{e,u,d}$ rely on the values of the $V_{\rm CKM}$ as  input, while the values of the SM Yukawa couplings in turn influence the running of the SMEFT Wilson coefficients and thus alter the prediction for various physical observables. 
Conversely the determination of the values of CKM matrix from the low energy flavor observables are affected in the presence of the SMEFT Wilson coefficients, this can be explicitly seen from the $W$ boson couplings to the charged quark current:
\begin{eqnarray}
-\frac{\bar{g}}{\sqrt{2}} W_{\mu}^{+}\left\{ \left[V_{\rm CKM}\left(\mathbbm{1}+v^{2} C^{(3)}_{H q}\right)\right]_{p r}\overline{u}_{L,p} \gamma^{\mu} d_{L,r}+\left(\frac{v^{2}}{2}  C_{H ud} \right)_{p r} \overline{u}_{R,p}\gamma^\mu d_{R,r}\right\},
\end{eqnarray}
the presence of $C^{(3)}_{H q}$ and $C_{H ud}$ definitely alter the matrix elements for those meson decay processes that relevant to the extraction of the value of $V_{\rm CKM}$. A computationally practical and consistent way to treat the value of the CKM matrix and the predictions for those low energy observables is nicely illustrated in Ref.~\cite{Descotes-Genon:2018foz}, and a philosophically identical algorithm is implemented in the python package \texttt{smelli}~\cite{Aebischer:2018iyb} built on \texttt{wilson}~\cite{Aebischer:2018bkb} and \texttt{flavio}~\cite{Straub:2018kue}.
In the \texttt{smelli}, the tree parameterization of the CKM matrix has been implemented, which takes four free parameters $|V_{us}|$, $|V_{cb}|$, $|V_{ub}|$ and $\delta$ due to the unitarity of $V_{\rm CKM}$  determined by four observables: 
\begin{eqnarray}\label{eq:CKMobs}
{\cal O}^{\rm CKM}_{1,2,3,4}=\Gamma\left(K \rightarrow \mu \nu_{\mu}\right) / \Gamma\left(\pi \rightarrow \mu \nu_{\mu}\right),\ {\rm Br}(B\to X_c e \nu),\ {\rm Br}(B^+\to \tau \nu),\  \Delta M_d/\Delta M_s.\nonumber \\
\end{eqnarray}
For certain input SMEFT Wilson coefficients (in the Warsaw down basis) at the new physics scale $\Lambda$, the \texttt{smelli} will iteratively solve the correct value of $V_{\rm CKM}$, such that the predicted observables ${{\cal O}}^{\rm CKM}_i$ calculated with the help of \texttt{flavio} match the experimental values.

So far we have explained the tree-level matching and subtleties related to the extraction of the SM parameters, e.g. gauge couplings, and Yukawa matrices.}
The above matching procedure should be done at the electroweak scale. When the UV physics scale $\Lambda$ is well above the electroweak scale, one would expect that the running effects in the SMEFT can be important.
\hl{In practice, we fisrt use the \texttt{smelli} to determine the correct CKM matrix for a given SMEFT Wilson coeffcient points and then use the \texttt{wilson} package~\cite{Aebischer:2018bkb} to implement the matching and running procedure for both the SMEFT and the LEFT. However,  the determination of  $V_{\rm CKM}$ in the presence of the SMEFT Wilson coefficients does not always work, because the iterative algorithm implemented in the \texttt{smelli} package needs the new physics contribution to the observables in eq.~\eqref{eq:CKMobs} not too large, otherwise the algorithm breaks down or there is actually no solution to such a SMEFT Wilson coefficient point. Moreover, we believe that even if there exists solution to the $V_{\rm CKM}$ using non-perturbative other than the one in the \texttt{smelli} package, it is very likely with the solved $V_{\rm CKM}$, the theoretical predictions to other observables will deviate from its experimental value too much. Therefore, in this sense the validity of the iterative algorithm would implies a ``perturbativity'' bounds on the Wilson coefficients. Such a bound is sometimes extremely high, for example the operator ${\cal O}^{(8)}_{qd1313}=(\bar{q}_1\gamma^\mu T^A q_3)(\bar{d}_1\gamma^\mu T^A d_3)$ generates the flavor changing neutral current at tree level contributing to $\Delta M_b$, and the validity of the \texttt{smelli} implies an lower bound on the $\Lambda$ around 1466 TeV. 
For this types of operators we find the sensitivity of neutrino experiments cannot compare with the validity bounds form \texttt{smelli} and do not list them in our results. We list part of the operators with validity bounds larger than $20$ TeV in table.~\ref{table:smelli} assuming the dimensionless Wilson coefficients defined in eq.~\eqref{eq:defsmeft} $c_{i_D}=1$.}   Starting from the UV scale $\Lambda$ where the Wilson coefficients are initially defined, the package will first run the SMEFT Wilson coefficients down to the electroweak scale, then perform the matching onto the Wilson coefficients in the LEFT, and further run them down to the 2 GeV scale where we finally match them onto the QM NSIs using the relations in table~\ref{table:matching}. In this sense, we have obtain the numerical functions for the QM NSIs with SMEFT Wilson coefficients and the UV scale $\Lambda$ as inputs, and it is based on this function we build the interface between the packages \texttt{wilson} and \texttt{GLoBES}.

\begin{table}[ht]
\centering 
\begin{tabular}{ccccccc} 
\hline\hline 
Operator & ${\cal O}^{(1)}_{qq_{1313}}$  & ${\cal O}^{(1)}_{qq_{2323}}$  & ${\cal O}^{(3)}_{qq_{1313}}$ & ${\cal O}^{(3)}_{qq_{2323}}$  & ${\cal O}^{}_{dd_{1313}}$  & ${\cal O}^{}_{dd_{2323}}$\\
$\Lambda$ valid (TeV) & $>365$ & $>51$ & $>365$ & $>51$ & $>383$ & $>53$ \\
\hline 
Operator & ${\cal O}^{(1)}_{qd_{1213}}$  & ${\cal O}^{(1)}_{qd_{1313}}$  & ${\cal O}^{(1)}_{qd_{2323}}$ & ${\cal O}^{(8)}_{qd_{1213}}$  & ${\cal O}^{(8)}_{qd_{1313}}$  & ${\cal O}^{(8)}_{dd_{2323}}$\\
$\Lambda$ valid (TeV) & $>23$ & $>1383$ & $>178$ & $>25$ & $>1466$ & $>188$ \\
\hline\hline
\end{tabular}
\caption{List of operators with very high validity bounds from the \texttt{smelli} package. }
\label{table:smelli} 
\end{table}

\section{\label{sec:experiments}Description of the neutrino oscillation experiments}
In the present work, we investigate the prospects of the next-generation neutrino oscillation experiments in searching new interactions. We mainly consider two types of experiments: (1) The long-baseline accelerator experiments Tokai-to-HyperKamiokande (T2HK)~\cite{Abe:2015zbg} in Japan and the Deep Underground Neutrino Experiment (DUNE)~\cite{Acciarri:2016crz} in the United States; (2) as well as the reactor experiment Jiangmen Underground Neutrino Observatory (JUNO)~\cite{An:2015jdp} in China. All three experiments are currently under construction and each will start taking data in the near future. The experiments are simulated with the {\tt GLoBES} package augmented with the calculation of NSI effects~\cite{NuPhys}. In our analysis we take both the charged and the neutral current effects into account in neutrino production, detection and propagation phases. In this section, we describe the simulation and analysis methods of the aforementioned neutrino experiments, and provide their expected sensitivity to the QM-NSI parameters at the end of this section. These results are then translated consistently onto the dimension-6 SMEFT operators with the {\tt wilson} package in section\,\ref{sec:results}.

\subsection{Long-baseline experiment T2HK}
\label{sec:pheno:T2HK}

Tokai-to-HyperKamiokande (T2HK) is the next-generation long-baseline neutrino oscillation experiment currently under construction in Japan. Being a successor to the presently running T2K experiment, T2HK will be an experiment with 295-km-long baseline, an accelerator facility in J-PARC in Tokai and a massive Water Cherenkov neutrino detector in Kamioka. There will also be near and intermediate detectors ND280 and INGRID, of which ND280 is located 280~m from the neutrino source. Both far and near detectors have 2.5$^\circ$ off-axis angle. The T2HK experiment will be commissioned upon the construction of the HyperKamiokande detector, which is envisioned to be a Water Cherenkov vessel with 187~kt of fiducial mass. The JHF beam in J-PARC will be upgraded to 1.3~MW beam power for T2HK~\cite{Abe:2015zbg}. An upgrade for the near detector facility in ND280~\cite{Lux:2019daw} is also underway. There have furthermore been plans to construct a second detector of 187~kt at the far detector site~\cite{Ballett:2016daj}.

Several proposals for the design and the size of the HyperKamiokande detector have been suggested. In this work, we assume the configuration described in Refs.~\cite{Abe:2016tii,Abe:2018uyc}. We assume an operation of 2.5 years in neutrino mode and 3.5 years in antineutrino mode with the first detector, which will be complemented by another 4 years of operating in antineutrino mode with the two-detector setup. We shall therefore assume a 2.5+3.5--year run with single-detector configuration (fiducial mass 187~kt) and a subsequent 0+4--year run with two-detector configuration (fiducial mass 374~kt). The upgraded ND280 facility is assumed to take data with 1,529 kg fiducial mass throughout the 10-year life cycle of the experiment.\footnote{There have also been discussions for adopting the $\nu$PRISM technique in the near detector for the T2HK experiment~\cite{Bhadra:2014oma}. Both detector facilities are located 2.5$^\circ$ off-axis from the neutrino beam. It has also been proposed that a second detector is to be built in South Korea~\cite{Abe:2016ero}. {There are furthermore talks on establishing an additional Water Cherenkov detector of 1~kton mass in the proximity of the T2HK source.}} The 1:3 ratio in the total running time is chosen to balance the difference in neutrino and antineutrino cross sections. This setup has been shown to be optimal in searching \emph{CP} violation in neutrino oscillations as well as in the precision measurement of the neutrino oscillation parameters~\cite{Ballett:2016daj}.

Our simulation of the T2HK experiment is constructed as follows. The starting points for the simulation are the data samples that have been collected in the T2K experiment before July 2020~\cite{Abe:2019vii,Zarnecki:2020yag,T2K:2020lrr}. The expected neutrino oscillation data in the T2HK experiment is simulated by extrapolating the T2K samples assuming a similar setup with 1.3~MW beam power and the fiducial masses and running time as described for the HyperKamiokande detector. The near detector samples from ND280 are also simulated to commensurate the 10-year total running time of T2HK. The data analysis technique is described in detail in Ref.~\cite{Du:2020dwr}.

The simulated T2HK data consists of 9 different samples, of which 5 are collected in HyperKamiokande and 4 in ND280, respectively. The events are categorized into different samples based on their event topology. We include both the electron appearance channels $\nu_\mu \rightarrow \nu_e$ and $\bar{\nu}_\mu \rightarrow \bar{\nu}_e$ and the muon disappearance channels $\nu_\mu \rightarrow \nu_\mu$ and $\bar{\nu}_\mu \rightarrow \bar{\nu}_\mu$ in the analysis of the HyperKamiokande and ND280 data~\cite{T2K:2020lrr}. The $\nu_\mu \rightarrow \nu_e$ events are reconstructed in HyperKamiokande in 23 equi-sized bins in [0.1, 1.25] GeV for charged-current quasi-elastic events (CCQE $\nu_e$ and CCQE $\bar{\nu}_e$), 16 bins in [0.45, 1.25] GeV for charged-current resonant pion production events ($\nu_e$ CC $\pi^+$) and charged-current quasi-elastic events (CCQE $\nu_\mu$ and CCQE $\bar{\nu}_\mu$), respectively.
The muonic samples are divided 28 and 19 even-sized energy bins in the interval [0.2, 3.0] GeV. All HyperKamiokande events are reconstructed at 2.5$^\circ$ off-axis angle 295~km from the source. The data samples of ND280 on the other hand consist of CCQE events of $\nu_\mu \rightarrow \nu_\mu$ and $\bar{\nu}_\mu \rightarrow \bar{\nu}_\mu$, which are reconstructed within 11 and 9 energy bins within [0, 2]~GeV interval. We also consider the electron appearance channels $\nu_\mu \rightarrow \nu_e$ and $\bar{\nu}_\mu \rightarrow \bar{\nu}_e$ in this work, scrutinizing events CCQE $\nu_e$ and CCQE $\bar{\nu}_e$ events within the energy interval [0, 5.5]~GeV with 18 equi-distant bins. As we shall see in section\,\ref{sec:pheno:nsi}, the electron appearance channels reconstructed in the ND280 facility provide the most stringent constraint on the new physics generating $\epsilon_{\mu e}^s$ in pion decay.

\subsection{Long-baseline experiment DUNE}
\label{sec:pheno:DUNE}

The Deep Underground Neutrino Experiment (DUNE)~\cite{Acciarri:2016crz,Acciarri:2015uup,Abi:2018rgm,Acciarri:2016ooe} is another world-class next-generation long-baseline neutrino oscillation experiment that is currently being established in the USA. The civil construction for the experiment site broke ground in 2017 and the operation is expected to begin at about the same time with T2HK in Japan. Taking over from the NuMI beamline presently used in the NO$\nu$A experiment, DUNE is designed to send neutrinos from the Fermilab National Accelerator Laboratory (FNAL) in Batavia, IL. The far detector facility is currently being built at the Sanford Underground Research Facility (SURF) in Homestake, SD, which is located about 1300~km from the source. The experiment will make use of the NuMI beamline which will have undergone PIP-II upgrade at FNAL. The far detector facility of DUNE will host four cryogenic vessels based on the novel Liquid Argon Time Projection Chamber (LArTPC) technology. Each vessel will have fiducial mass of 10~kt, forming a total active mass of 40~kt. DUNE will also have a near detector facility located 547~m from the NuMI beamline at FNAL. The near detector will consist of three different sub-detectors, which will include a liquid argon vessel akin to a LArTPC with 50~t fiducial mass~\cite{AbedAbud:2021hpb}. Further details about the detector configuration can be found in Ref.~\cite{Abi:2020evt}. 

There have been various proposals to configure the experimental setup for DUNE. In our simulation of DUNE, we will follow the configuration that was used in the DUNE Conceptual Design Report (CDR)~\cite{Acciarri:2015uup} as the reference design.\footnote{Recently, the simulation files used to produce the results in the Technical Design Reports were also published~\cite{Abi:2021arg}. In the present work, however, we retain the simulation files used in the Conceptual Design Report~\cite{Alion:2016uaj}, which defines the official configuration for DUNE.} The simulation details for the neutrino beam and far detector setup are given in Ref.~\cite{Alion:2016uaj}. We will also assume a near detector facility adopting the same detector response as in the far detector, but with a total active mass of 58.7~t.\footnote{We note that the total exposure of the three sub-detectors is consistent with 58.7~t fiducial mass of a single LArTPC detector. This number accounts roughly the contributions to the muon neutrino and antineutrino events as are currently expected from the high-pressure gaseous argon (HPgTPC) as well as from the System for on-Axis Neutrino Detection (SAND) sub-system.} It has been noted that a total running time of 7 years, split evenly between the neutrino and antineutrino beam modes, is sufficient for the discovery of mass hierarchy. We hence assume 3.5+3.5 years of running time with both near and far detectors online.

In this study, we consider the neutrino and antineutrino events in the DUNE far and near detectors within the energy range [0, 20]~GeV, which is divided into 64 bins of 0.125~GeV, 2 bins of 1.0~GeV and 5 bins of 2.0~GeV widths, respectively. The data samples are simulated for both near and far detectors corresponding to signal events from $\nu_\mu \rightarrow \nu_\mu$, $\bar{\nu}_\mu \rightarrow \bar{\nu}_\mu$, $\nu_\mu \rightarrow \nu_e$ and $\bar{\nu}_\mu \rightarrow \bar{\nu}_e$ channels with relevant backgrounds.\footnote{There have recently been several studies that have highlighted the importance of tau neutrino samples in DUNE~\cite{deGouvea:2019ozk,Ghoshal:2019pab}. In the present work, we only consider the electron appearance and muon disappearance channels in the near and far detectors of DUNE, disregarding any sensitivity to $\nu_\tau$ and $\bar{\nu}_\tau$.} The full description regarding the background composition as well as the detector response is given in Ref.~\cite{Alion:2016uaj}. Similarly to the T2HK experiment, DUNE is going to be able to probe neutrino NSIs in pion decay as well as in the propagation.

\subsection{Reactor experiment JUNO}
\label{sec:pheno:JUNO}

Jiangmen Underground Neutrino Observatory (JUNO) is the next-generation reactor neutrino experiment currently preparing its operations in Guangdong, China~\cite{An:2015jdp}. The experimental configuration involves a medium-baseline system of 53~km length. The neutrino source will be comprised of nuclear reactors in Yangjiang and Taishan power plant complexes, amounting to a combined \juno{26.6~GW$_{\rm th}$ thermal power}. The neutrino detector will be a single monolith with 20~kton of liquid scintillator. The experiment will aim to establish the mass hierarchy at \juno{3-4~$\sigma$} CL within the first six years and perform precision measurements on the neutrino oscillation parameters in the interplay region of $\Delta_{21}$ and $\Delta_{31}$~\cite{Xing:2018zno,Huber:2019frh}. The experiment will detect neutrinos with 3\%/$\sqrt{E\text(MeV)}$ energy resolution. \juno{The data taking will begin in the near future.} Also planned is the high-resolution near detector facility Taishan Antineutrino Observatory (TAO)~\cite{Abusleme:2020bzt}, which will be located 30~m from Taishan power plant.

In the present work, we simulate the electron antineutrino spectrum as expected for the JUNO detector from the Yangjiang and Taishan nuclear power plants. We also simulate the TAO detector, which is a near detector of \juno{1~t} fiducial mass located next to the Taishan reactor core~\cite{Abusleme:2021zrw}. TAO will measure the reactor neutrino spectrum in order to constrain the source-related systematic uncertainties, giving also an opportunity to search for source-related NSI effects with a very short baseline. The design of TAO follows a gadolinium-doped liquid scintillator with 1.7\%/$\sqrt{E}$ energy resolution, therefore having close resemblance with the detector design of Daya Bay experiment~\cite{An:2012bu}. Both the JUNO and TAO samples enclose inverse beta decay events from $\bar{\nu}_e \rightarrow \bar{\nu}_e$ channel reconstructed within [1.8, 9.0]~MeV energy range, which is divided into a total of 198 bins. The backgrounds of the JUNO detectors comprise of geo-neutrinos, accidental backgrounds, fast neutrons, ($\alpha$, $n$) backgrounds as well as of $^9$Li-$^8$He background. A more complete description of the signal and background composure in JUNO is provided in Ref.~\cite{An:2015jdp}. The same background composition is also assumed for the TAO detector in our simulations\footnote{{ There are in fact certain differences in the JUNO and TAO backgrounds. A more realistic description of TAO backgrounds involve the fast-neutron background as the primary component.}}.

Our simulation of the JUNO experiment adopts the configuration described in Ref.~\cite{An:2015jdp}, with key parameters updated to match Ref.~\cite{Abusleme:2021zrw}. \juno{We assume 6 years of data taking}. A liquid scintillating detector is adopted with TAO characteristics following the details given in Ref.~\cite{Capozzi:2020cxm}. In the latter, only the reactor spectrum from the Taishan nuclear power plant is taken into account.\footnote{We consider that the contribution from the Yangjiang power plant in TAO is of the order of ${\cal O}$(1\%) in terms of NSI sensitivity. It is therefore safe to ignore the reactor neutrinos other than the ones from Taishan in the simulation of TAO spectrum.} We assume a correlated \juno{5\% flux normalization error\footnote{\juno{We adopt 5\% flux error to account for the convoluted uncertainty in the Huber-Mueller model in presence of NSIs and forbidden decays.}}, uncorrelated 2\% flux errors, uncorrelated 1\% detection errors as well as uncorrelated shape errors of 0.8\% for both JUNO and TAO.}

\subsection{Numerical methods and constraints on NSI}
\label{sec:pheno:nsi}

The analysis presented in this work is performed with GLoBES~\cite{Huber:2004ka,Huber:2007ji} complemented with the probability engine for non-standard interactions~\cite{NuPhys}. This section will present the details of the numerical analysis with the NSI sensitivity as an example. The basic details of the simulation profiles are to be provided in this subsection, as well as a table of constraints on the CC and NC NSI.

The simulation details of the considered neutrino experiments T2HK, DUNE and JUNO/TAO are summarized in Table\,\ref{LBLExperiments}. Whereas the long-baseline accelerator experiments T2HK and DUNE will be able to probe neutrino NSIs in the pion decay and in propagation, the reactor experiment JUNO can be used to search neutrino NSIs mainly in beta decay and inverse beta decay.

\begin{table}[!t]
\caption{\label{LBLExperiments} Long-baseline neutrino experiments simulated in this study. The expected number of events are shown for the signal(background) events in each experiment. Each experiment employs either Water Cherenkov (W.C.) and Liquid Argon Time Projection Chamber (LAr) technologies in their near and far detectors. The fiducial mass of T2HK is assumed to double after 2.5+3.5~years. JUNO experiment includes TAO detector with location reported with respect to Taishan.}
\begin{center}
\begin{tabular}{cccc}
\hline\hline
\rule{0pt}{3ex}Experiment & T2HK & DUNE & JUNO/TAO  \\ \hline
\rule{0pt}{3ex}Type & superbeam & superbeam & reactor \\
\rule{0pt}{3ex}Source location & Japan & USA & China  \\ 
\rule{0pt}{3ex}Beam/reactor power & 1.3~MW & 1.07~MW & \juno{26.6~GW$_{\rm th}$} \\ 
\rule{0pt}{3ex}Running time & 2.5+7.5~yrs & 3.5+3.5~yrs & 6~yrs \\ 
\rule{0pt}{3ex}Detector technology & W.C. & LAr & L.Sc. \\ 
\rule{0pt}{3ex}Fiducial mass (far) & 187~kt (374~kt) & 40~kt & 20~kt \\ 
\rule{0pt}{3ex}Fiducial mass (near) & 1.529~t & 67.2~t & \juno{1~t} \\ 
\rule{0pt}{3ex}Baseline length (far) & 295~km & 1300~km & 53~km \\ 
\rule{0pt}{3ex}Baseline length (near) & 280~m & 547~m & \juno{30~m} \\ 
\rule{0pt}{3ex}Off-axis angle & 2.5$^\circ$ & 0$^\circ$ & 0$^\circ$ \\ \hline
\rule{0pt}{3ex}References & Ref.~\cite{Abe:2015zbg} & Ref.~\cite{Acciarri:2016crz} & Ref.~\cite{An:2015jdp} \\ \hline\hline
\end{tabular}
\end{center}
\end{table}

In all considered experiments, the simulated neutrino oscillation data is analysed with $\chi^2$ functions, which span over energy bins $i =$ 1, 2, ... and detectors $d$. In the case of long-baseline experiments, the $\chi^2$ function used in the analysis is
\begin{equation}
\chi^2 = \sum_{d} \left( \sum_{i} 2\left[ T_{i,d} - O_{i,d} \left( 1 + \log\frac{O_{i,d}}{T_{i,d}} \right) \right] + \frac{\zeta_{\text{sg}}^2}{\sigma_{\zeta_{\text{sg}}}^2} + \frac{\zeta_{\text{bg}}^2}{\sigma_{\zeta_{\text{bg}}}^2} \right) + {\rm priors},
\label{LBLChi2}
\end{equation}
where $O_{i,d}$ and $T_{i,d}$ stand for the observed and theoretical (predicted) events in the near and far detectors, which are denoted with {\sl d = N} and {\sl F}, respectively. The systematic uncertainties are addressed with the so-called pull-method~\cite{Fogli:2002pt}. We consider normalization errors for signal and background events with nuisance parameters $\zeta_{\text{sg}}$ and $\zeta_{\text{bg}}$, which influence the predicted events $T_{i,d}$ in near and far detectors with a simple shift: $T_{i,d} = (1+\zeta_{\text{sg}})N^{\text{sg}}_{i,d} + (1+\zeta_{\text{bg}})N^{\text{bg}}_{i,d}$ where $N^{\text{sg}}_{i,d}$ and $N^{\text{bg}}_{i,d}$ are signal and background events, respectively. The priors used in equation\,(\ref{LBLChi2}) are defined as Gaussian distributions of each oscillation parameter.

In the case of the reactor neutrino experiments, it is sufficient to use the $\chi^2$ function
\begin{equation}
\begin{split}
\chi^2 &= \sum_{d} \sum_{i} \frac{(O_{d,i}-T_{d,i} (1 + a_{\rm norm} + \sum_{r} w_{r}^{d} \alpha_{r} + \zeta_d)))^2}{O_{d,i}}\\ &+ \sum_{r} \frac{\alpha_{r}^2}{\sigma_{r}^2} + \sum_{d} \frac{\zeta_{d}^2}{\sigma_{d}^2}  + \frac{a_{\rm norm}^2}{\sigma_{a}^2} + {\rm priors},
\end{split}
\label{ReactorChi2}
\end{equation}
where $d$ runs through all detectors and $r$ through all reactor cores in the reactor experiment. The observed and theoretical events are given for detector $d$ and energy bin $i$ by $O_{d,i}$ and $T_{d,i}$, respectively. The nuisance parameters $\alpha_{r}$ and $\zeta_d$ and their Gaussian widths are assigned to address the systematic uncertainties in the reactors and detectors, respectively. The nuisance parameter $a_{\rm norm}$ is related to the overall normalization error in the reactor rates. Here $w_{r}^{d}$ gives the fractional contribution of the reactor $r$ reactor to the number of events in detector $d$.

In the analysis of the simulated data, we make use of the experimental constraints that have been provided on the standard neutrino oscillation parameters in the previous experiments. The standard three-neutrino oscillations and the oscillation parameters therein are consistently studied in the NuFit collaboration~\cite{NuFIT:5-0}. In the present work, we utilize the recent fit from the world data~\cite{Esteban:2020cvm} performed in the standard parameterization. The central values and standard deviations of the priors used in this work are summarized in Table~\ref{tab:bestfits}, where the fit values are provided assuming normal ordering for the neutrino masses.
\begin{table}[!t]
\caption{\label{tab:bestfits} The best-fit values and 1$\,\sigma$ confidence level (CL) uncertainties in the standard three-neutrino mixing~\cite{Esteban:2020cvm}. The values are shown for normally ordered neutrino masses.}
\begin{center}
\begin{tabular}{ccc}\hline\hline
Parameter & Central value & Input error (1\,$\sigma$~CL) \\ \hline
\rule{0pt}{3ex}$\theta_{12}$ ($^\circ$) & 33.440 & 0.755 \\ 
\rule{0pt}{3ex}$\theta_{13}$ ($^\circ$) & 8.570 & 0.120 \\ 
\rule{0pt}{3ex}$\theta_{23}$ ($^\circ$) & 49.200 & 1.050 \\ 
\rule{0pt}{3ex}$\delta_\text{CP}$ ($^\circ$) & 197.000 & 25.500 \\ 
\rule{0pt}{3ex}$\Delta m_{21}^2$ (10$^{-5}$ eV$^2$) & 7.420 & 0.205 \\ 
\rule{0pt}{3ex}$\Delta m_{31}^2$ (10$^{-3}$ eV$^2$) & 2.517 & 0.027 \\ \hline\hline
\end{tabular}
\end{center}
\end{table}
In all experiments, the simulated data is analysed while applying the prior function
\begin{equation}
    \label{eq:chi2_prior}
    \begin{split}
    \chi^2_{\rm prior} &= \left( \frac{\theta_{12} - (\theta_{12})_0}{\sigma_{(\theta_{12})_0}} \right)^2 + \left( \frac{\theta_{13} - (\theta_{13})_0}{\sigma_{(\theta_{13})_0}} \right)^2 + \left( \frac{\theta_{23} - (\theta_{23})_0}{\sigma_{(\theta_{23})_0}} \right)^2 + \left( \frac{\delta_{CP} - (\delta_{CP})_0}{\sigma_{(\delta_{CP})_0}} \right)^2 \\
    &+ \left( \frac{\Delta m_{21}^2 - (\Delta m_{21}^2)_0}{\sigma_{(\Delta m_{21}^2)_0}} \right)^2 + \left( \frac{\Delta m_{31}^2 - (\Delta m_{31}^2)_0}{\sigma_{(\Delta m_{31}^2)_0}} \right)^2 + \left( \frac{\hat{\rho} - 1}{\sigma_{(\hat{\rho})_0}} \right)^2,
    \end{split}
    \end{equation}
where $\omega$ runs through the standard oscillation parameters $\theta_{12}$, $\theta_{13}$, $\theta_{23}$, $\delta_{CP}$, $\Delta m_{21}^2$ and $\Delta m_{31}^2$ as well as the normalized matter density $\hat{\rho}$, whilst $\omega_0$ and $\sigma_{\omega_0}$ denote the central values and standard deviations pertaining to the a priori constraints. Throughout this study, an average matter density of 5\% uncertainty is assumed as calculated from the corresponding baseline length.

An important part of the analysis is the treatment of the NSI parameters. In principle, SMEFT operators may contribute to one or more source, detection or matter NSI parameter. A naive estimate on the sensitivity expected on the individual NSI parameters can be obtained by computing $\Delta \chi^2 = \chi^2_{\rm SI} - \chi^2_{\rm NSI}$, where $\chi^2_{\rm NSI}$ is computed with the appropriate $\chi^2$ function assuming one NSI parameter to be non-zero and $\chi^2_{\rm SI}$ with all NSI parameters set to zero. By requiring $\Delta \chi^2 = 3.8416$, the sensitivity to NSI parameter can be extracted at 95\% CL. Following this procedure, we calculate the experiment sensitivity at 95\% CL for each neutrino interaction process we consider in the neutrino experiments. The results are summarized in Table\,\ref{tab:nsi-limits} {and illustrate the general sensitivity to the magnitude of each parameter}\footnote{{ We emphasize that the sensitivities presented in Table\,\ref{tab:nsi-limits} are not used directly to derive constraints to SMEFT parameters. In the main analysis of our work, the values of all NSI parameters are computed from SMEFT according to $\Lambda$. Correlations between different NSI parameters and neutrino energies are appropriately taken into account.}}. \juno{The effect of neutrino NSIs in JUNO has also been studied in Refs.~\cite{Li:2014mlo,An:2015jdp}.}

\begin{table}[!t]
\caption{\label{tab:nsi-limits} The expected upper constraints on the neutrino NSI parameters associated with specific interaction processes in long-baseline experiments T2HK\,\cite{Abe:2015zbg} and DUNE\,\cite{Acciarri:2016crz} and reactor experiment JUNO\,\cite{An:2015jdp}. The limits are provided for the combined setup of T2HK and DUNE and for JUNO at 95\% CL assuming normally ordered neutrino masses.}
\begin{center}
\begin{tabular}{lccl}\hline\hline
Process & NSI parameter & Constraint & Experiment(s) \\ \hline
\multirow{3}{*}{Pion decay} & $\left|\epsilon^s_{\mu e}\right|$ & 9$\times$10$^{-6}$ & \multirow{3}{*}{T2HK, DUNE} \\ 
{} & $\left|\epsilon^s_{\mu \mu}\right|$ & 2$\times$10$^{-2}$ & {} \\ 
{} & $\left|\epsilon^s_{\mu \tau}\right|$ & 5$\times$10$^{-2}$ &  {}\\ 
\hline\hline
\multirow{5}{*}{Propagation in matter} & $\left(\epsilon^m_{e e} - \epsilon^m_{\mu \mu}\right)$ & (-0.3, 0.3) & \multirow{5}{*}{T2HK, DUNE} \\ 
{} & $\left(\epsilon^m_{\tau \tau} - \epsilon^m_{\mu \mu}\right)$ & (-0.2, 0.2) & {} \\ 
{} & $\left|\epsilon^m_{e \mu}\right|$ & 2$\times$10$^{-2}$ & {} \\ 
{} & $\left|\epsilon^m_{e \tau}\right|$ & 5$\times$10$^{-2}$ & {} \\ 
{} & $\left|\epsilon^m_{\mu \tau}\right|$ & 2$\times$10$^{-2}$ & {} \\ 
\hline\hline
\multirow{3}{*}{Beta decay} & $\left|\epsilon^s_{ee}\right|$ & \juno{2$\times$10$^{-3}$} & 
\multirow{3}{*}{\juno{JUNO/TAO}} \\ 
{} & $\left|\epsilon^s_{e\mu}\right|$ & \juno{5$\times$10$^{-2}$} &  \\ 
{} & $\left|\epsilon^s_{e\tau}\right|$ & \juno{5$\times$10$^{-2}$} &  \\ 
\hline\hline
\multirow{3}{*}{Inverse beta decay} & $\left|\epsilon^d_{e e}\right|$ & \juno{2$\times$10$^{-3}$} & \multirow{3}{*}{\juno{JUNO/TAO}} \\ 
{} & $\left|\epsilon^d_{\mu e}\right|$ & \juno{4$\times$10$^{-2}$} &  \\ 
{} & $\left|\epsilon^d_{\tau e}\right|$ & \juno{5$\times$10$^{-2}$} &  \\ 
\hline\hline
\end{tabular}
\end{center}
\end{table}

The following remarks can be made on the sensitivity shown in Table\,\ref{tab:nsi-limits}. Owing to its long baseline length, DUNE is expected to provide the most stringent constraints on the matter NSI parameters $\epsilon^m_{\ell \ell'}$, where $\ell$, $\ell' = e$, $\mu$ and $\tau$. The constraints on the diagonal parameters $\epsilon^m_{e e}$, $\epsilon^m_{\mu \mu}$ and $\epsilon^m_{\tau \tau}$ are only shown by the relative differences $\epsilon^m_{e e} - \epsilon^m_{\mu \mu}$ and $\epsilon^m_{\tau \tau} - \epsilon^m_{\mu \mu}$ as neutrino oscillations are insensitive to the absolute scale of diagonal matter NSI. The far detectors of T2HK and DUNE play the dominant part in constraining the matter NSIs through neutrino propagation. { Regarding the source NSI parameters $\epsilon^s_{\ell \ell'}$, the most stringent constraints can be expected from near detector data, which constitutes from DUNE and T2HK near detector data in pion decay and TAO detector data in beta decay. The strictest constraint is obtained on $\epsilon^s_{\mu e}$ thanks to large statistics in T2HK near detector.}

\section{Results from neutrino oscillation experiments}\label{sec:results}

The results from the systematic scan of the new physics originating from dimension-6 SMEFT operators is studied in this section. The primary results consist of the lower constraints to the UV scale $\Lambda$ of SMEFT operators according to the simulated T2HK, DUNE and JUNO data.\footnote{It should be noted that the results are projected only with one operator at the time. In a more usual scenario, UV-complete models entail multiple operators, which could have an adverse effect on the experimental sensitivity to the UV scale $\Lambda$ and the relevant Wilson coefficients~\cite{Du:2020dwr}.} The analysis covers 1635 SMEFT operators in dimension-6 and it is divided into two categories: Whilst the operators associated with noticeable neutral current (NC) neutrino NSIs are reviewed in section\,\ref{sec:results_NC_NSI}, operators contributing to charged current (CC) NSIs are studied in section\,\ref{sec:results_CC_NSI}. We furthermore provide the complete list of the lower boundaries on the UV scale $\Lambda$ from the considered neutrino experiments in Appendix~\ref{app:A}.

\subsection{Operators relevant for NC NSI}
\label{sec:results_NC_NSI}

Neutrino oscillations in matter can be studied in experiments with long baseline lengths, as the neutrinos and antineutrinos can traverse deep in the underground. As the electron number density rises, the impact of the matter NSI parameters $\epsilon^m_{\alpha \beta}$ ($\alpha$, $\beta = e$, $\mu$, $\tau$) is enhanced.  One may therefore expect the most stringent constraint on the matter NSIs arise from DUNE, with sub-leading contributions from T2HK. The medium-baseline reactor experiment JUNO on the other hand is found to yield non-negligible sensitivity, though outclassed by the long-baseline accelerator experiments T2HK and DUNE.

In the present section, we review the dimension-6 SMEFT operators which are associated with matter NSI. We note that there are nearly two hundred operators that lead to non-zero sensitivity in the matter NSI parameters, however, the majority of these parameters can be probed only up to $\Lambda \gtrsim$ 300~GeV in the considered neutrino experiments. We therefore focus on the operators where a lower bound higher than 0.3\,TeV can be achieved in DUNE. To see the full list of tabulated sensitivity, we refer the reader to Appendix\,\ref{app:A}. The lower bounds on $\Lambda$ are calculated by setting the corresponding Wilson coefficient to unity and finding the parameter region where the scale can be excluded by 95\% CL or greater significance. We furthermore consider only one operator at a time.
  \begin{figure}[!t]
        \center{\includegraphics[width=\textwidth]
        {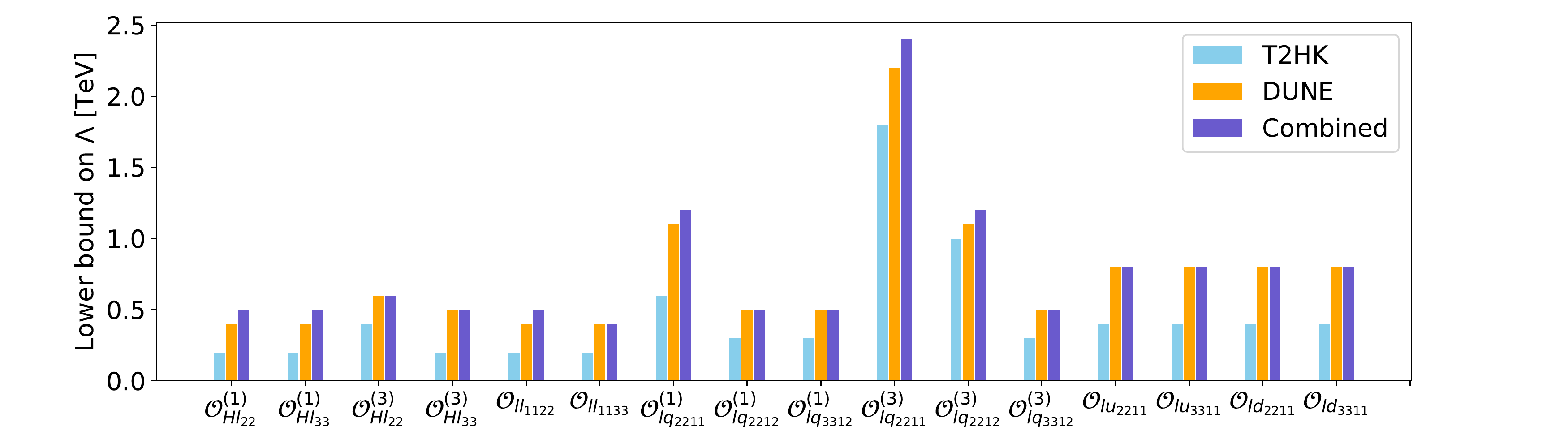}}
        \caption{\label{fig:epsmee} Sensitivity to the operators with notable (\textbf{$\epsilon^m_{ee} - \epsilon^m_{\mu \mu}$}) in the long-baseline experiments T2HK and DUNE. The lower constraints on the UV scale $\Lambda$ are shown while assuming only one operator at a time. The operators selected for this figure were chosen by demanding a sensitivity of more than 300~GeV in DUNE.}
      \end{figure}
      
We begin the review of the expected sensitivity to the UV scale $\Lambda$ in the long-baseline experiments T2HK and DUNE. The SMEFT operators that lead to notable contributions to ($\epsilon^m_{e e} - \epsilon^m_{\mu \mu}$) are shown in figure\,\ref{fig:epsmee} at 95\% CL. The results show sensitivity to $\Lambda$ in the range [0.4,~2.1]\,TeV in the combined analysis of T2HK and DUNE. We find that the T2HK experiment generally yields weaker constraints on $\Lambda$ compared with DUNE, owing to the significantly shorter baseline length compared to DUNE. An exception is formed in the case of ${\cal O}_{lq_{2211}}^{(3)}$, where a stronger constraint on $\Lambda$ is obtained from T2HK with $\Lambda \gtrsim$ 1.8\,TeV at 95\% CL. The difference in sensitivity arises in that case from the simultaneous contribution to $\epsilon^s_{\mu e}$ which can be probed with significantly higher precision with the electron appearance channels in the T2HK near detector. We find that operators ${\cal O}_{{ll}_{1133}}$, ${\cal O}_{{lq}_{3312}^{(1)}}$, ${\cal O}_{{lu}_{2211}}$, ${\cal O}_{{lu}_{3311}}$,  ${\cal O}_{{ld}_{2211}}$ and  ${\cal O}_{{ld}_{3311}}$ are associated with only NC NSIs, whereas other operators also lead to CC NSIs.

    \begin{figure}[!t]
        \center{\includegraphics[width=\textwidth]
        {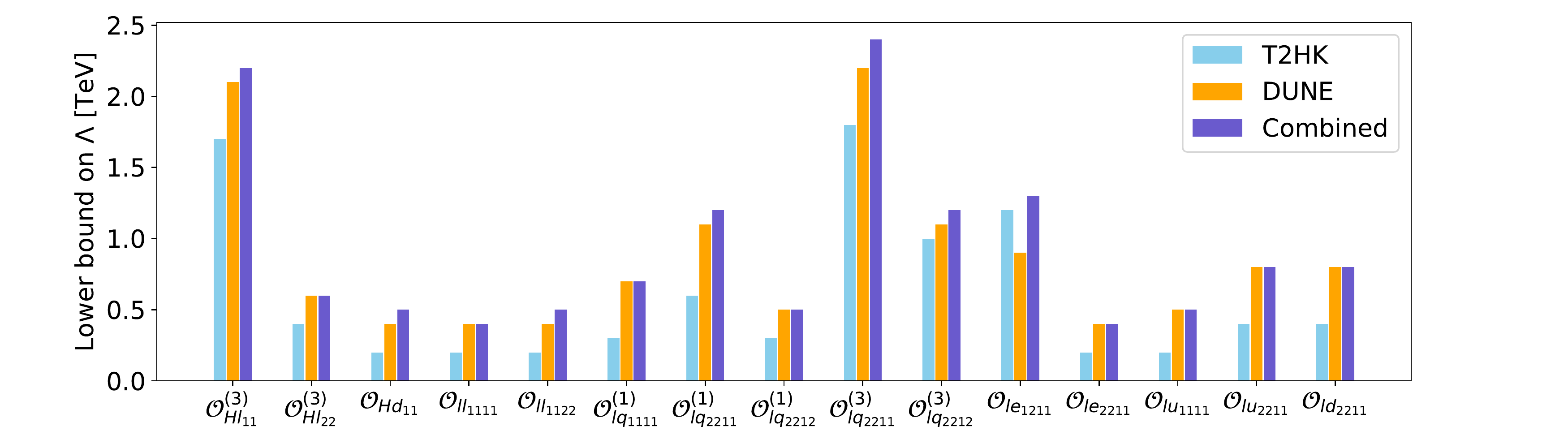}}
        \caption{\label{fig:epsmtautau} Sensitivity to the operators with notable (\textbf{$\epsilon^m_{\tau \tau} - \epsilon^m_{\mu \mu}$}) in the long-baseline experiments T2HK and DUNE. The lower constraints on the UV scale $\Lambda$ are shown while assuming only one operator at a time. The operators selected for this figure were chosen by demanding a sensitivity of more than 300~GeV in DUNE.}
    \end{figure}

The second important quantity that can be probed in the diagonal of the matter NSI matrix is ($\epsilon^m_{\tau \tau} - \epsilon^m_{\mu \mu}$). The operators with significant values in ($\epsilon^m_{\tau \tau} - \epsilon^m_{\mu \mu}$) are shown in figure\,\ref{fig:epsmtautau}. As before, the operators selected for this figure can be constrained by their associated UV scale by $\Lambda \gtrsim$ 300~GeV or better in DUNE. As can be seen in the figure, combined sensitivity in T2HK and DUNE falls between [0.4,~2.1]\,TeV. Again, most operators shown in the figure are most stringently constrained by DUNE, whereas operators ${\cal O}_{{Hl}_{11}}^{(3)}$, ${\cal O}_{{lq}_{2211}}^{(3)}$ and ${\cal O}_{{le}_{1211}}$ are provided with a higher constraint in T2HK due to the non-negligible values in source NSI parameters $\epsilon^s_{\mu e}$. The dominant constraint arises from the far detector of DUNE through ($\epsilon^m_{\tau \tau} - \epsilon^m_{\mu \mu}$), though. Of the considered operators, ${\cal O}_{{le}_{1211}}$, ${\cal O}_{{le}_{2211}}$, ${\cal O}_{{lu}_{1111}}$, ${\cal O}_{{lu}_{2211}}$ and ${\cal O}_{{ld}_{2211}}$ contribute to NC NSIs, whilst others also yield non-negligible effects in CC NSI.
    \begin{figure}[!b]
        \center{\includegraphics[width=\textwidth]
        {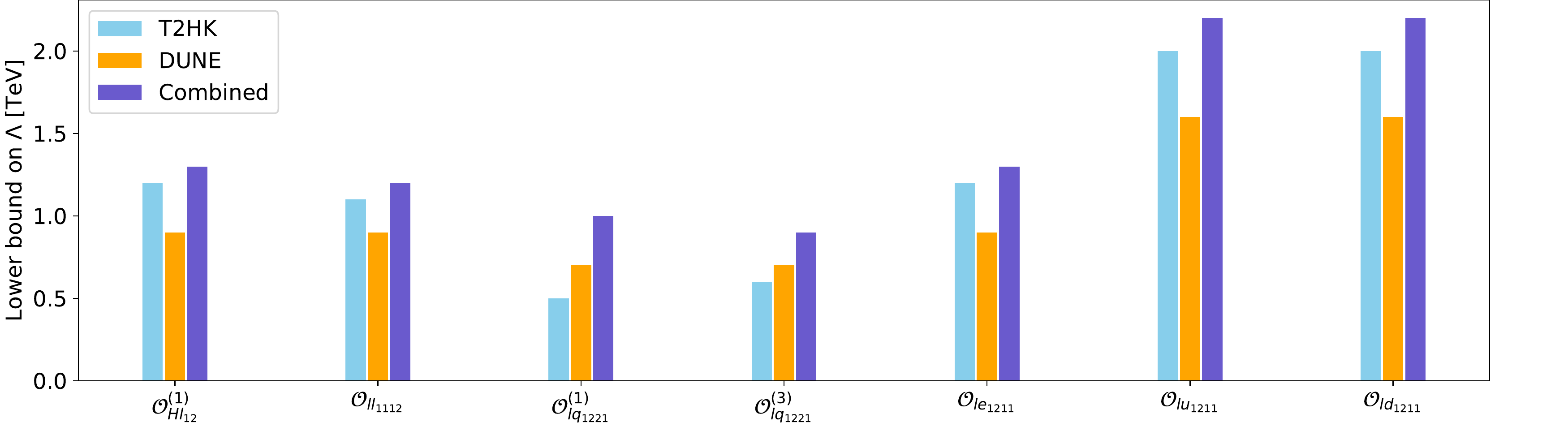}}
        \caption{\label{fig:epsmemu1} Sensitivity to the operators with notable \textbf{$\epsilon^m_{e \mu}$} in the long-baseline experiments T2HK and DUNE. The lower constraints on the UV scale $\Lambda$ are shown while assuming only one operator at a time. The operators selected for this figure were chosen by demanding of more than 300~GeV in DUNE and $\Lambda_{\rm comb} \in$ [0.3,~5.0]\,TeV from the combined sensitivity.}
      \end{figure}
      
\begin{figure}[!t]
        \center{\includegraphics[width=\textwidth]
        {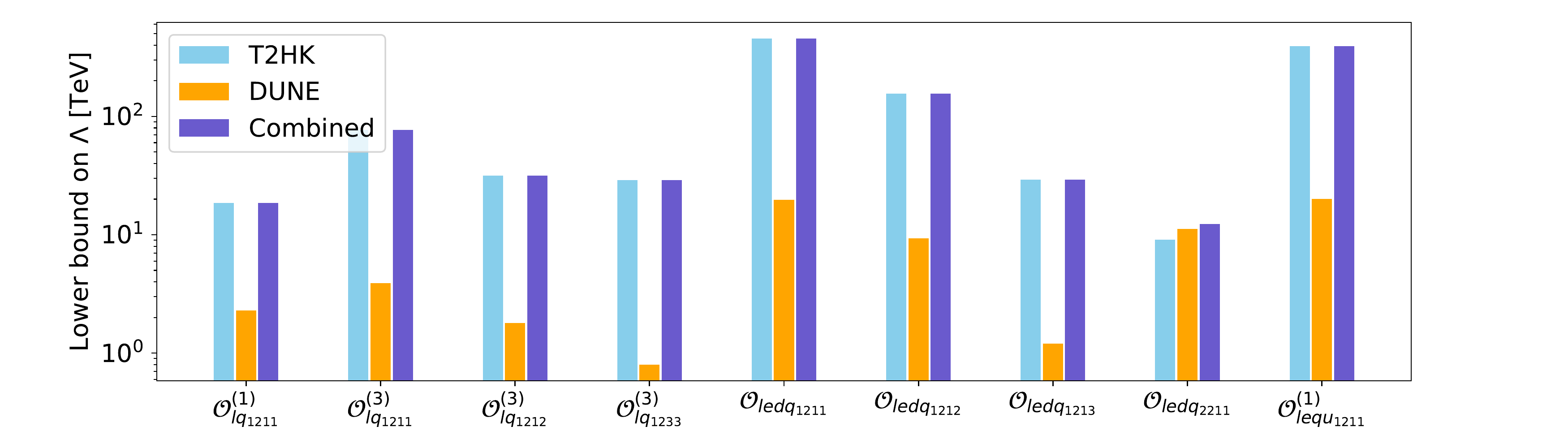}}
        \caption{\label{fig:epsmemu2} Sensitivity to the operators with notable \textbf{$\epsilon^m_{e \mu}$} in the long-baseline experiments T2HK and DUNE. The lower constraints on the UV scale $\Lambda$ are shown while assuming only one operator at a time. The operators selected for this figure were chosen by demanding of more than 300~GeV in DUNE and 10\,TeV from the combined sensitivity.}
      \end{figure}

Long-baseline neutrino experiments T2HK and DUNE are particularly sensitive in the measurement of the off-diagonal matter NSI parameters $\epsilon^m_{e \mu}$, $\epsilon^m_{e \tau}$ and $\epsilon^m_{\mu \tau}$. In figure\,\ref{fig:epsmemu1}, we present the dimension-6 SMEFT operators with link to $\epsilon_{e \mu}^m$. The combined sensitivity to UV scale $\Lambda$ is required to fall between [0.3,~5.0]\,TeV in addition to DUNE sensitivity being greater than 300~GeV. Basing on the results, we find the UV scale associated with the operators to be testable up to $\Lambda_{\rm comb} \in$ [0.9,~2.2]\,TeV, when the data from both T2HK and DUNE is taken into account. In this case, the leading sensitivity is provided by T2HK instead of DUNE, indicating a simultaneous contribution from the source NSI parameters $\epsilon^s_{\mu e}$ and $\epsilon^s_{\mu \mu}$ in pion decay. This behaviour is even stronger in operators listed in figure\,\ref{fig:epsmemu2}, where even the most significant values arising in $\epsilon^m_{e \mu}$ are shadowed by the source NSI parameter $\epsilon^s_{\mu e}$ associated with pion decay. This is particularly true for the operators ${\cal O}_{{ledq}_{1211}}$ and ${\cal O}_{{ledq}_{1212}}$ as well as for ${\cal O}_{{lequ}_{1211}}^{(1)}$ and ${\cal O}_{{lq}_{2211}}^{(3)}$, where the sensitivity belongs to the interval [10,~500]\,TeV. The most stringent constraint from the simulated data is obtained for ${\cal O}_{{ledq}_{1211}}$, which can be probed up to $\Lambda \simeq$ 454.2\,TeV in T2HK and DUNE. {  The impact of electron appearance data in T2HK ND280 is further discussed in appendix\,\ref{app:B}.}

\begin{figure}[!b]
        \center{\includegraphics[width=\textwidth]
        {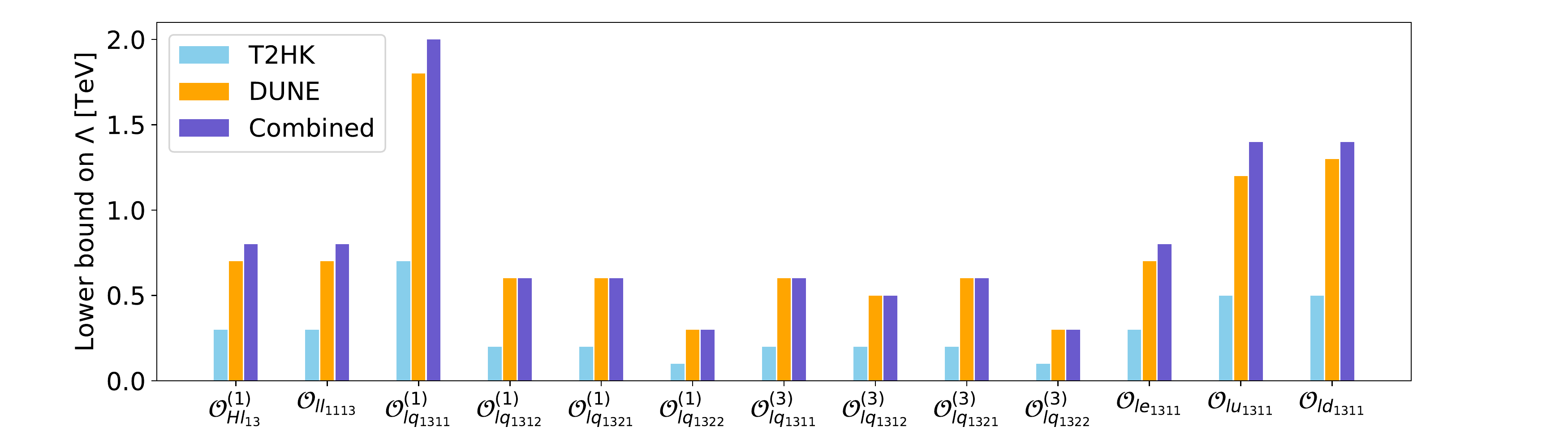}}
        \caption{\label{fig:epsmetau} Sensitivity to the operators with notable \textbf{$\epsilon^m_{e \tau}$} in the long-baseline experiments T2HK and DUNE. The lower constraints on the UV scale $\Lambda$ are shown while assuming only one operator at a time. The operators selected for this figure were chosen by demanding a sensitivity of more than 300~GeV in DUNE.}
      \end{figure}

The second off-diagonal matter NSI parameter that can be probed effectively is $\epsilon_{e \tau}^m$. We present the sensitivity to the UV scale of the operators where notable contributions to $\epsilon_{e \tau}^m$ are found. The corresponding results are shown for T2HK and DUNE at 95\% CL in figure\,\ref{fig:epsmetau}, where the lower bounds of $\Lambda$ achievable for each considered operator are shown. The operators selected for this sample can be probed in the studied long-baseline experiments such that a constraint of $\Lambda \sim$ 300\,TeV or higher can be set after seven years of running with DUNE. This time the highest constraint on the scale of the new physics is set by DUNE in each operator. This behaviour highlights the exceptional sensitivity to $\epsilon^m_{e \tau}$ at the DUNE far detector, which provides a stronger constraint than T2HK. In the same time, the strength of the source NSI parameters $\epsilon^s_{\mu \beta}$ ($\beta = e$, $\mu$, $\tau$) originating from pion decay is found to be relatively small. In the given set of SMEFT operators, DUNE sets the most stringent constraints.

\begin{figure}[!t]
        \center{\includegraphics[width=\textwidth]
        {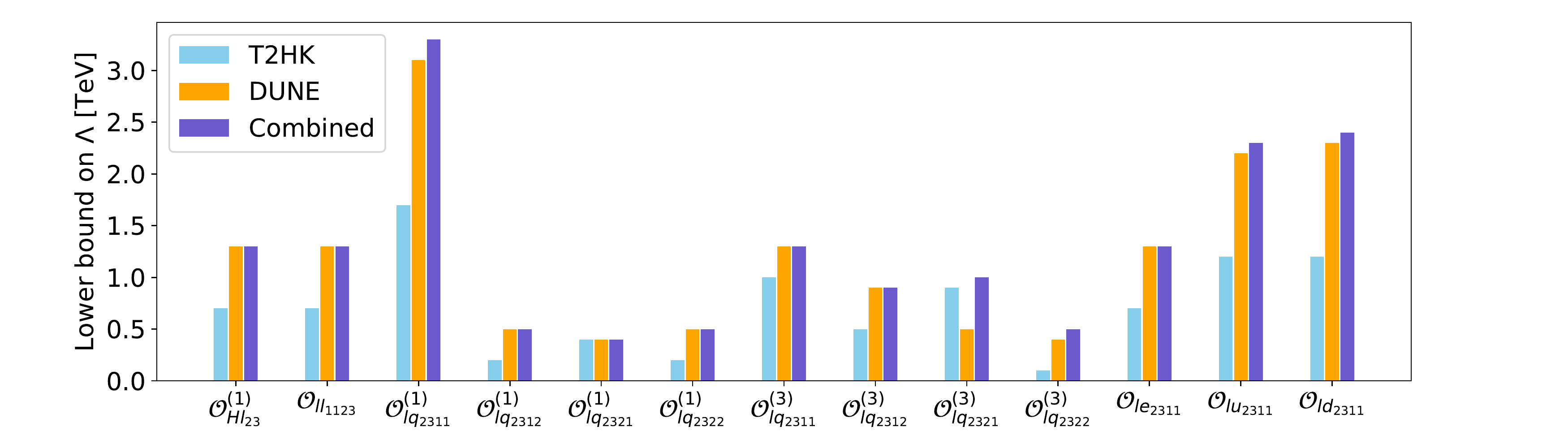}}
        \caption{\label{fig:epsmmutau} Sensitivity to the operators with notable \textbf{$\epsilon^m_{\mu \tau}$} in the long-baseline experiments T2HK and DUNE. The lower constraints on the UV scale $\Lambda$ are shown while assuming only one operator at a time. The operators selected for this figure were chosen by demanding a sensitivity of more than 300~GeV in DUNE.}
      \end{figure}

The third off-diagonal matter NSI parameter to be studied in this work is $\epsilon^m_{\mu \tau}$, which can also be probed with the highest sensitivity set by DUNE. In figure\,\ref{fig:epsmmutau}, we list the operators that can lead to large values of $\epsilon^m_{\mu \tau}$. In this set of operators, the minimum sensitivity to the UV scale expected from DUNE is required to be 300~GeV. The general behavior of the sensitivity implies that the predicted values for $\epsilon^m_{\mu \tau}$ are much higher than those of other neutrino NSI parameters, giving DUNE an advantage in constraining the UV scale of individual operators.

Of the operators presented in figures\,\ref{fig:epsmetau} and \ref{fig:epsmmutau}, we find CC and NC NSIs in every operator except
${\cal O}_{{ll}_{1113}}$, ${\cal O}_{{ll}_{1123}}$, ${\cal O}_{{lq}_{2312}^{(1)}}$, ${\cal O}_{{lq}_{2322}^{(1)}}$, ${\cal O}_{{lq}_{1312}^{(3)}}$, ${\cal O}_{{lq}_{1322}^{(3)}}$, ${\cal O}_{{le}_{1311}}$, ${\cal O}_{{le}_{2311}}$, ${\cal O}_{{lu}_{1311}}$, ${\cal O}_{{lu}_{2311}}$, ${\cal O}_{{ld}_{1311}}$ and ${\cal O}_{{ld}_{2311}}$, which can be associated with only NC NSI.

In this section, we have shown that with several exceptions DUNE yields the most stringent constraints on the dimension-6 SMEFT operators we have considered so far. These operators are associated with relatively large values of the matter NSI parameters, which can be better constrained in DUNE thanks to its baseline length. We have found that T2HK and DUNE can provide sensitivities up to $\Lambda_{\rm comb} \sim$ 3.3\,TeV, with DUNE making the most significant contribution to the overall sensitivity. Furthermore, several operators can be probed up to 454.2\,TeV in these two experiments, as the parallel contribution to the source NSI parameter $\epsilon^s_{\mu e}$ can be simultaneously tested in the T2HK near detector. We therefore find the long-baseline experiments T2HK and DUNE to have synergy in testing the new physics associated with dimension-6 SMEFT operators.

\subsection{Operators relevant for CC NSI}
\label{sec:results_CC_NSI}

The second type of operators we discuss in this study are the ones that lead to significant values of the source and detection NSI parameters $\epsilon^s_{\ell \ell'}$ ($\ell$, $\ell' = e$, $\mu$, $\tau$). As it was discussed in section\,\ref{sec:smeft-to-nsi}, there are mainly three processes in neutrino oscillation experiments where source and detection NSIs can be generated in presence of SMEFT operators: pion decay, beta decay and inverse beta decay. In this section, we investigate the impact of the source NSIs testable in pion decays in long-baseline experiments~\footnote{  In superbeam experiments, detector-related NSIs are still hindered by relatively large systematic uncertainties. As hybrid detectors are developed for T2HK and DUNE to bring systematic uncertainties under control, it deserves a more careful treatment to scrutinize new physics in near detectors. From practical point of view, it is expected to implement detector NSIs within neutrino event generators appropriately in the same SMEFT framework.} and the source and detection NSIs arising from the beta and inverse beta decays respectively in reactor experiments.

\begin{figure}[!t]
        \center{\includegraphics[width=\textwidth]
        {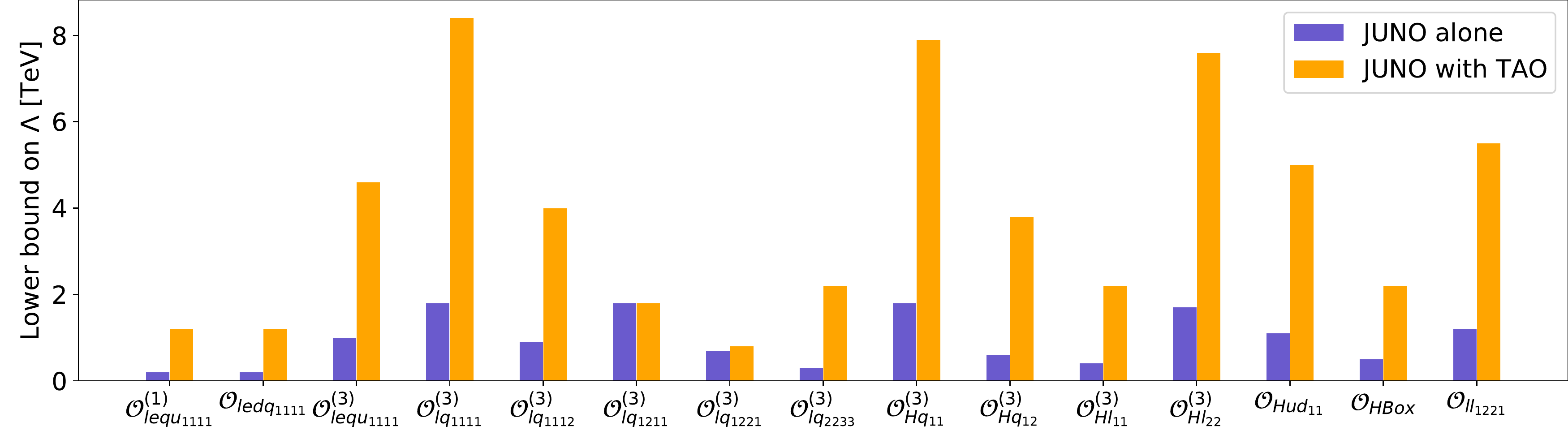}}
        \caption{\label{fig:juno_and_tao} Sensitivity to the dimension-6 SMEFT operators in JUNO. Shown is the sensitivity to the UV-scale $\Lambda$ for individual operators in both JUNO data and the combined data of JUNO and TAO.}
      \end{figure}

The comparison between the sensitivity of JUNO and TAO is shown in figure\,\ref{fig:juno_and_tao}. The results comprise of the lower constraints derived from the JUNO experiment for the UV scale $\Lambda$. Presented are the expected sensitivity in the scenarios where the near detector TAO is taken into account as well as without it. All results are given at 95\% CL. The operators selected for this figure are connected to notable CC NSIs in reactor experiments like JUNO.

The sensitivity shown in figure\,\ref{fig:juno_and_tao} illuminates the power at which JUNO experiment is able to constrain the UV scale of dimension-6 SMEFT operators with notable NSI effects at 95\% CL. The presented results comprise of both the two-detector setup of JUNO as well as the case where simulated data from JUNO detector is analysed alone. We note that the impact of TAO is significant. In case of ${\cal O}_{{lq}_{1111}}^{(3)}$ for example, the inclusion of TAO data increases sensitivity in JUNO experiment from \juno{1.8\,TeV to 8.4\,TeV}. Similar improvements can be found for every operator that can be probed with the JUNO experiment. In figure\,\ref{fig:juno_and_tao}, find NC NSIs with every ${\cal O}_{lq}^{(3)}$ and ${\cal O}_{Hq}^{(3)}$ operator as well as ${\cal O}_{{H\Box}}$ and ${\cal O}_{{ll}_{1221}}$. Other operators can be studied only via CC NSIs.

We finally make a remark on the effect of including multiple operators in the study. In a typical realization of UV-complete theory, there could exist many operators that contribute to neutrino interactions at the same time. Such effects may manifest in neutrino oscillation experiments either by suppressing or enhancing the sensitivity to the UV-scale. While one probes the scale for operator ${\cal O}_{{lequ}_{1111}}^{(1)}$, for example, the presence of a second operator ${\cal O}_{{ledq}_{1111}}$ increases the lower bound from \juno{1.2\,TeV to 1.7\,TeV} in the combined analysis of JUNO and TAO data. Conversely, considering operator ${\cal O}_{{lq}_{1111}}^{(3)}$ together with ${\cal O}_{{Hq}_{11}}^{(3)}$ causes the sensitivity in JUNO to drop from \juno{8.4\,TeV to 2.2\,TeV}. Knowledge of the operators and Wilson coefficients involved in a UV-complete model is therefore a prerequisite for correct evaluation of the sensitivity to its scale in neutrino experiments.

\section{Constraints from the COHERENT experiment}
\label{sec:coherent}

Coherent Elastic neutrino-Nucleus Scattering (CE$\nu$NS) is a very rare process predicted in the SM, which was firstly observed from the COHERENT collaboration using a CsI detector in 2017\,\cite{Akimov:2017ade} and has recently been confirmed by the same group using the LAr detector\,\cite{Akimov:2020pdx}. In this section, utilizing the data publicly available as released in Refs.\,\cite{Akimov:2018vzs,Akimov:2020czh}, we discuss the implications of the CE$\nu$NS signal observed by the COHERENT collaboration on UV physics in the SMEFT framework up to dimension-6. \yong{We also present the comparison between our results obtained in this work with existing ones from previous studies in the end of this section.}

\subsection{Theoretical setup}
Neutrino-nucleus scattering can be parameterized as\,\cite{Freedman:1973yd}
\eqal{\frac{d\sigma_{\nu N}}{dE_R} = \frac{G_F^2M_N}{\pi}\left( 1 - \frac{M_N E_R}{2E_\nu^2} \right) \left( Z g^V_p F_p(q^2) + N g^V_n F_n(q^2) \right)^2,}
with $M_N$ the target nucleus mass, $E_R$ its recoil energy from the scattering, $E_\nu$ the incident neutrino energy, $q^2$ the momentum transfer, and $Z\,(N)$ the proton (neutron) numbers of the nucleus. It fulfills our needs to take the tree-level results for $g_{p(n)}^V$, which takes the following forms, respectively,
\eqal{g_{p}^V = \frac{1}{2}-2\sin^2\theta_W,\quad g_n^V = -\frac{1}{2}.}
$F_{p(n)}$ are the nucleon form factors, for which we will use the Helm parameterization\,\cite{Helm:1956zz} in this work and comment on that in practice, other forms of parameterizing the form factors lead to consistent results.

From the differential scattering rate, the event rate in the $i$-th bin can be computed as follows:
\eqal{N_i = N_N\sum_{\alpha=e,\mu,\bar{\mu}}\int_{E_R^i}^{E_R^{i+1}}\epsilon(E_R)\cdot dE_R\int_{E_\nu^{\rm min}}^{E_\nu^{\rm max}}dE_\nu \frac{d\mathcal{F}_{\nu_\alpha}}{dE_\nu}\frac{d\sigma_{\nu N}}{dE_R},}
where $N_N$ is the total number of nuclei inside the detector, and $\epsilon(E_R)$ is the detector efficiency, for which we use the data released in Refs.\,\cite{Akimov:2018vzs,Akimov:2020czh}. $\mathcal{F}_{\nu_\alpha}$'s are the normalized neutrino flux, for which we will use the analytical approximation from kinematics as presented in Ref.\,\cite{Denton:2020hop}:\footnote{As will see below and also pointed out in Ref.\,\cite{Denton:2020hop}, this approximation leads to agreement with the more realistic simulation of the flux from Spallation Neutrino Source at the Oak Ridge National Laboratory.}
\eqal{\frac{d\mathcal{F}_{\nu_e}}{dE_\nu}&=\frac{fN_{\rm POT}}{4\pi L^2}\cdot\frac{192E_\nu^2}{m_\mu^3}\cdot\left(\frac{1}{2} -\frac{E_\nu}{m_\mu} \right),\\
\frac{\mathcal{F}_{{\bar{\nu}_\mu}}}{dE_\nu}&=\frac{fN_{\rm POT}}{4\pi L^2}\cdot\frac{64E_\nu^2}{m_\mu^3}\cdot\left(E_\nu -\frac{m_\pi^2-m_\mu^2}{2m_\pi} \right),\\
\frac{\mathcal{F}_{{\nu_\mu}}}{dE_\nu}&=\frac{fN_{\rm POT}}{4\pi L^2}\cdot\delta\left(E_\nu -\frac{m_\pi^2-m_\mu^2}{2m_\pi} \right),}
with $m_{\pi,\mu}$ the masses of pions and muons, $f$ the number of neutrinos produced for each proton on target (POT), $N_{\rm POT}$ the total number of POT, and $L$ the distance between the source and detector. The lower and the upper bounds of $E_\nu$ can be determined from the nuclear recoil energy $E_R$ and from kinematics of the decay chain $\pi^+\to\mu^+(\to e^+\nu_e\bar{\nu}_\mu)\nu_\mu$, respectively, which read
\eqal{E_\nu^{\rm min} = \sqrt{\frac{M_N E_R}{2}},\quad E_\nu^{\rm max} = \frac{m_\mu}{2}\simeq52.8\,\rm MeV.}

\begin{table}[!t]
\caption{\label{tab:cevnsconfiguration} Summary of the configurations of the COHERENT experiments with the CsI and the LAr detectors\,\cite{Akimov:2017ade,Akimov:2018vzs,Akimov:2020czh,Akimov:2020pdx}, and uncertainties for the nuisance parameters in our analysis.}
\centering{
\begin{adjustbox}{max width = \textwidth}
\begin{tabular}{c|c|c|c|c|c|c|c|c|c}
\hline\hline
Configurations  & L (cm)  & fiducial mass (kg) & $N_{\rm POT}$  & f & $N_{\rm meas.}$ & $N_{\rm bkg.}$ & $N_{\rm s.s.}$ & $\sigma_\alpha$ & $\sigma_\beta$\\ 
\hline
CsI & 1930 & 14.6 & $5\times10^{20}$ & 0.08 & 142 & 6 & 405 & 0.28 & 0.25 \\
\hline
LAr (analysis A) & 2750 & 24 & $1.37\times10^{23}$ & 0.09 & 159 & 563 & 3131 & 0.052 & 0.067\\
LAr (analysis B) & 2750 & 24 & $1.37\times10^{23}$ & 0.09 & 121 & 222 & 1112 & 0.070 & 0.107\\
\hline\hline
\end{tabular}
\end{adjustbox}
}
\end{table}

\begin{figure}[!t]
\center{\includegraphics[width=\textwidth]{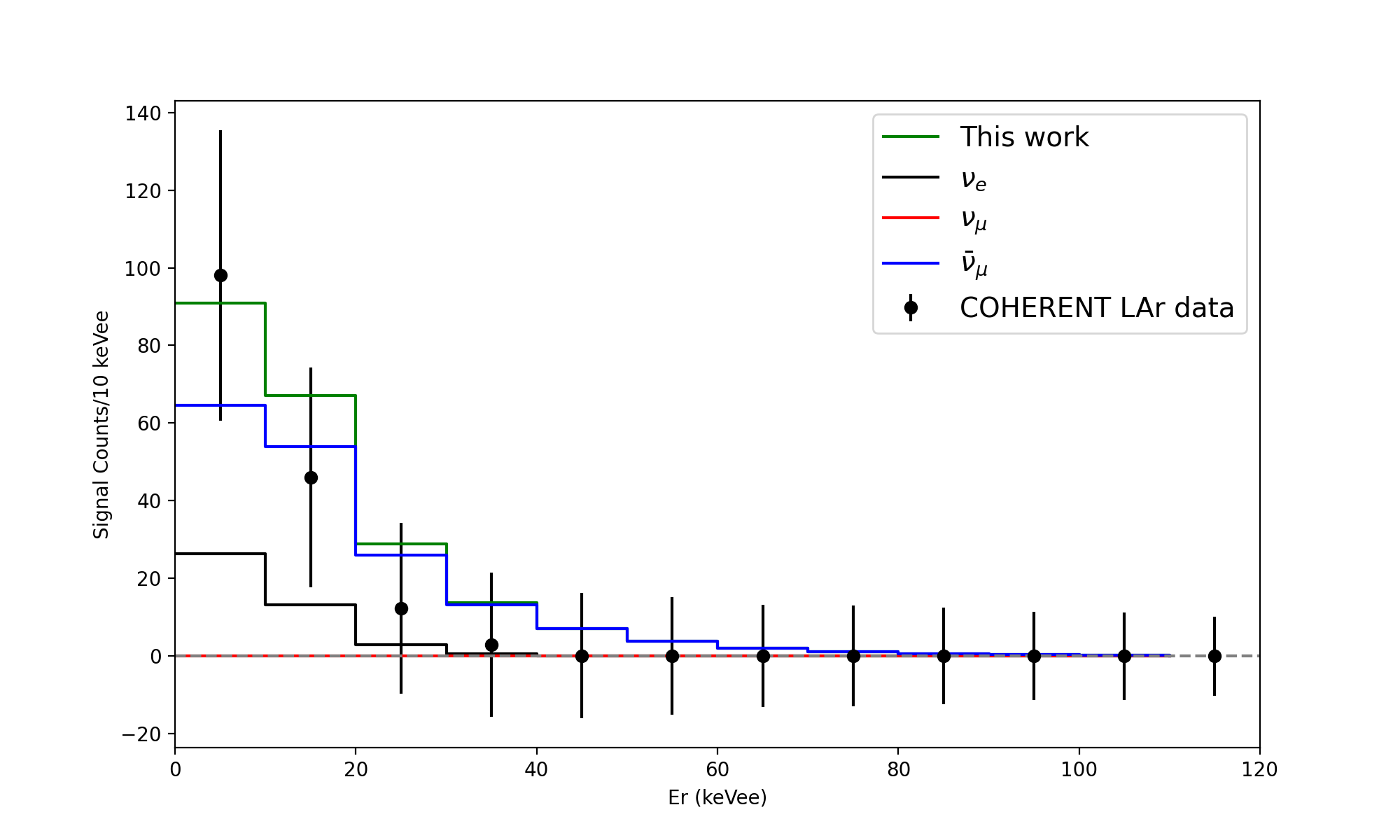}}
\caption{\label{fig:coherent} Reproduction of the COHERENT result with the liquid Argon (LAr) target. The black dots represent the results reported in Ref.\,\cite{Akimov:2020czh,Akimov:2020pdx}, the green one is the total number of events from our theoretical calculation, whose three components are given by the black, the red and the blue, respectively.}
\end{figure}

The experimental configurations for both the CsI and the LAr detectors are summarized in table\,\ref{tab:cevnsconfiguration} based on Refs.\,\cite{Akimov:2017ade,Akimov:2018vzs,Akimov:2020czh,Akimov:2020pdx}, with which the number of theoretical prediction of CE$\nu$NS signal in each bin over the electron-equivalent energy is obtained and shown in figure\,\ref{fig:coherent} for the LAr detector. The black dots are the observed number of events as reported in Refs.\,\cite{Akimov:2020czh,Akimov:2020pdx}. The red, the blue and the black curves represent the predicted number of events in each bin with an incident $\nu_\mu$, $\bar{\nu}_\mu$ and $\nu_e$ neutrino flavor respectively, together with the green curve as the sum of them in each bin. Excellent agreement with experimental results is obtained as seen from the figure.

\subsection{Constraints on SMEFT operators}

Since dimension-6 SMEFT operators would induce NC neutrino NSIs in the low energy regime, neutrino-nucleus scattering could also be affected by these NSIs, which would in turn modify the predicted number of events in each bin. We then ask ourselves the following question: With the reported results on the CE$\nu$NS events, how could we infer their implications on the potential UV physics model independently? To answer this question, we will perform a simple single-bin $\chi^2$ analysis defined as follows:
\eqal{\chi^2=\left(\frac{N_{\rm meas.}-N_{\rm th.}(1+\alpha)-N_{\rm bkg.}(1+\beta)}{\sigma_{\rm stat}}\right)^2 + \left(\frac{\alpha}{\sigma_\alpha}\right)^2 + \left(\frac{\beta}{\sigma_\beta}\right)^2,}
where $N_{\rm meas.}$ is the measured number of events, $N_{\rm th.}$ is the theoretical prediction, and $N_{\rm bkg.}$ the number of beam-on background events. $\sigma_{\rm stat.}=\sqrt{N_{\rm meas.}+N_{\rm bkg.}+2N_{s.s.}}$ represents the statistical uncertainty with $N_{s.s.}$ the number of steady-state background events. To obtain constraints on the UV physics at 95\% C.L., we minimize the $\chi^2$ function over the nuisance parameters $\alpha$ and $\beta$, whose uncertainties are denoted by $\sigma_\alpha$ and $\sigma_\beta$ respectively, and require $\Delta\chi^2\equiv\left|\chi^2-\left(\chi^2\right)_{\rm min.}^{\rm SM}\right|>3.84$.\footnote{Theoretically, NSIs could also modify the flux of incident neutrino beams. In our analysis, we assume this effect to be fully captured by the statistical uncertainties. {  A more detailed analysis of such effects can be seen in Ref.~\cite{Khan:2021wzy}, for example.}}

\begin{figure}[!t]
\center{\includegraphics[width=\textwidth]{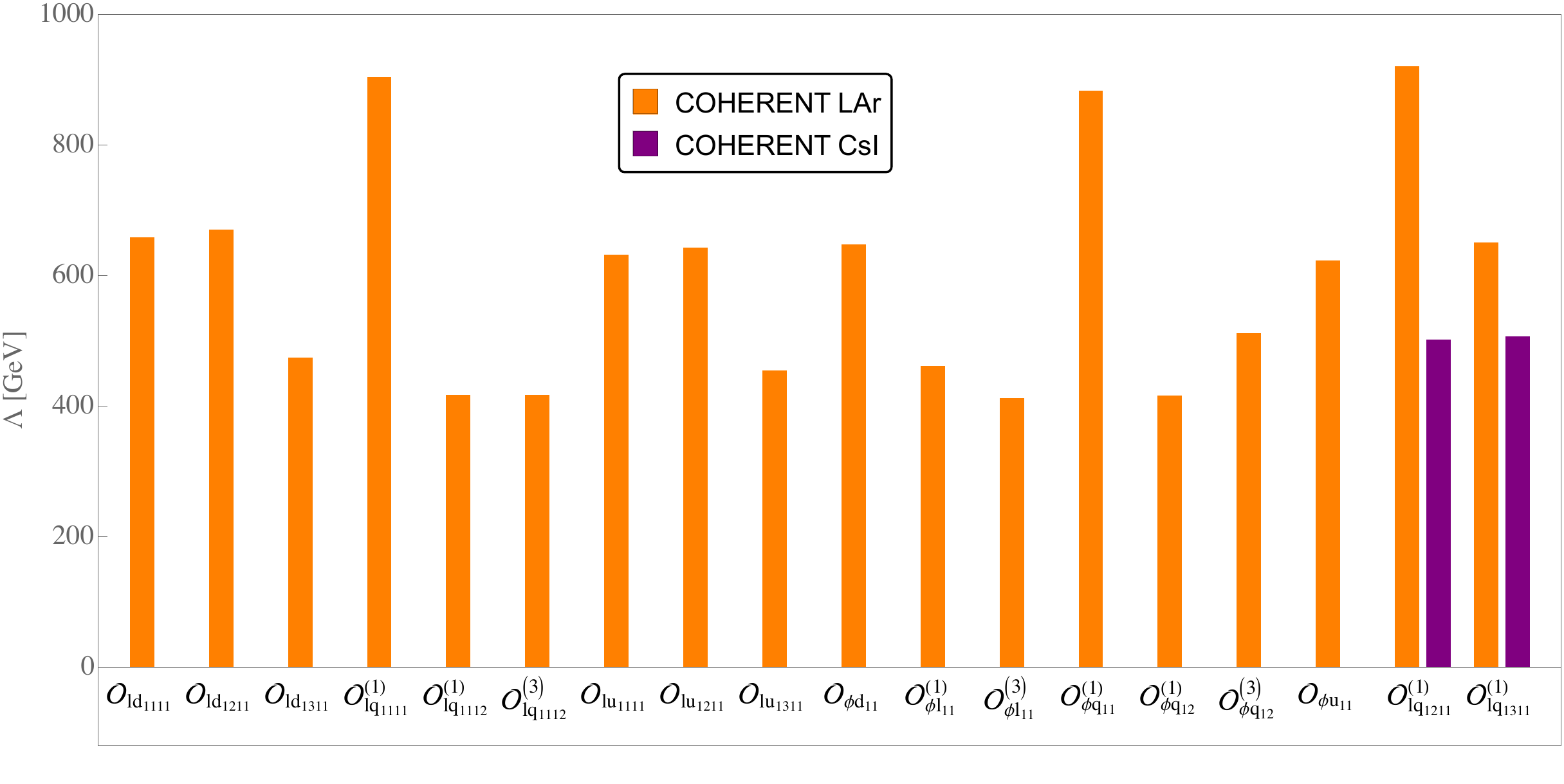}}
\caption{\label{fig:coherentuv} Constraints on the dimension-6 SMEFT operators from the COHERENT experiment. The purple is obtained using the results reported in Refs.\,\cite{Akimov:2017ade,Akimov:2018vzs} with the CsI detector, and the orange in Refs.\,\cite{Akimov:2020czh,Akimov:2020pdx} with the LAr detector.}
\end{figure}

\begin{table}[!t]
\caption{\label{tab:operator-6-cevns} Lower bounds on dimension-6 SMEFT operators at 95\% CL from the COHERENT experiment by fixing the dimensionless Wilson coefficients at one. Numbers in the second and fourth columns are obtained for the LAr detector, while numbers in parenthesis are for the CsI detector.}
\centering{
\begin{tabular}{cc||cc}\hline\hline
Operators & Lower bounds on $\Lambda$\,(TeV) & Operators & Lower bounds on $\Lambda$\,(TeV) \\ \hline
${\cal O}_{{ld}_{1111}}$ & 0.659 & ${\cal O}_{{\phi d}_{11}}$ & 0.647 \\
${\cal O}_{{ld}_{1211}}$ & 0.671 & ${\cal O}_{{\phi l}_{11}}^{(1)}$ & 0.461\\
${\cal O}_{{ld}_{1311}}$ & 0.474 & ${\cal O}_{{\phi l}_{11}}^{(3)}$ & 0.412\\
${\cal O}_{{lq}_{1111}}^{(1)}$ & 0.904$^{*}$ & ${\cal O}_{{\phi q}_{11}}^{(1)}$ & 0.883$^{*}$\tablefootnote{For the ${\cal O}_{{lq}_{1111}}^{(1)}$ and the ${\cal O}_{{\phi q}_{11}}^{(1)}$ operators, we find $\Lambda<0.460$\,TeV and $\Lambda<0.453$\,TeV, respectively, would be free from any COHERENT constraints due to cancellation among neutrino NSIs in the LEFT.} \\
${\cal O}_{{lq}_{1112}}^{(1)}$ & 0.417 & ${\cal O}_{{\phi q}_{12}}^{(1)}$ & 0.416\\
${\cal O}_{{lq}_{1112}}^{(3)}$ & 0.417 & ${\cal O}_{{\phi q}_{12}}^{(3)}$ & 0.512\\
${\cal O}_{{lu}_{1111}}$ & 0.632 & ${\cal O}_{{\phi u}_{11}}$ & 0.623\\
${\cal O}_{{lu}_{1211}}$ & 0.643 & ${\cal O}_{{l q}_{1211}}^{(1)}$ & 0.921 (0.502)\\
${\cal O}_{{lu}_{1311}}$ & 0.455 & ${\cal O}_{{l q}_{1311}}^{(1)}$ & 0.651 (0.507)\\
\hline\hline
\end{tabular}}
\end{table}

We summarize all the parameters in our analysis in table\,\ref{tab:cevnsconfiguration} for both the CsI and the LAr detectors, and report our results in figure\,\ref{fig:coherentuv}.\footnote{We find consistent constraints on the dimension-6 SMEFT operators by following either analysis A or analysis B in Ref.\,\cite{Akimov:2020pdx}.} The results are obtained by fixing the dimensionless Wilson coefficients at one and considering only one dimension-6 SMEFT operator to be non-zero at a time. Constraints on those SMEFT operators are shown in purple for the CsI target and in orange for the LAr target. After scanning over the 1635 SMEFT operators in the Warsaw basis, we find that only two of them, $\mathcal{O}_{lq_{1211}}^{(1)}$ and $\mathcal{O}_{lq_{1311}}^{(1)}$, could be constrained with the CsI detector, whose lower bounds on $\Lambda$ are about 502\,GeV and 507\,GeV respectively. Using the latest result reported in Refs.\,\cite{Akimov:2020czh,Akimov:2020pdx} with the LAr detector, the lower bounds on these two operators are improved to about 920\,GeV and 651\,GeV respectively. Furthermore, the number of constrained dimension-6 SMEFT operators is now improved to eighteen, with the strongest bounds obtained for the $\mathcal{O}_{lq_{1211}}^{(1)}$, $\mathcal{O}_{lq_{1111}}^{(1)}$ and $\mathcal{O}_{\phi q_{11}}^{(1)}$ operators around 900\,GeV. The quantitative results are also summarized in table\,\ref{tab:operator-6-cevns}.

We also comment on that, though we only consider one SMEFT operator to dominate at the UV scale, at the low energy scale where neutrino oscillation experiments are performed, this single SMEFT operator could induce multiple neutrino NSIs in the LEFT. These induced neutrino NSIs could interfere with each other either constructively or destructively, leading to either an enhancement or a suppression in the predicted number of CE$\nu$NS events. The ${\cal O}_{{\phi q}_{11}}^{(1)}$ operator for example. In addition, the induced neutrino NSIs could also vary non-trivially with the UV scale $\Lambda$ through renormalization group running that correlates different SMEFT operators through the anomalous dimension matrix. For instance, the ${\cal O}_{{lq}_{1111}}^{(1)}$ operators. For this reason, during our analysis for the COHERENT experiment, instead of considering only one non-vanishing neutrino NSIs at the low energy scale, we take all the induced neutrino NSIs from some SMEFT operator into account to obtain the results presented in figure\,\ref{fig:coherentuv} and summarized in table\,\ref{tab:operator-6-cevns}.

\yong{We also briefly comment on the comparison between our results from this work and those obtained from a global fitting as presented in \cite{Falkowski:2014tna,Efrati:2015eaa,Falkowski:2015krw,Falkowski:2017pss} and references therein. In the recent work\,\cite{Falkowski:2017pss}, the authors performed a global fitting of 4-fermion operators using data collected at $e^+e^-$ colliders and various low-energy precision measurements. Due to a larger set of precision observables used in their analysis, the constraints presented in Ref.~\cite{Falkowski:2017pss} are evidently stronger than the ones shown in this work. However, beyond those 4-fermion operators, constraints from the global fitting are still absent or at least exhibit some flat directions that need to be lifted with the inclusion of more precision observables. This would be the place where neutrino oscillation experiments could play a non-trivial role in investigating new physics model independently as presented in this work. Furthermore, one can already foresee the potential of a global fitting on SMEFT operators with the implementation of data from various precision experiments, including the neutrino oscillation ones, in the near future.}

\section{Conclusions}\label{sec:concl}
In this work, we have mapped the entire physical potential of the next-generation terrestrial neutrino oscillation experiments T2HK, DUNE and JUNO in probing neutrino NSIs from pion decay, beta decay and inverse beta decay as well as neutrino propagation in matter. Taking into account both charged and neutral current NSIs, the simulated neutrino oscillation data has been analysed in the SMEFT framework. The main results of this work consist of the complete set of lower bounds on the UV-scale $\Lambda$ arising from dimension-6 SMEFT operators. The projections on $\Lambda$ are presented at 95\% CL for T2HK, DUNE and JUNO complemented with TAO in figures\,\ref{fig:epsmee}-\ref{fig:juno_and_tao} for the operators with most influential contributions, as well as a full summary of all relevant operators pictorially presented and quantitatively summarized in Appendix\,\ref{app:A}.

We find the long-baseline accelerator experiments T2HK and DUNE able to constrain the UV scale $\Lambda$ from below at ${\sim \cal O}$(1) TeV level for the operators that can be associated with significant NSIs in neutrino propagation. The far detectors play a dominant role in constraining the parameters depicting matter NSI effects. DUNE is found to have the highest sensitivity due to its baseline length. Long-baseline experiments can furthermore probe operators with notable source NSIs in pion decay. We find the near detectors of T2HK and DUNE to have a significant potential in this regard. An important detail is the expected sensitivity to the NSI parameter $|\epsilon_{\mu e}^s|$, where the simulated electron neutrino samples from the ND280 detector lead to a superior sensitivity in T2HK. We find that operators linked to relatively large values of $|\epsilon_{\mu e}^s|$ can be constrained up to ${\cal O}$(100) TeV when the electron neutrino samples from ND280 are taken into account. Examples of such operators are ${\cal O}_{{ledq}_{1211}}$, ${\cal O}_{{ledq}_{1212}}$, ${\cal O}_{{lequ}_{1211}}^{(1)}$ and ${\cal O}_{{lequ}_{1211}}^{(3)}$.

Reactor experiment JUNO shall be able to probe operators which are mainly different from those that are accessible in T2HK and DUNE. TAO detector plays a crucial role in constraining the source and detection NSIs involved in the beta decay and inverse beta decay, respectively. The best sensitivity is achieved for the operator ${\cal O}_{{lq}_{1111}}^{(3)}$, where the UV scale can be constrained up to \juno{5.8\,TeV} at 95\% CL in JUNO detector. This limit is pushed to \juno{8.4\,TeV} when the simulated data from TAO is included in the analysis.\footnote{Medium-baseline oscillations in JUNO can also be used to study operators that lead to NSIs in neutrino propagation. While such operators were not in the focus of our study regarding the JUNO experiment, neutral current NSI effects are taken into account in all projections for JUNO and TAO.}

The very rare CE$\nu$NS process, observed by the COHERENT collaboration with both CsI and LAr detectors, is also taken into account in our analysis. The main results are shown in figure\,\ref{fig:coherentuv} and summarized in table\,\ref{tab:operator-6-cevns}. The results are obtained by directly utilizing the public data released in Refs.\,\cite{Akimov:2018vzs,Akimov:2020czh}. We find that with the CsI detector, only two dimension-6 SMEFT operators are bounded to be above $\sim$500\,GeV. With the LAr detector, the number of constrained operators would be increased to eighteen with lower bounds between $\sim$400\,GeV and 900\,GeV. With the accumulation of statistics, we expect the prospects in COHERENT to improve in the future.

Last but not least, we comment on that, though above-mentioned experiments are sensitive to new physics that can differ by orders of magnitude, they are complementary to each other in exploring different subsets of SMEFT operators. Our findings also exemplify the necessity of exploiting various experiments and their complementarity in searching for new physics.

\acknowledgments{
\juno{We are grateful to the JUNO internal review for the valuable comments on the manuscript.} YD thanks Kate Scholberg for helpful discussion and acknowledges computing support from the HPC Cluster of ITP-CAS. JHY was supported by the National Key Research and Development Program of China under Grant No. 2020YFC2201501. JT and SV were supported in part by National Natural Science Foundation of China under Grant No. 12075326 and No. 11881240247, by Guangdong Basic and Applied Basic Research Foundation under Grant No. 2019A1515012216. DY, HLL, and JHY were supported by the National Natural Science Foundation of China (NSFC) under Grants No. 12022514 and No. 11875003. JHY was also supported by the National Natural Science Foundation of China (NSFC) under Grant No. 12047503. SV was additionally supported by China Postdoctoral Science Foundation under Grant No.\,2020M672930. JT is grateful for the support from the CAS Center for Excellence in Particle Physics (CCEPP).  }

\appendix
\section{\label{app:A}A full list of constrained dimension-6 SMEFT operators}
In this section, we provides the complete list of the lower bounds obtained in this study for the UV scale $\Lambda$ associated with individual dimension-6 SMEFT operators. We consider constraints that can be expected at 95\% CL in the next-generation neutrino oscillation experiments. The sensitivity is presented for the long-baseline experiments T2HK and DUNE considering data from both near and far detectors, as well as for the medium-baseline reactor experiment JUNO augmented with short-baseline data from TAO. The simulation and analysis of the experiment data is detailed in section\,\ref{sec:experiments}. This review covers all 1635 dimension-6 SMEFT operators.

The projections are enclosed in figures\,\ref{CCAndNC1}-\ref{NC3} for dominant operators\footnote{Plots shown below, as well as some additional plots, are also available at \url{https://github.com/Yong-Du/SMEFT_NSIs}.} and quantitatively summarized for all the dimension-6 SMEFT operators in tables\,\ref{tab:operator-constraints-2}--\ref{tab:operator-constraints-6}. Presented are the lower constraints obtained from the simulation of the experiments T2HK, DUNE and JUNO as well as from the combined analysis of T2HK and DUNE. The results are presented at 95\% CL. The sensitivity was derived by considering the neutrino NSIs in pion decay, beta decay, inverse beta decay and neutrino propagation in the simulated experiments. The operators which could not be constrained by T2HK, DUNE or JUNO are excluded from the tables.  

We remind the reader about the precision of the results. The lower bounds are presented up to one digit, rounded down to nearest 0.1\,TeV. Sensitivities below this threshold should therefore be treated as unconstrained.

\subsection{Constraints on CC and NC NISs from T2HK, DUNE, and JUNO}
\begin{figure}[t]
\centering{
  \begin{adjustbox}{max width = \textwidth}
\begin{tabular}{cc}
\includegraphics[scale=0.5]{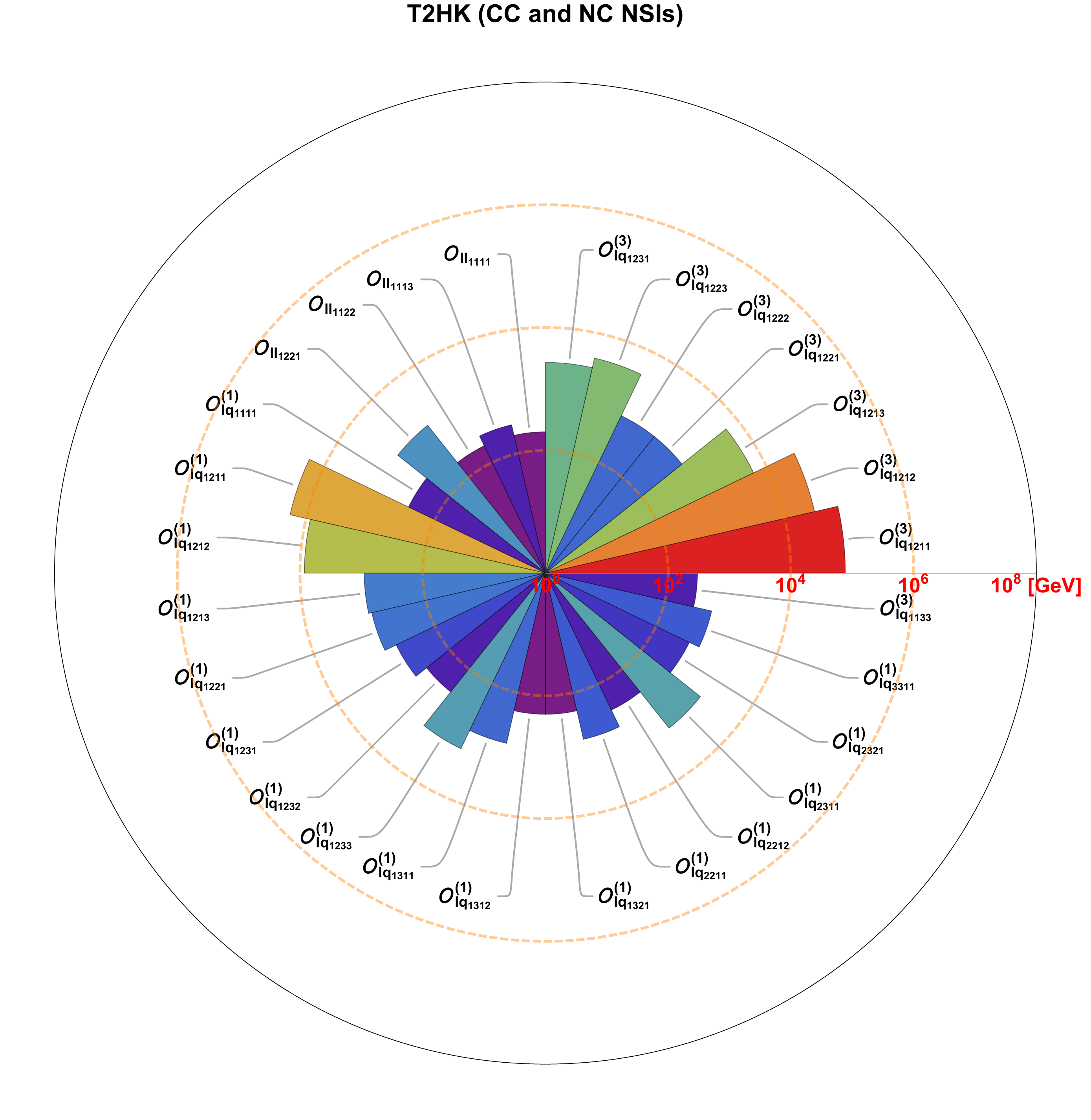} ~& ~ \includegraphics[scale=0.5]{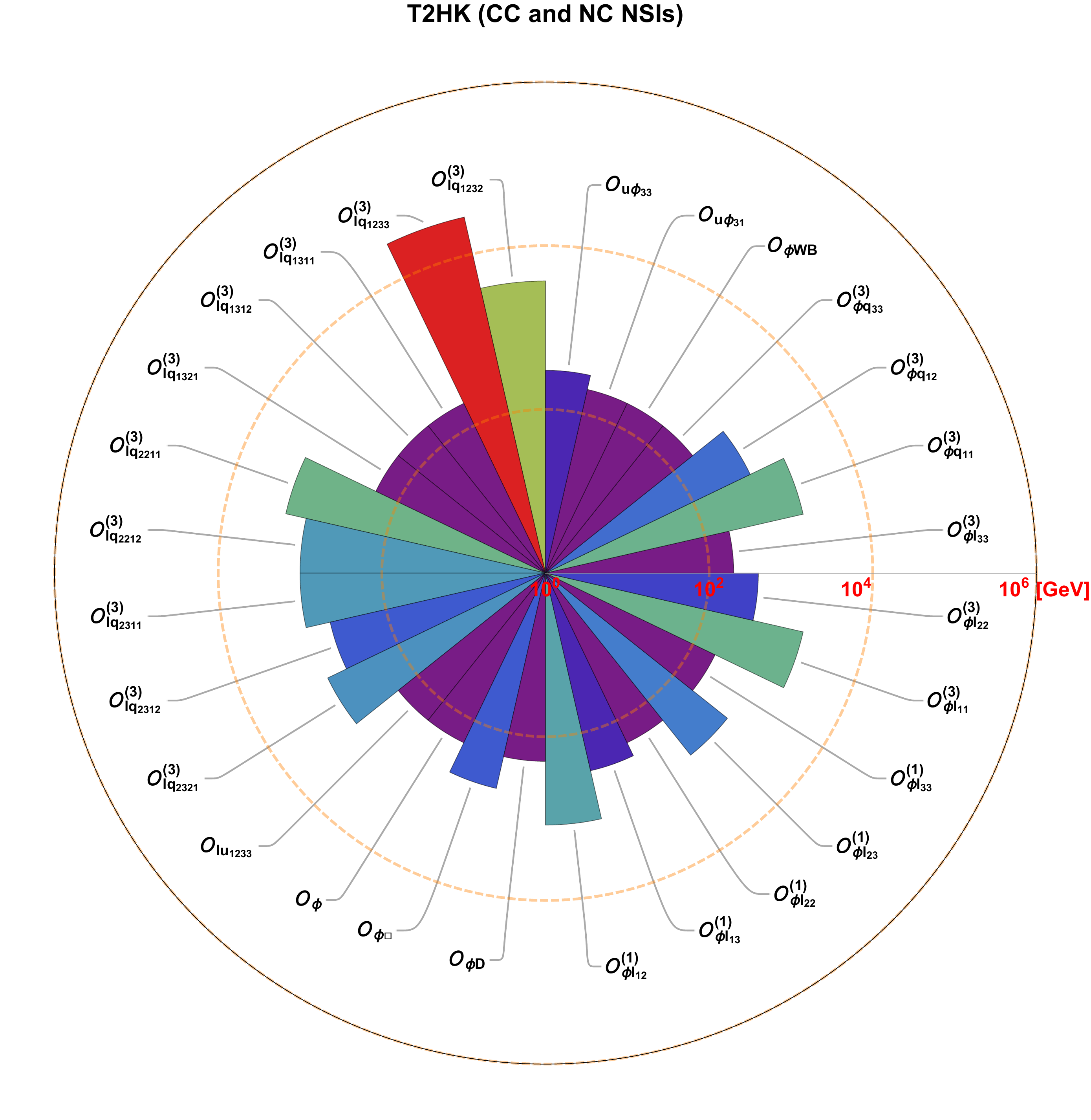}\\
\includegraphics[scale=0.5]{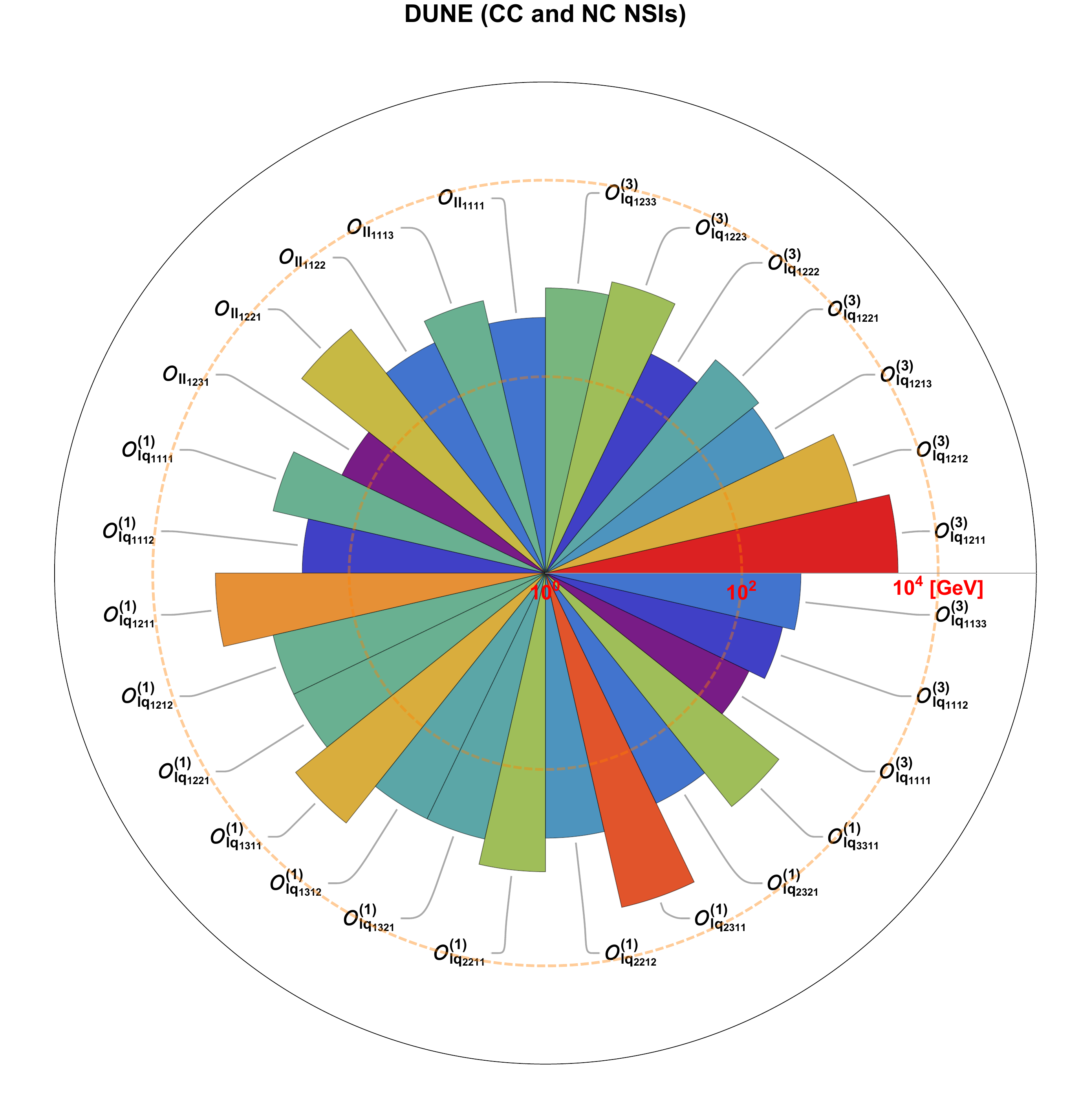} ~& ~ \includegraphics[scale=0.5]{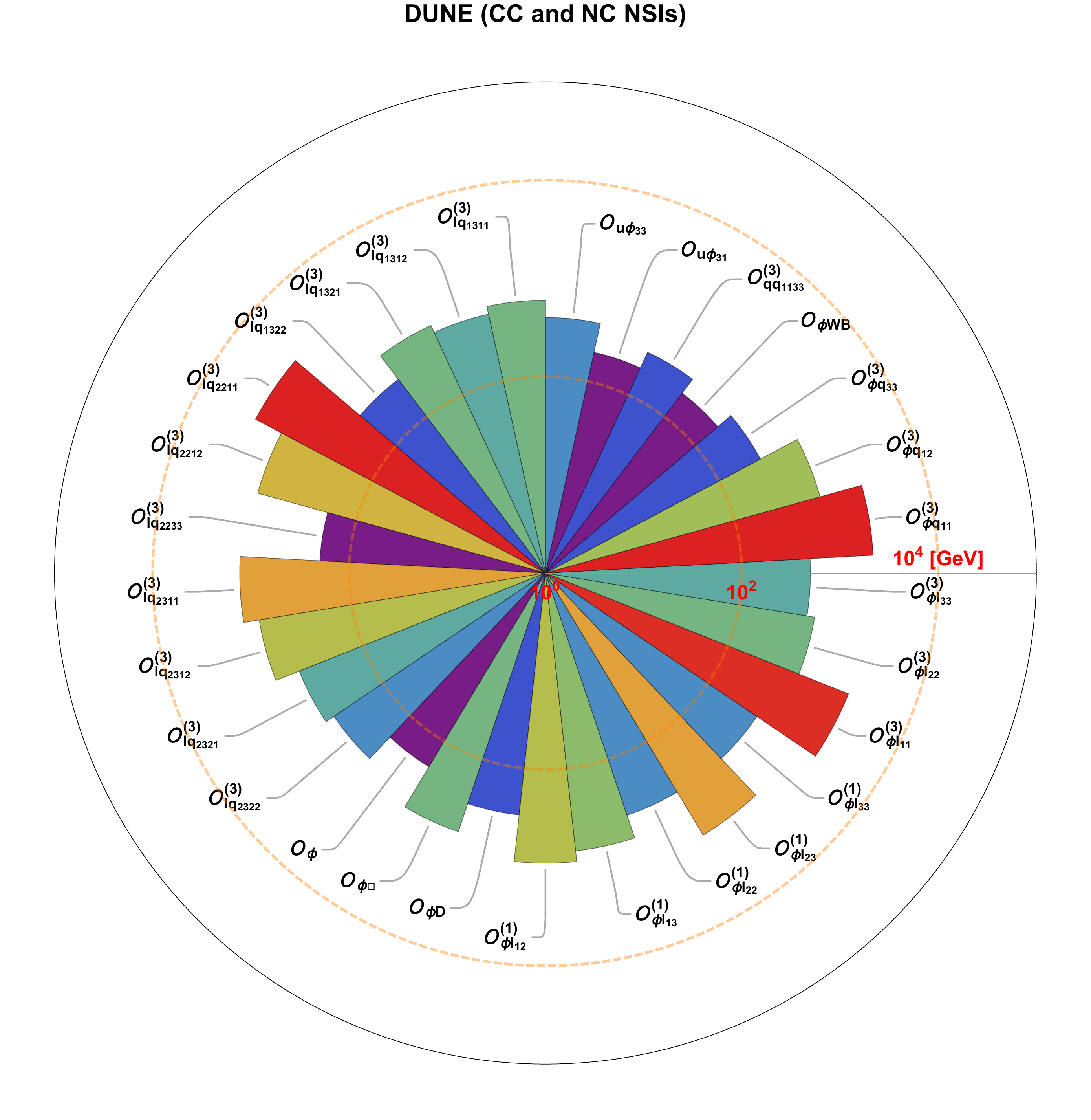}
\end{tabular}
  \end{adjustbox}}
\caption{Constraints on dimension-6 SMEFT operators that can induce significant neutral and charged current NSIs. The results are obtained from T2HK, DUNE, JUNO, as well as a combined analysis of T2HK and DUNE. We include the near detector for both T2HK and DUNE in our analysis. For JUNO, we present our results both with and without the near detector TAO.}\label{CCAndNC1}
\end{figure}

\begin{figure}[t]
\centering{
  \begin{adjustbox}{max width = \textwidth}
\begin{tabular}{cc}
\includegraphics[scale=0.5]{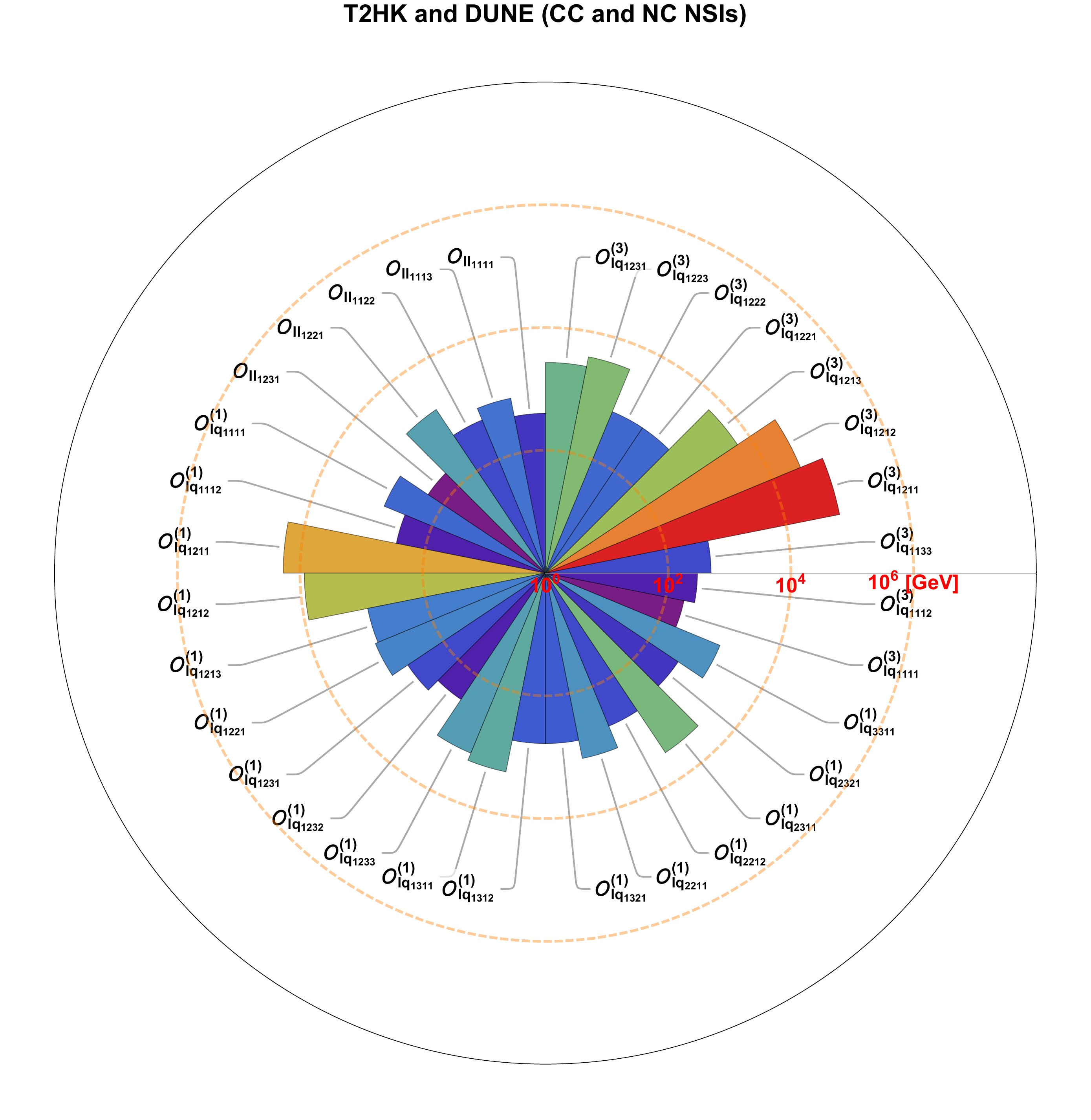} ~& ~ \includegraphics[scale=0.5]{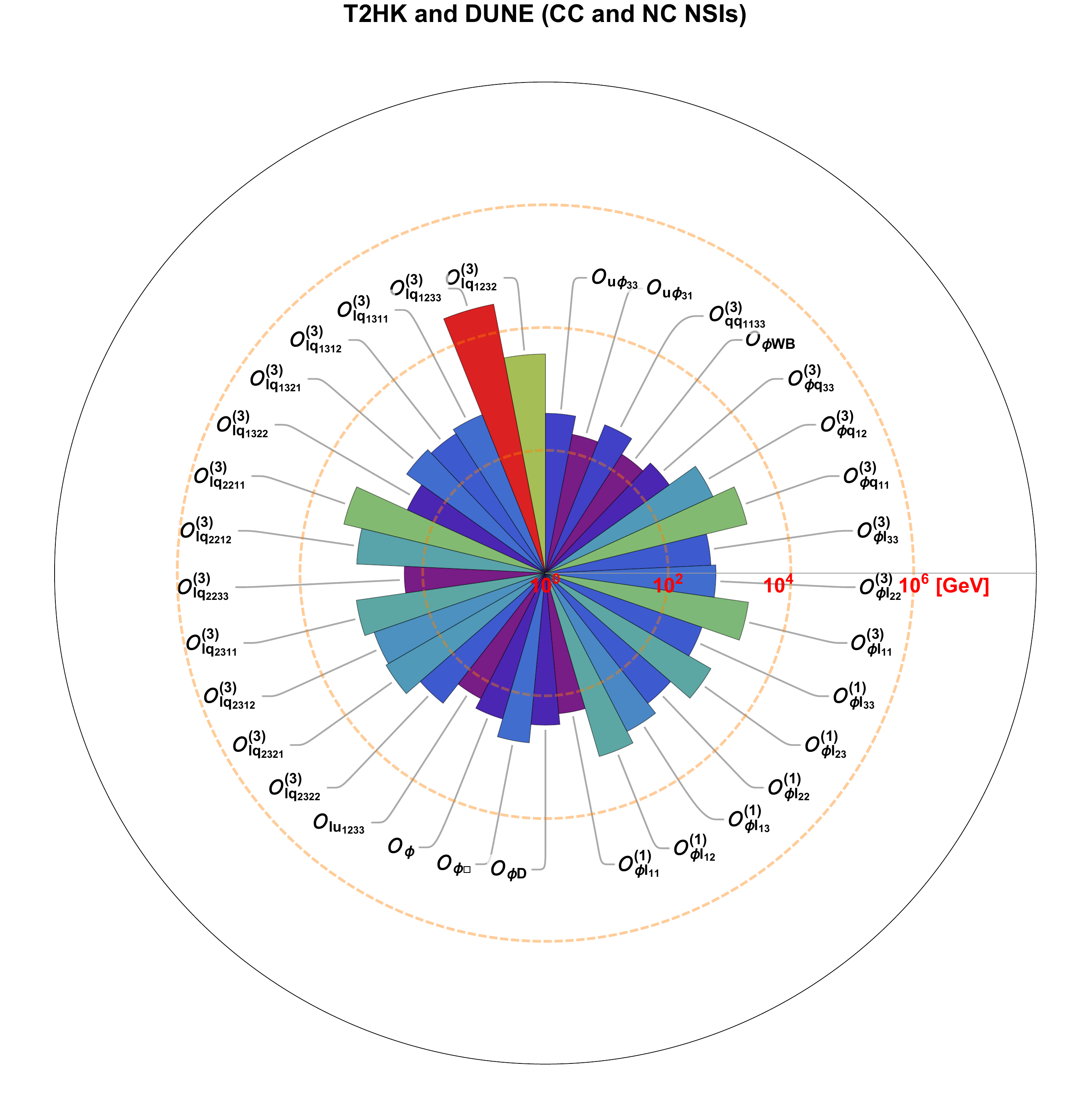}\\
\includegraphics[scale=0.5]{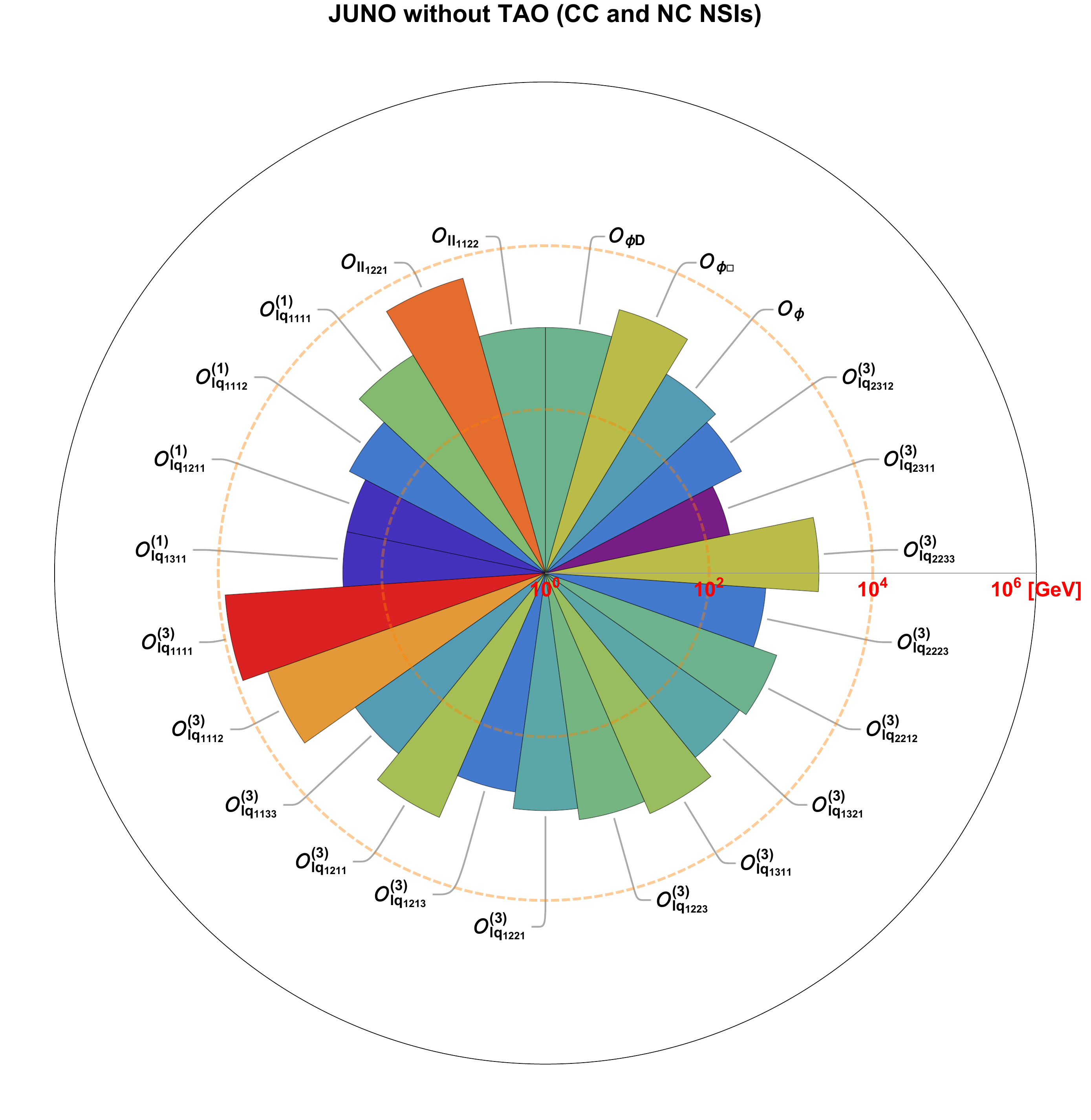} ~& ~ \includegraphics[scale=0.5]{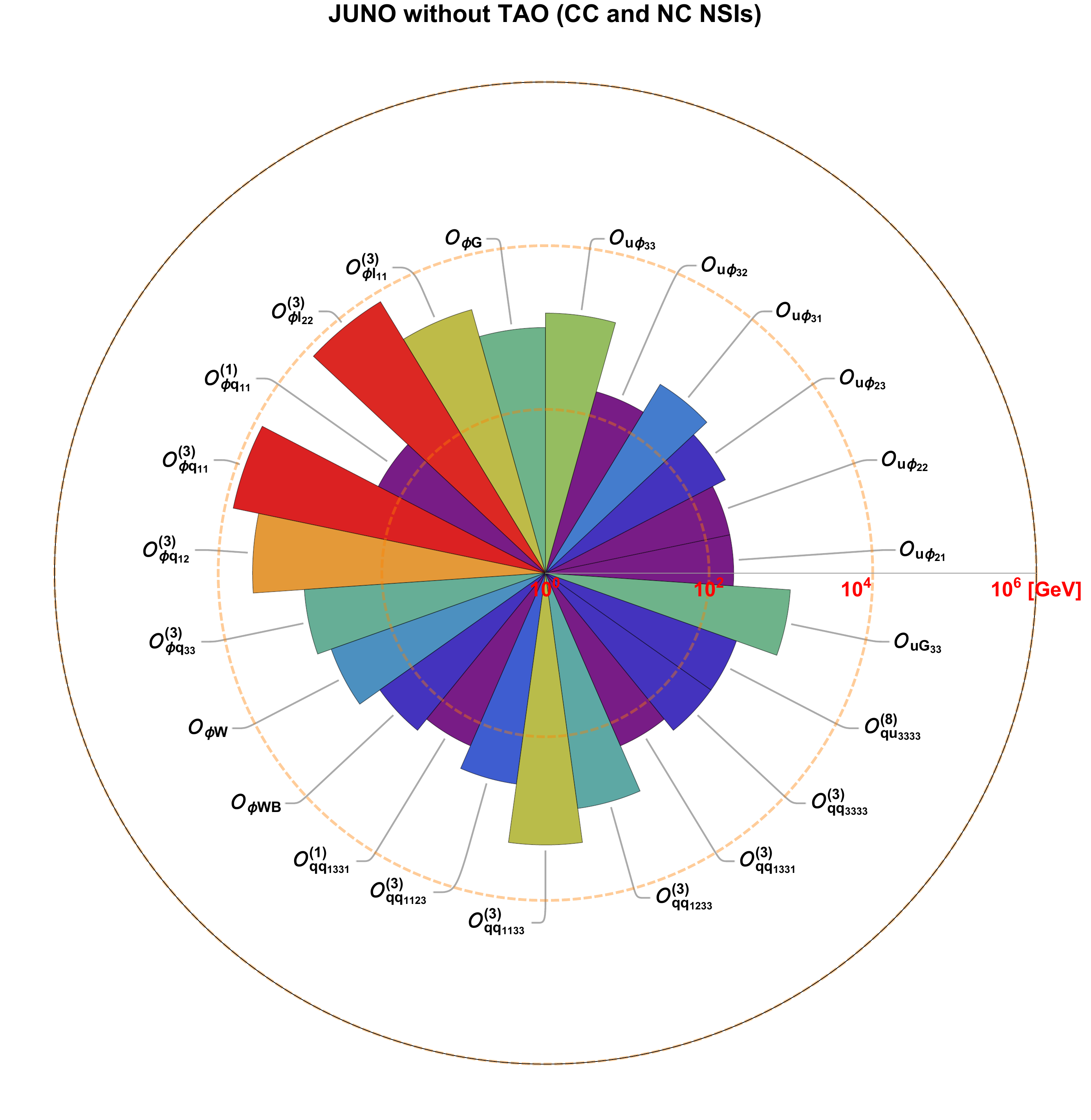}\\
\includegraphics[scale=0.5]{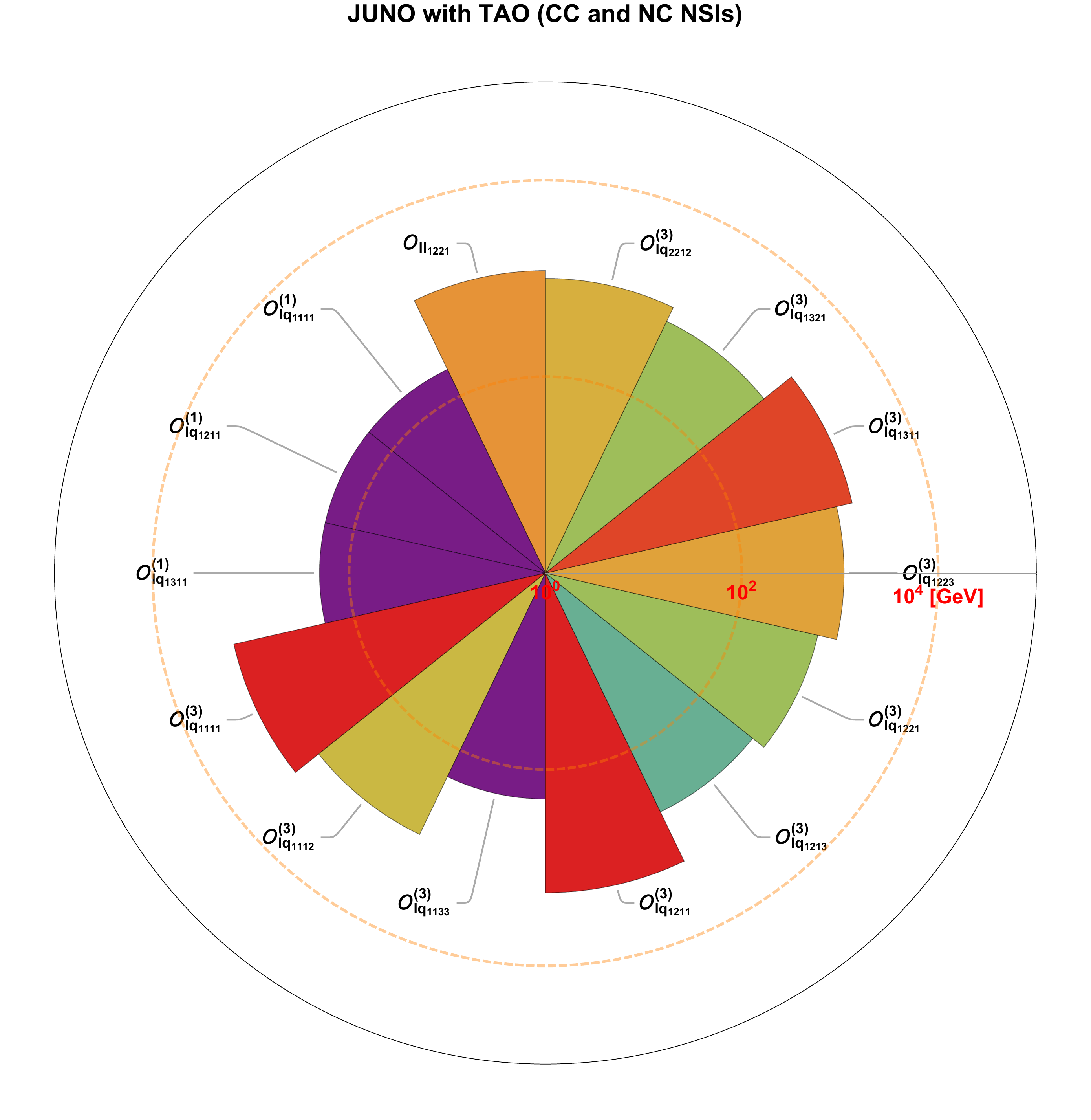} ~& ~ \includegraphics[scale=0.5]{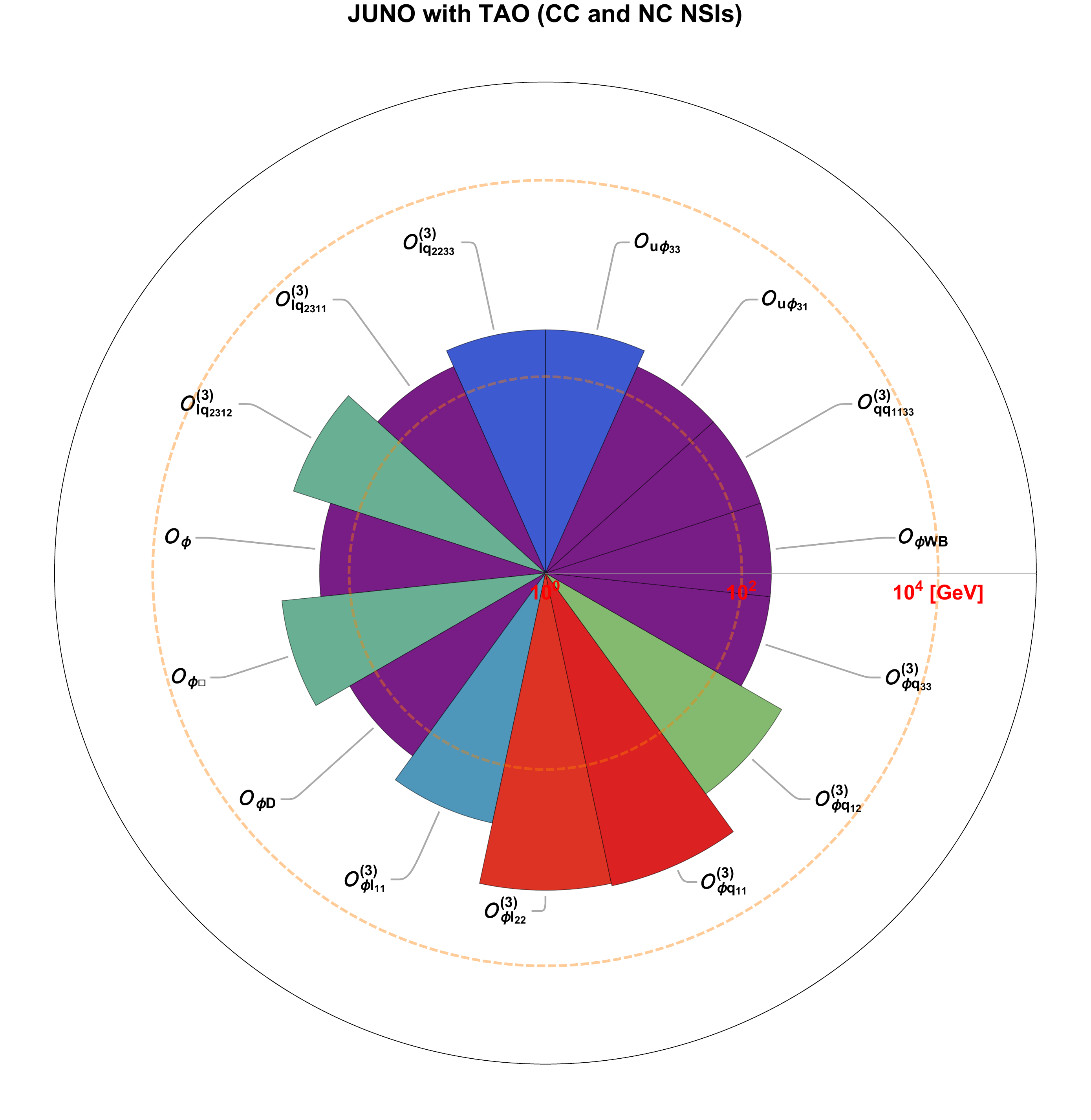}
\end{tabular}
  \end{adjustbox}}
\caption{Figure\,\ref{CCAndNC1} continued.}\label{CCAndNC2}
\end{figure}


\clearpage

\subsection{Constraints on CC NSIs from T2HK, DUNE, and JUNO}

\begin{figure}[!tbh]
\centering{
  \begin{adjustbox}{max width = \textwidth}
\begin{tabular}{cc}
\includegraphics[scale=0.5]{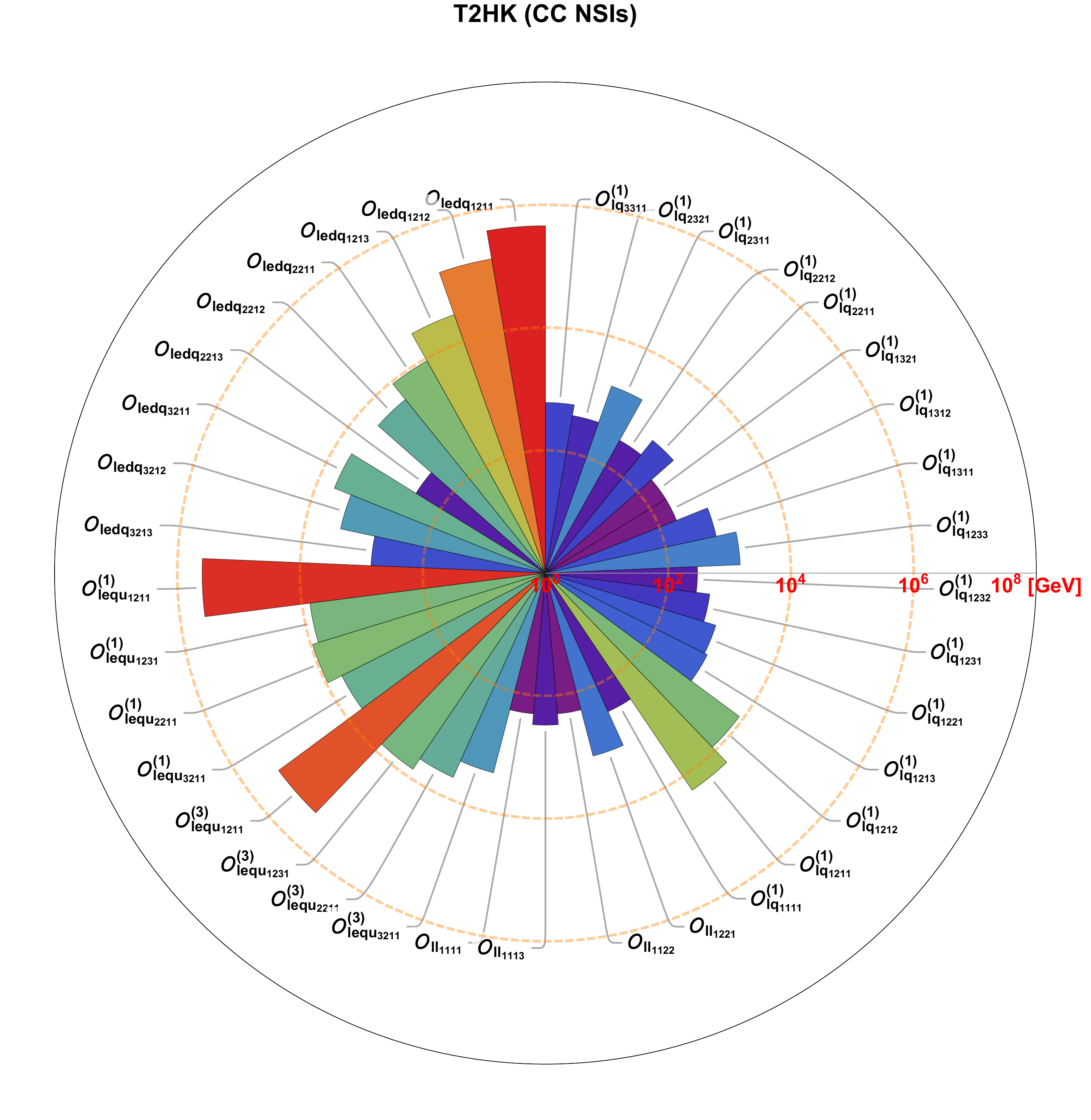} ~& ~ \includegraphics[scale=0.5]{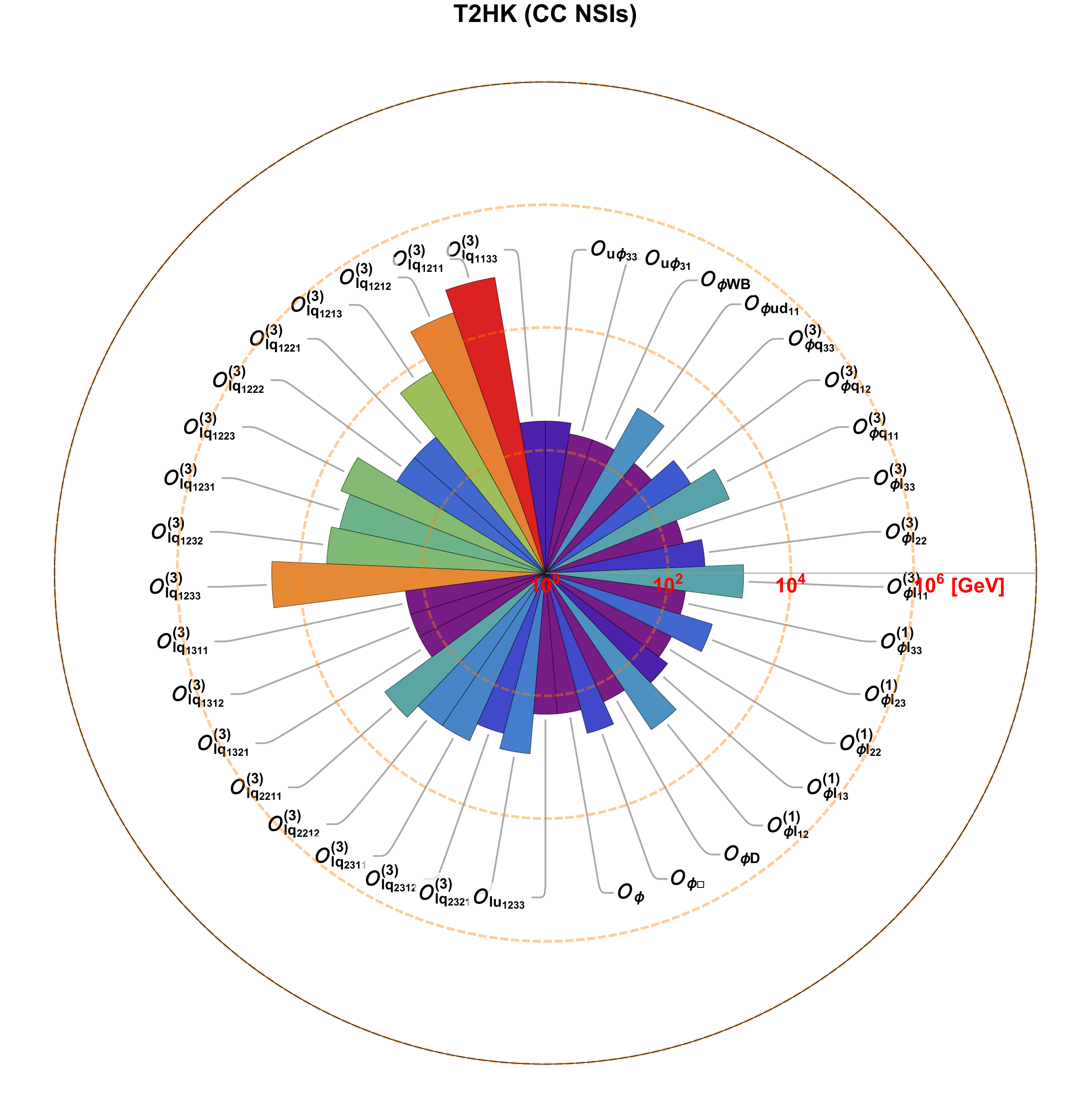}\\
\includegraphics[scale=0.5]{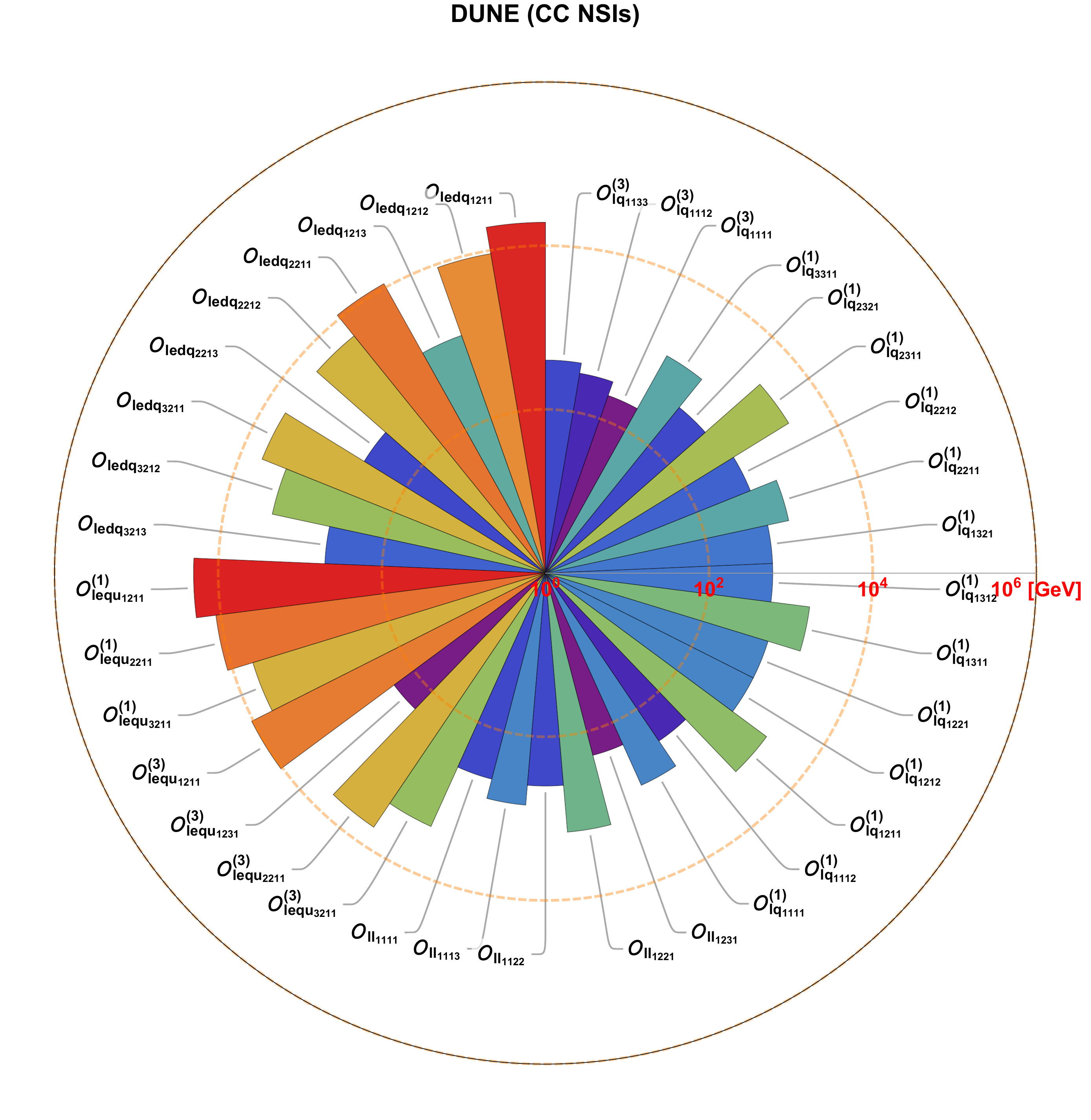} ~& ~ \includegraphics[scale=0.5]{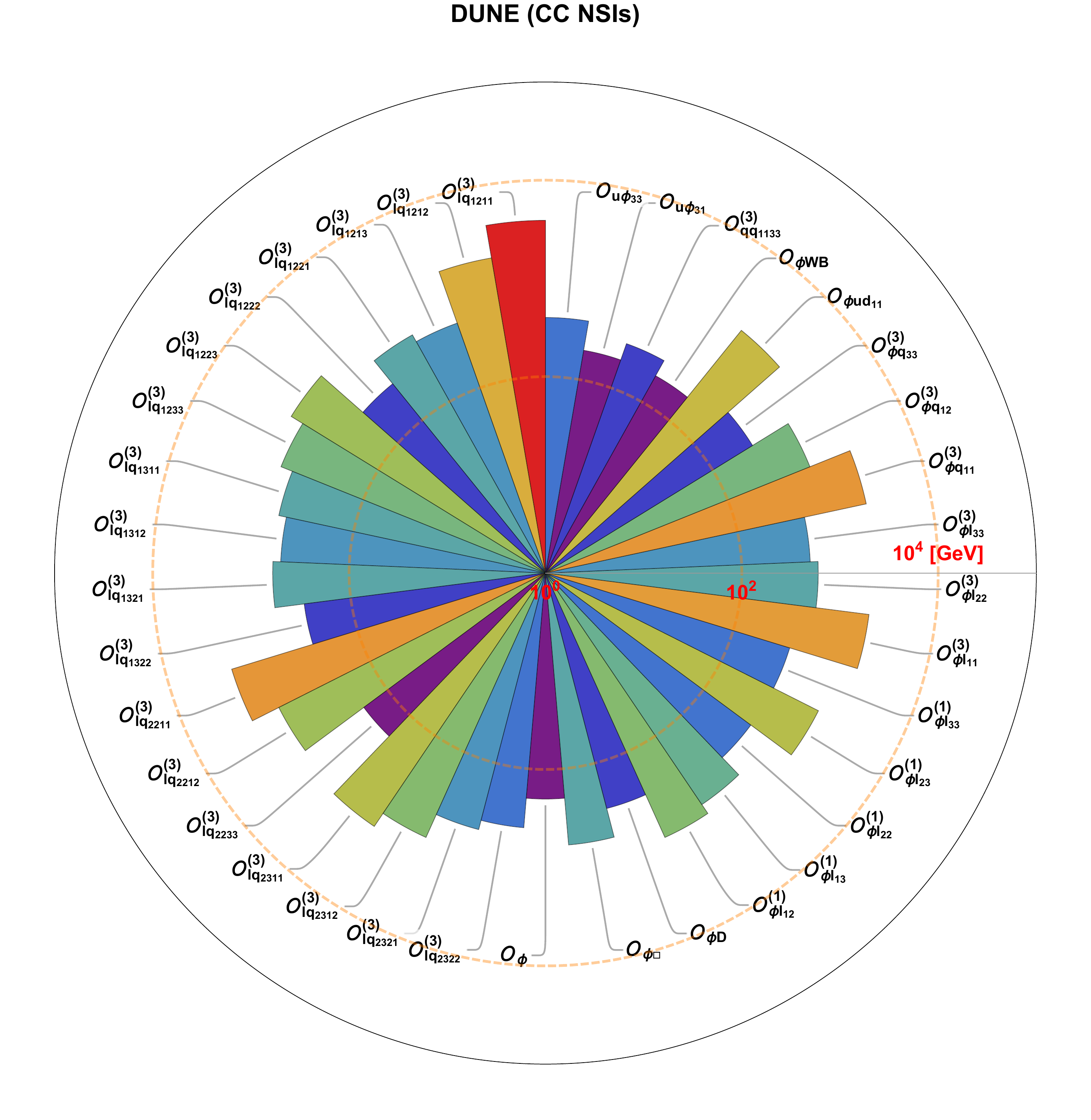}
\end{tabular}
  \end{adjustbox}}
\caption{Same as figure\,\ref{CCAndNC1} but for constraints on dimension-6 SMEFT operators that can only induce significant CC NSIs.}\label{CC0}
\end{figure}

\begin{figure}[t]
\centering{
  \begin{adjustbox}{max width = \textwidth}
\begin{tabular}{cc}
\includegraphics[scale=0.5]{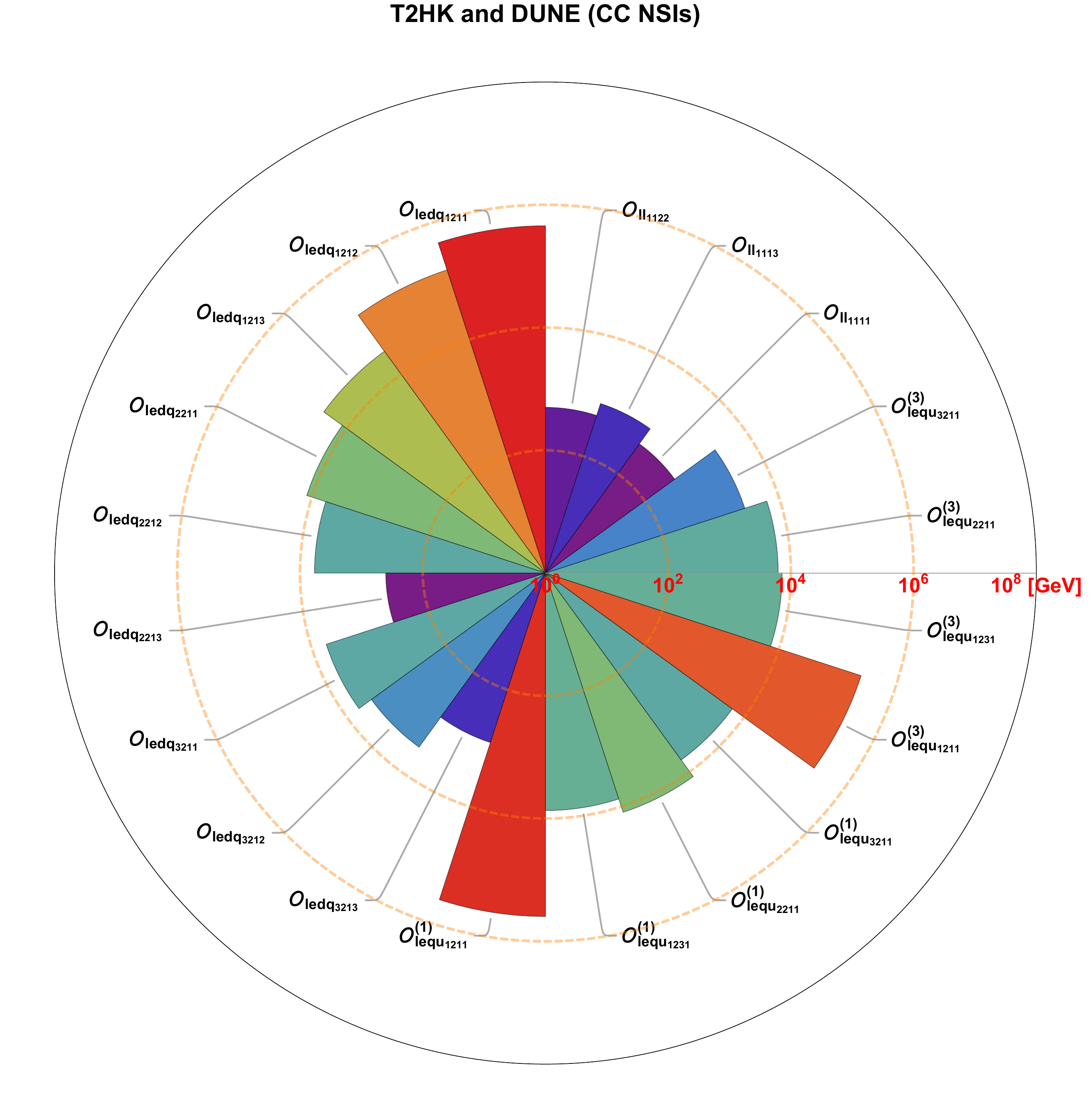} ~& ~ \includegraphics[scale=0.5]{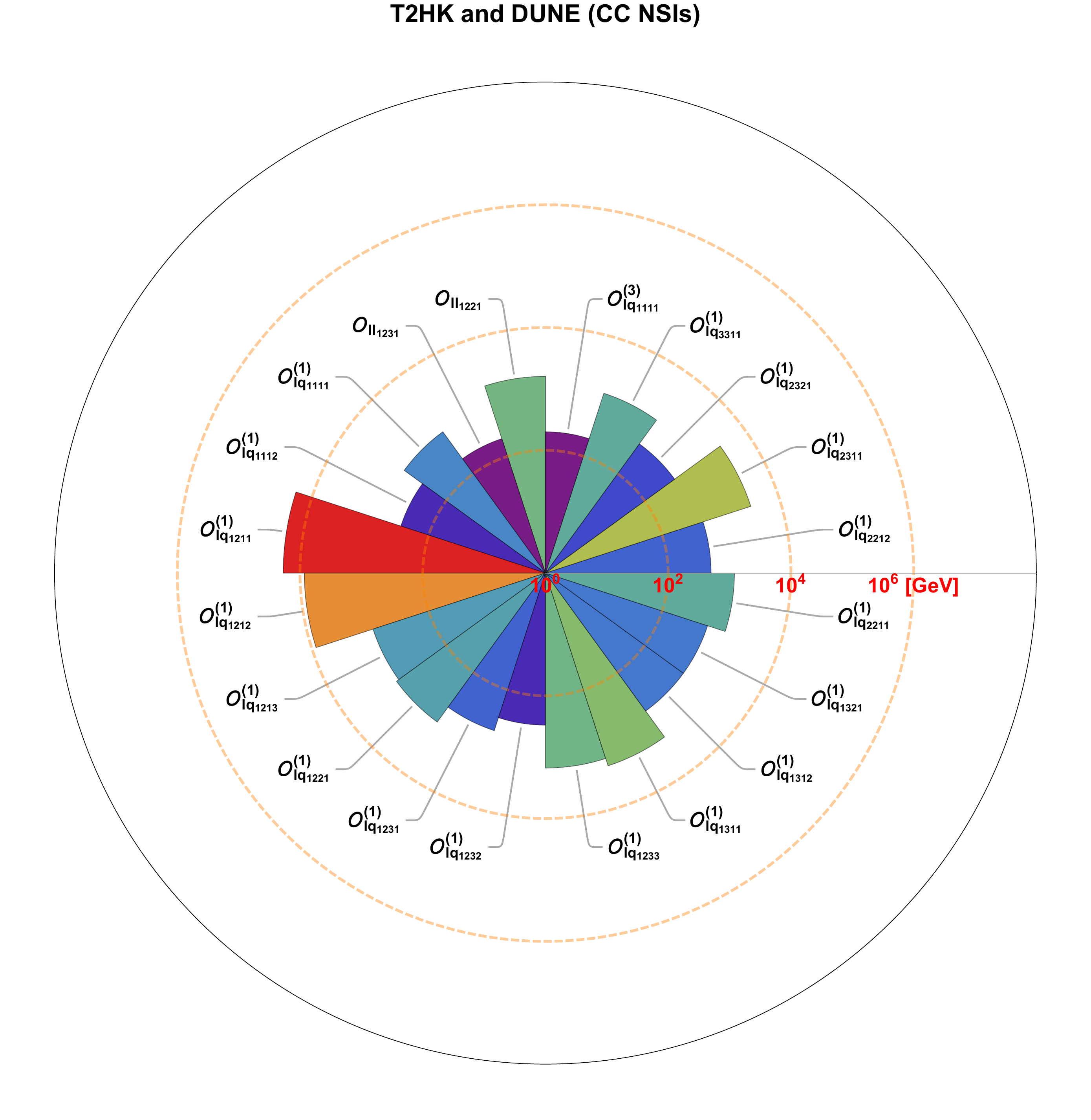}\\
\includegraphics[scale=0.5]{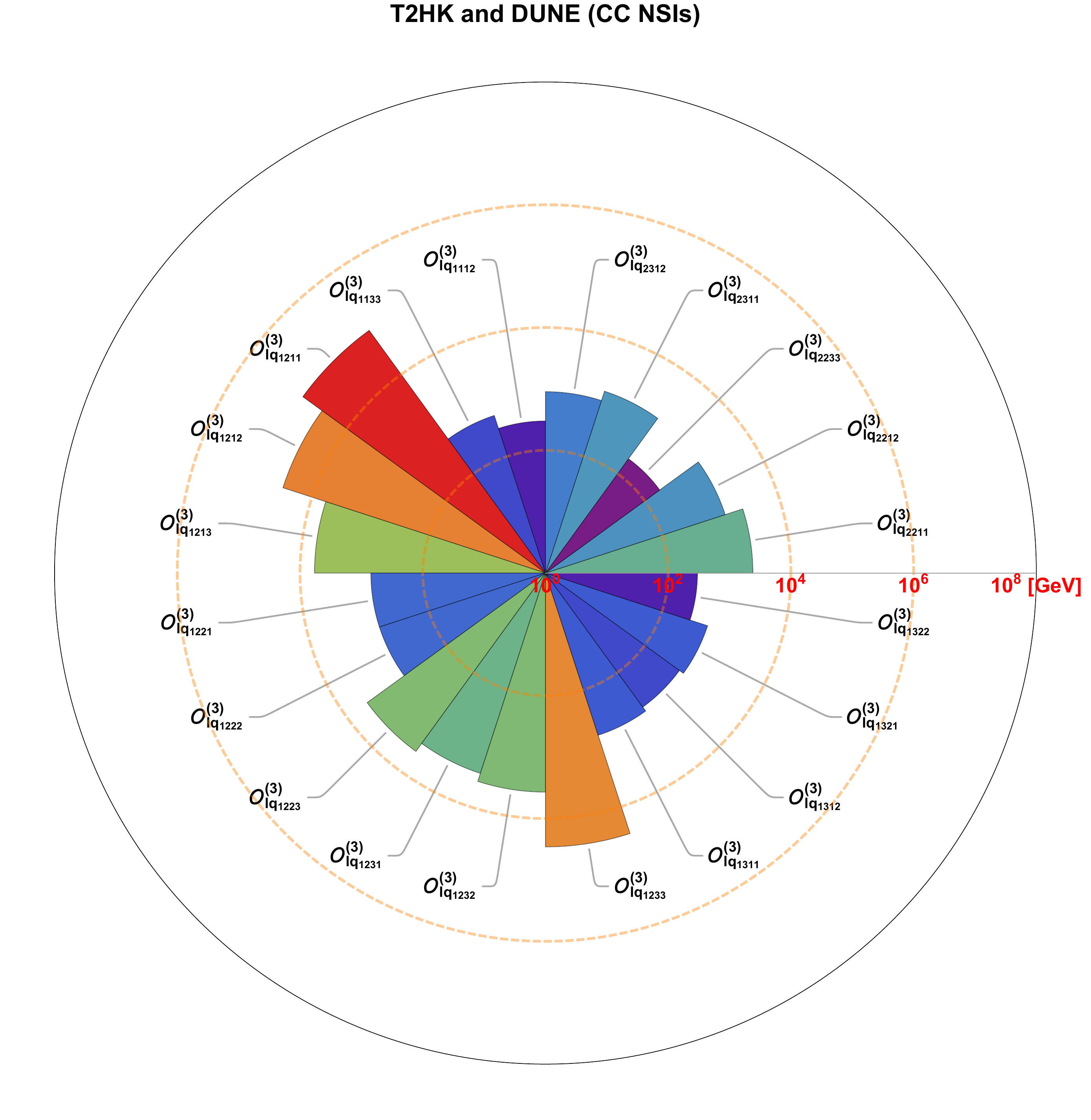} ~& ~ \includegraphics[scale=0.5]{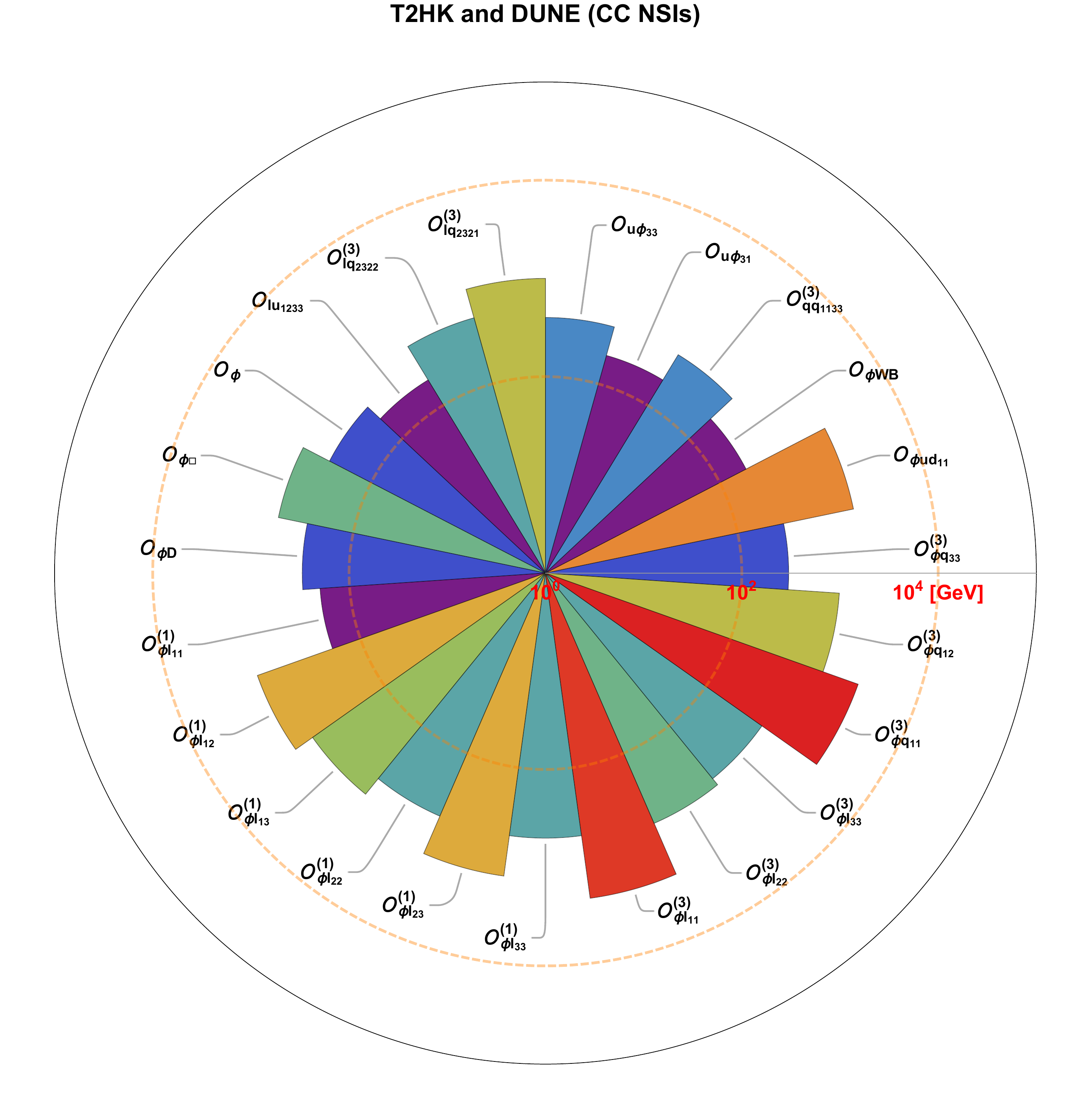}
\end{tabular}
  \end{adjustbox}}
\caption{Figure\,\ref{CC0} continued.}\label{CC1}
\end{figure}

\begin{figure}[t]
\centering{
  \begin{adjustbox}{max width = \textwidth}
\begin{tabular}{cc}
\includegraphics[scale=0.5]{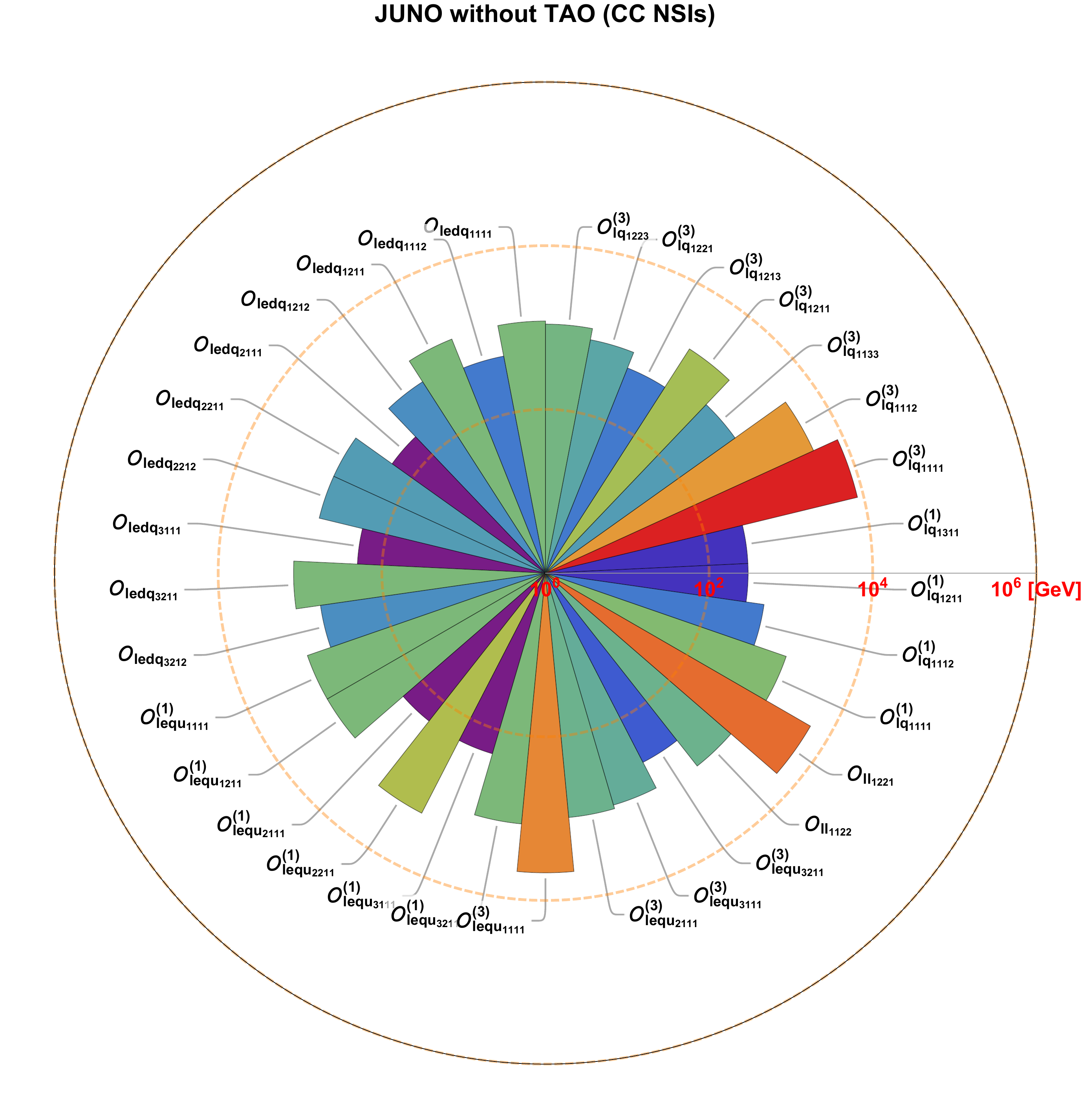} ~& ~ \includegraphics[scale=0.5]{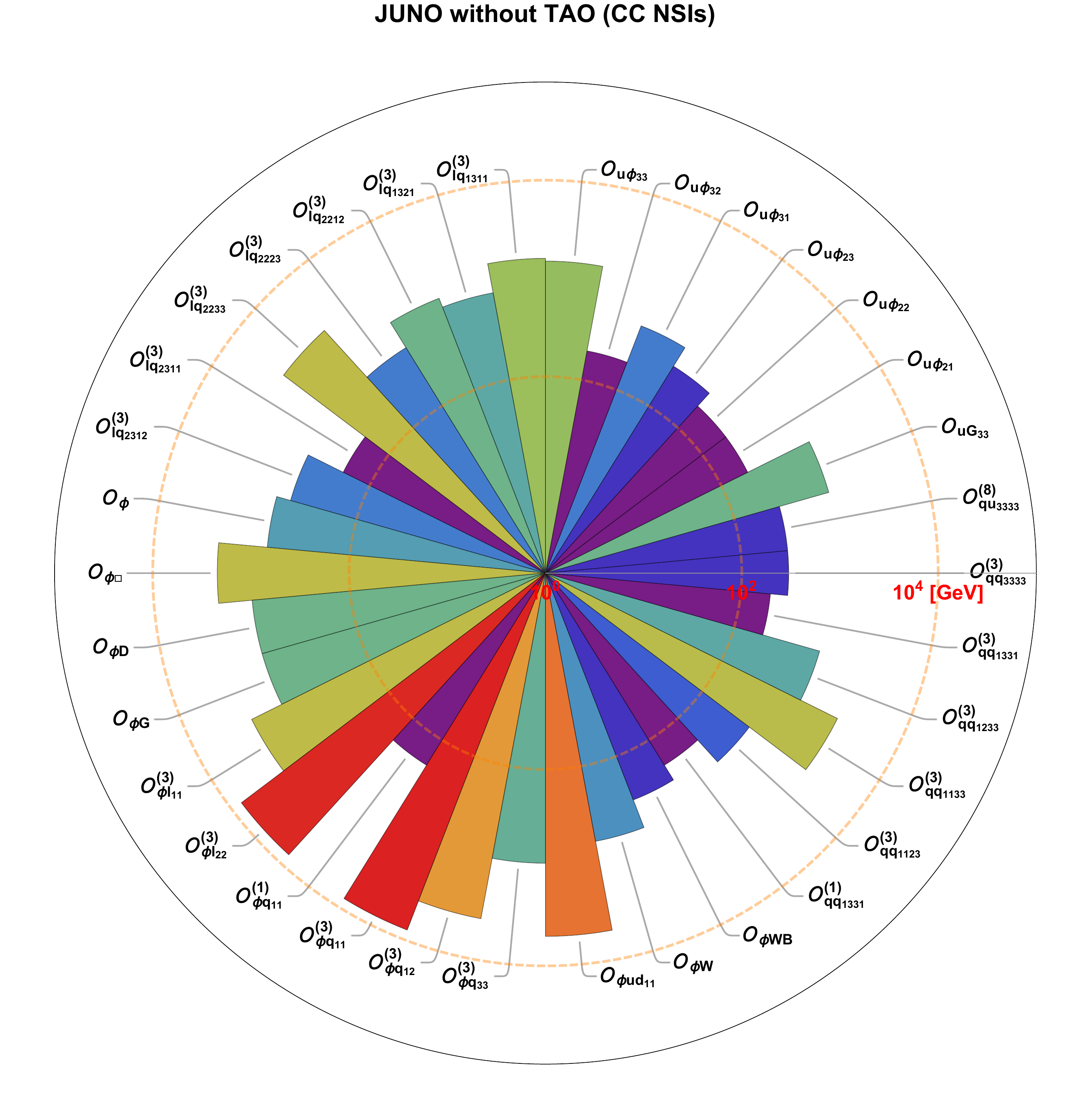}\\
\includegraphics[scale=0.5]{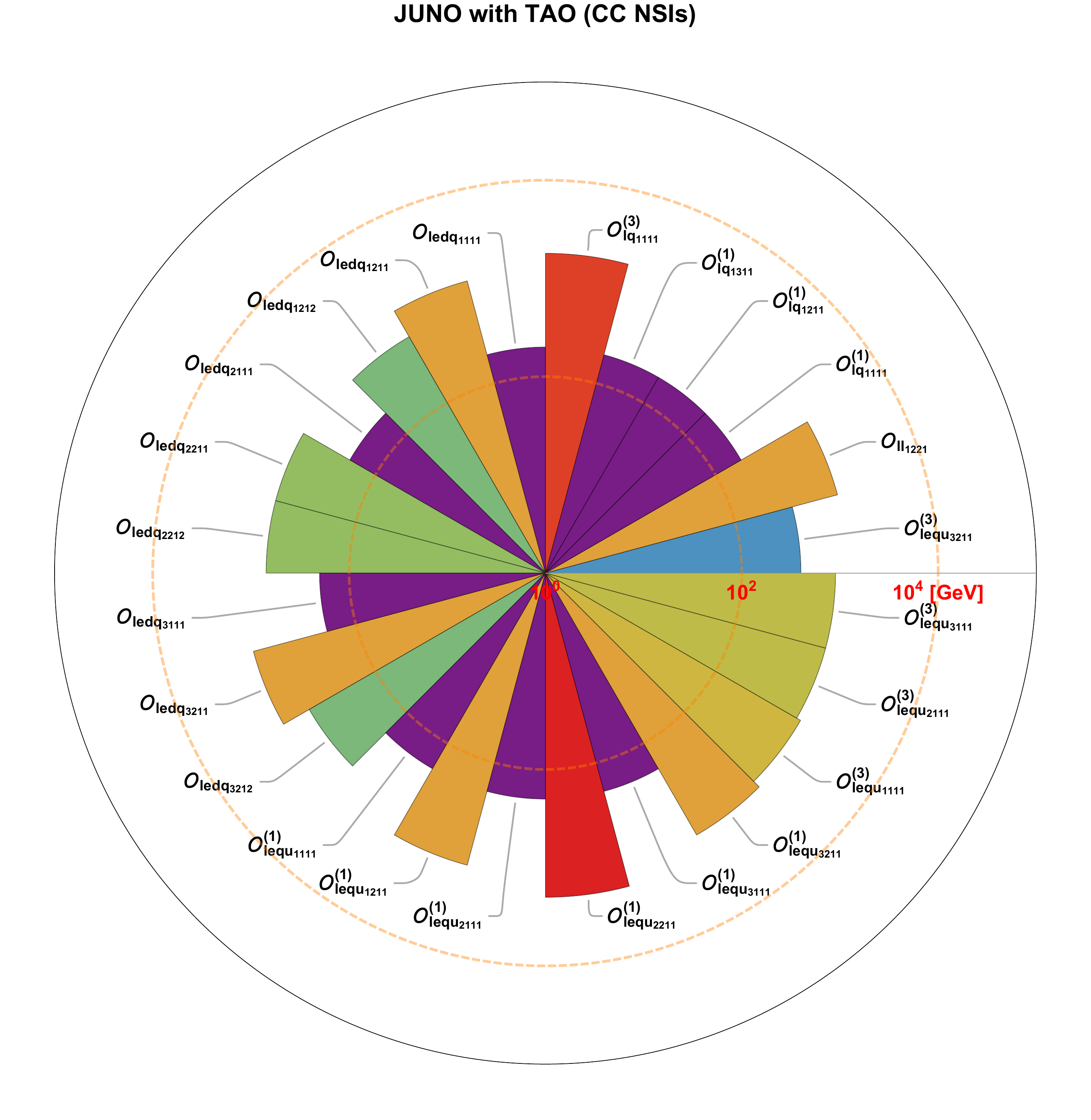} ~& ~ \includegraphics[scale=0.5]{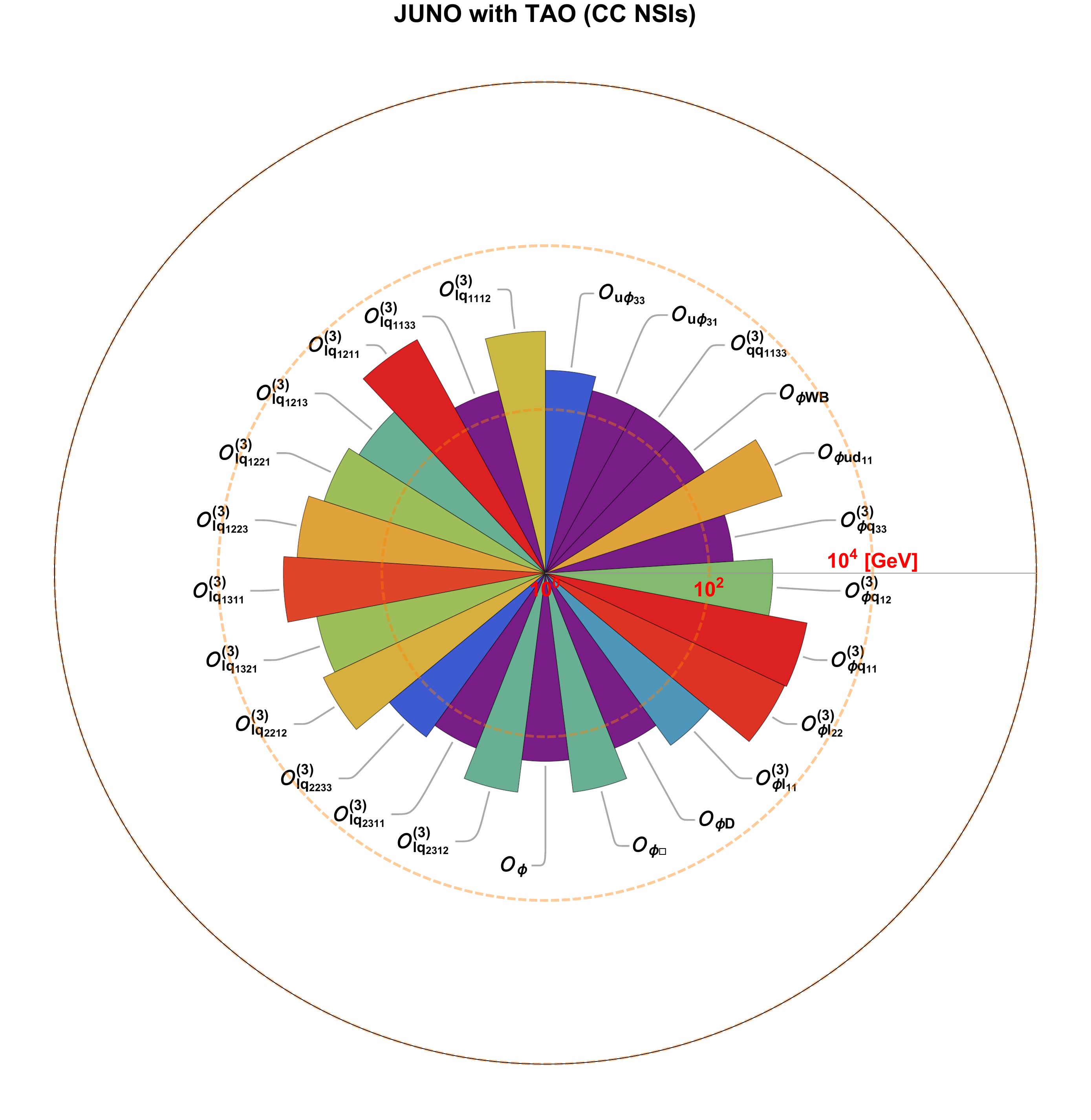}
\end{tabular}
  \end{adjustbox}}
\caption{Figure\,\ref{CC1} continued.}\label{CC2}
\end{figure}

\clearpage

\subsection{Constraints on NC NSIs from T2HK, DUNE, and JUNO}
\begin{figure}[!tbh]
\centering{
  \begin{adjustbox}{max width = \textwidth}
\begin{tabular}{cc}
\includegraphics[scale=0.5]{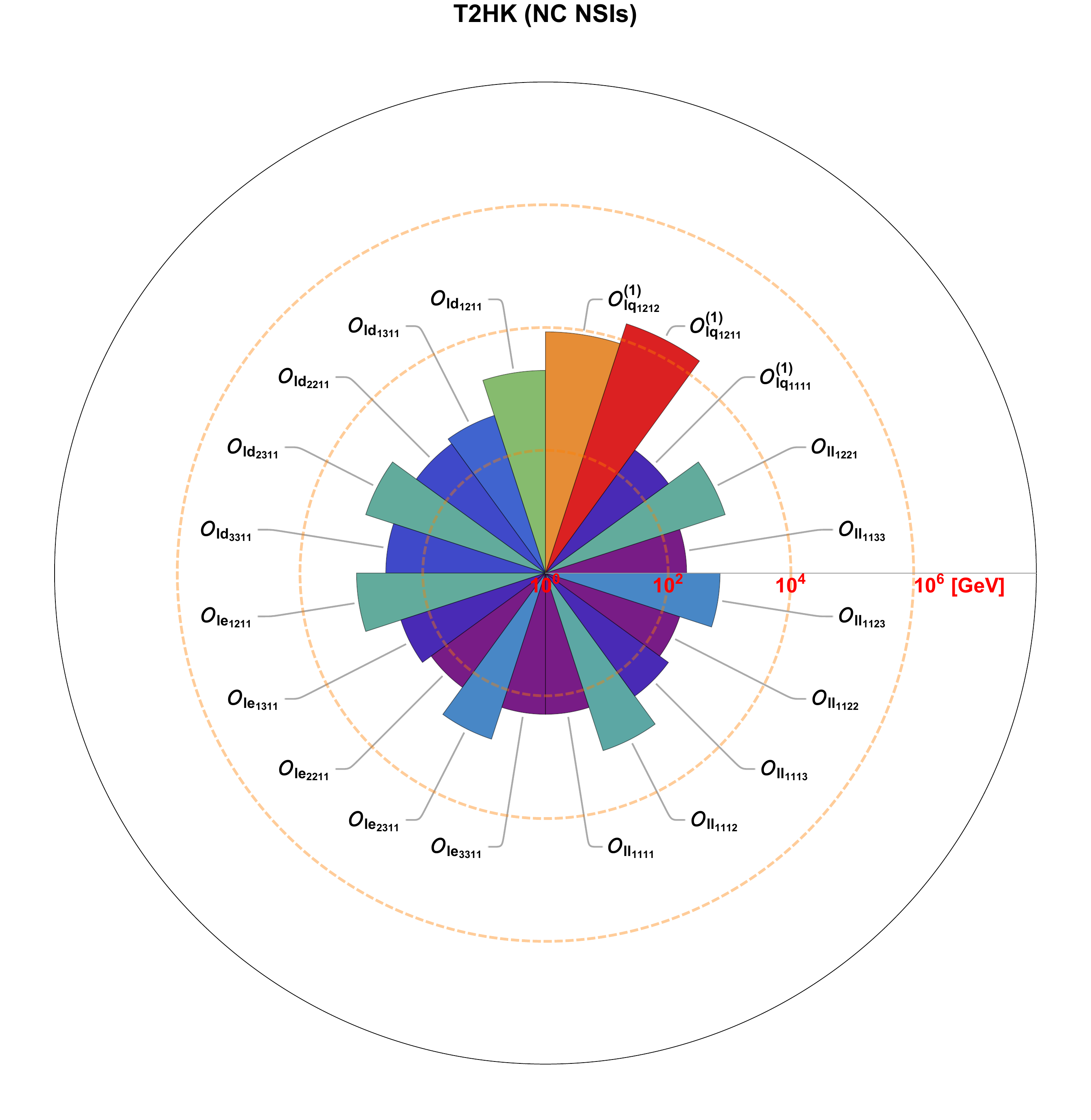} ~& ~ \includegraphics[scale=0.5]{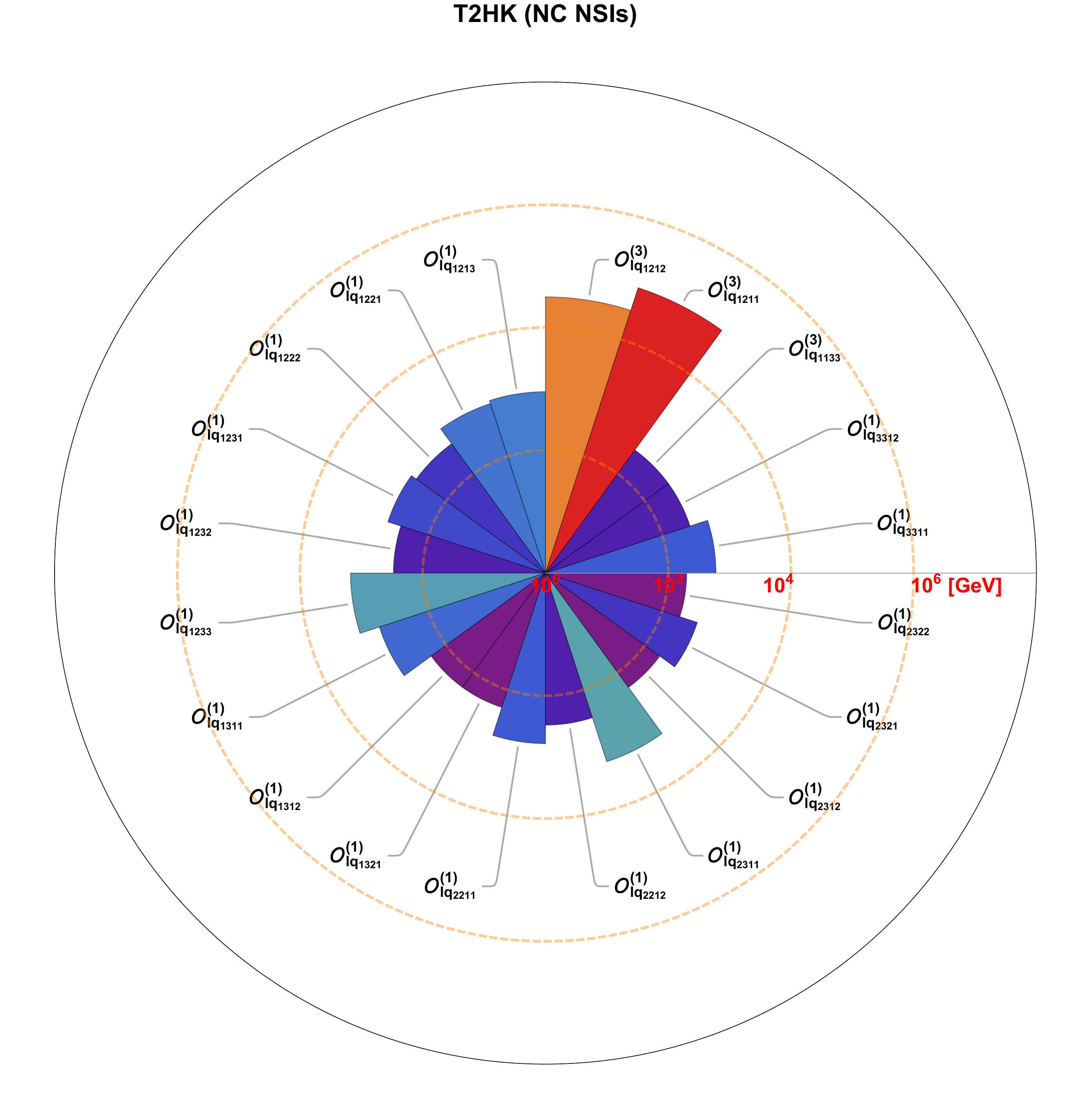}\\
\includegraphics[scale=0.5]{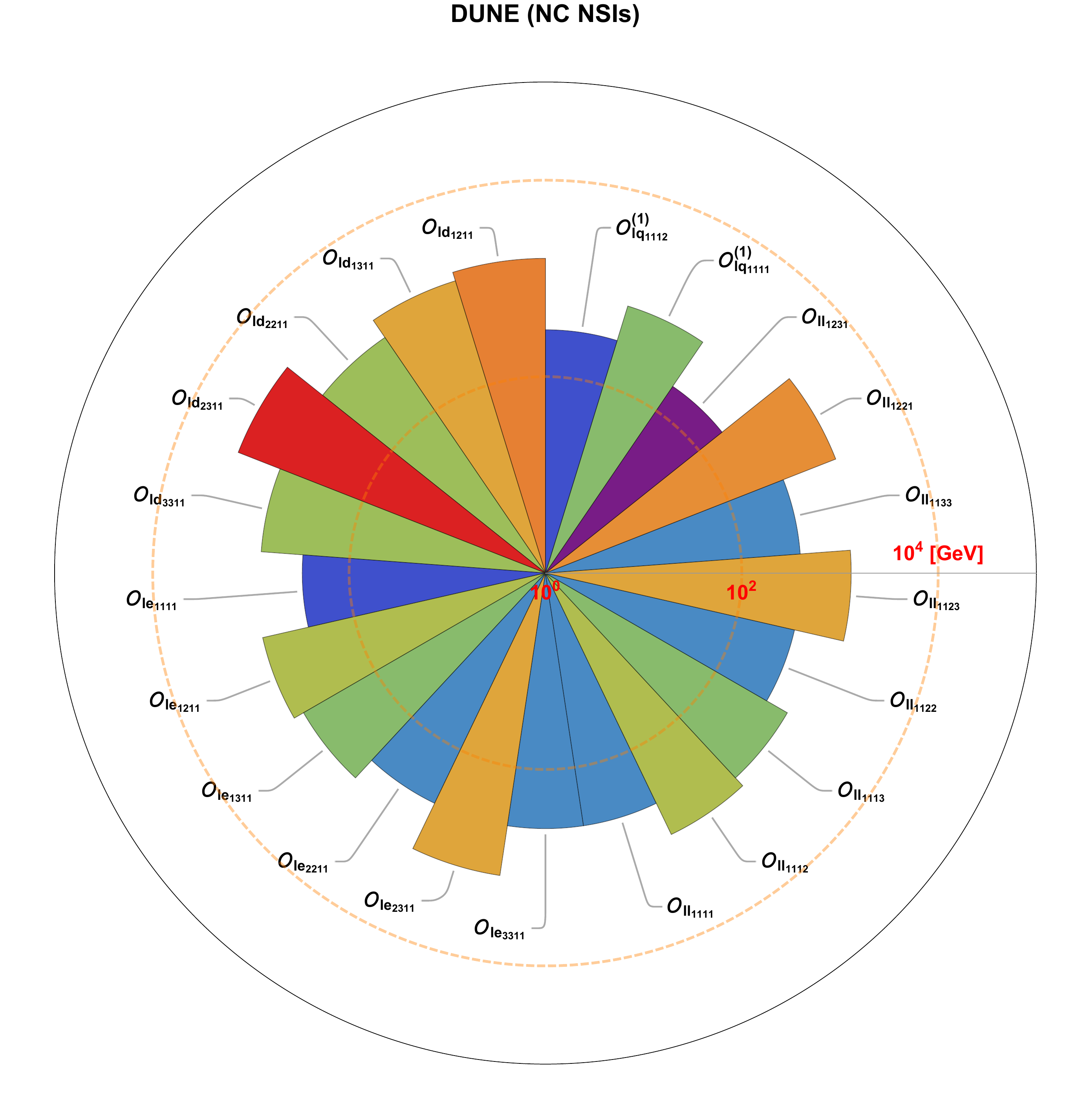} ~& ~ \includegraphics[scale=0.5]{plots/DUNENC1}
\end{tabular}
  \end{adjustbox}}
\caption{Same as figure\,\ref{CCAndNC1} but for constraints on dimension-6 SMEFT operators that can only induce significant NC NSIs.}\label{NC1}
\end{figure}

\begin{figure}[t]
\centering{
  \begin{adjustbox}{max width = \textwidth}
\begin{tabular}{cc}
\includegraphics[scale=0.5]{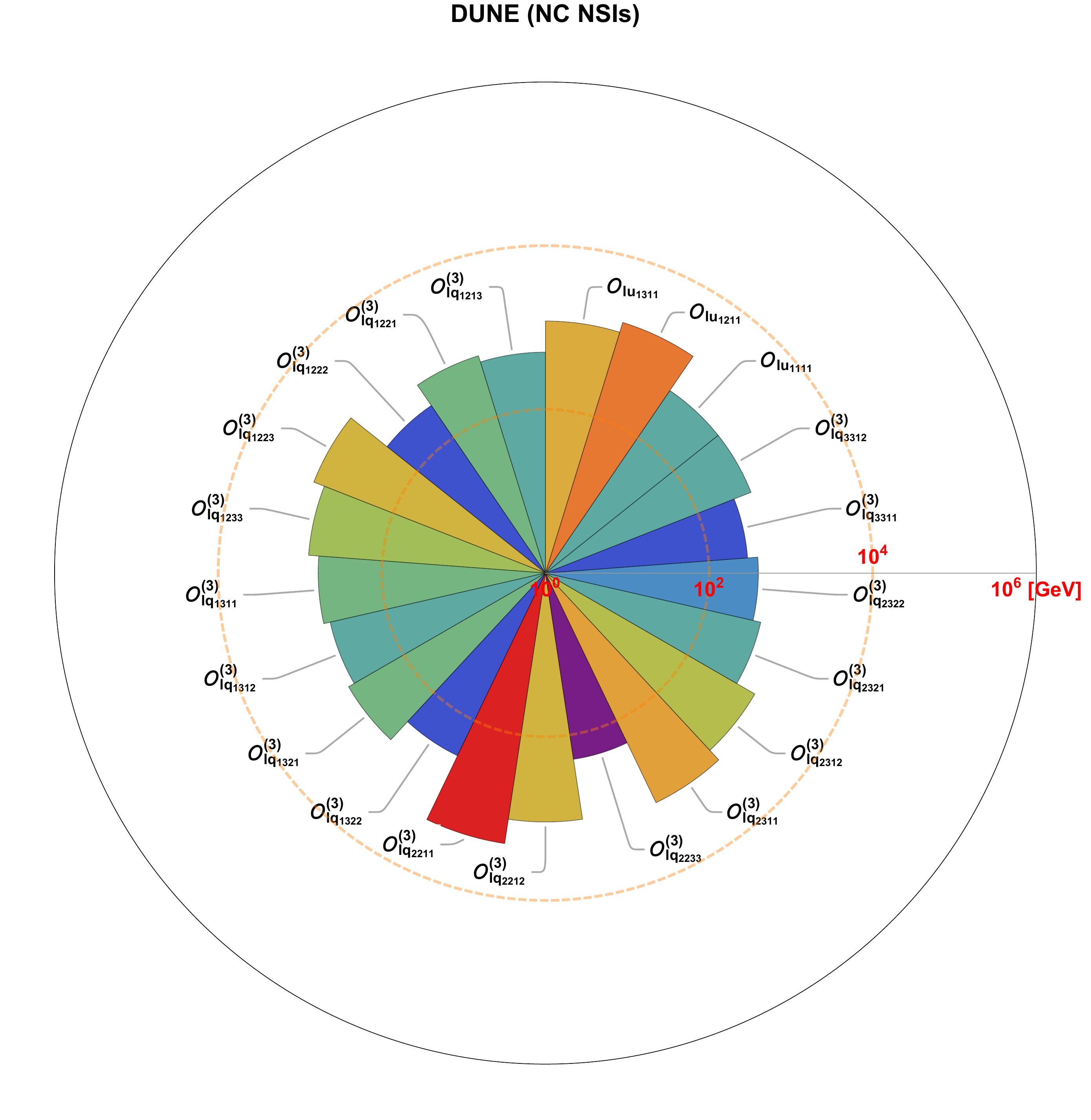} ~& ~ \includegraphics[scale=0.5]{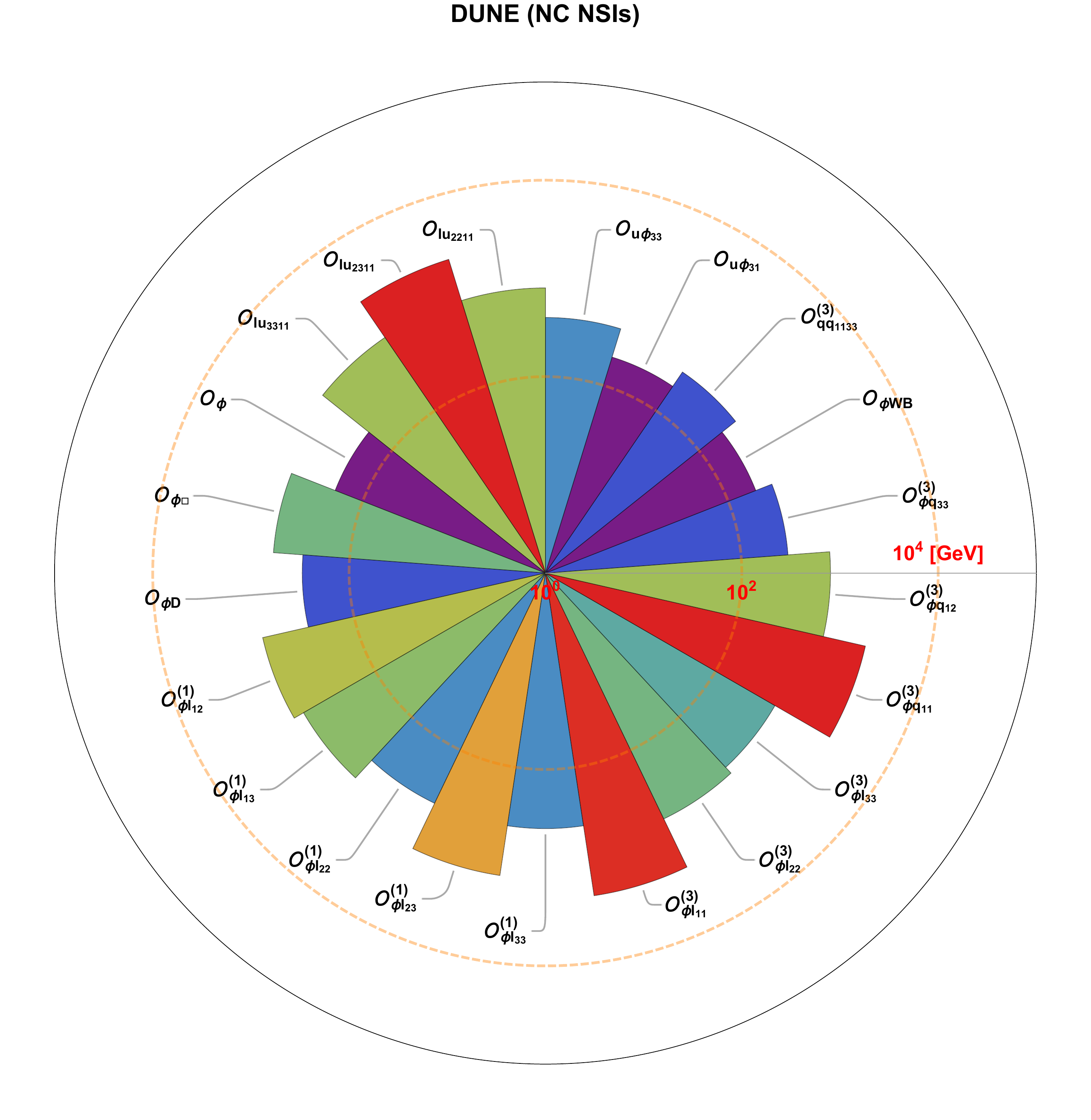}\\
\includegraphics[scale=0.5]{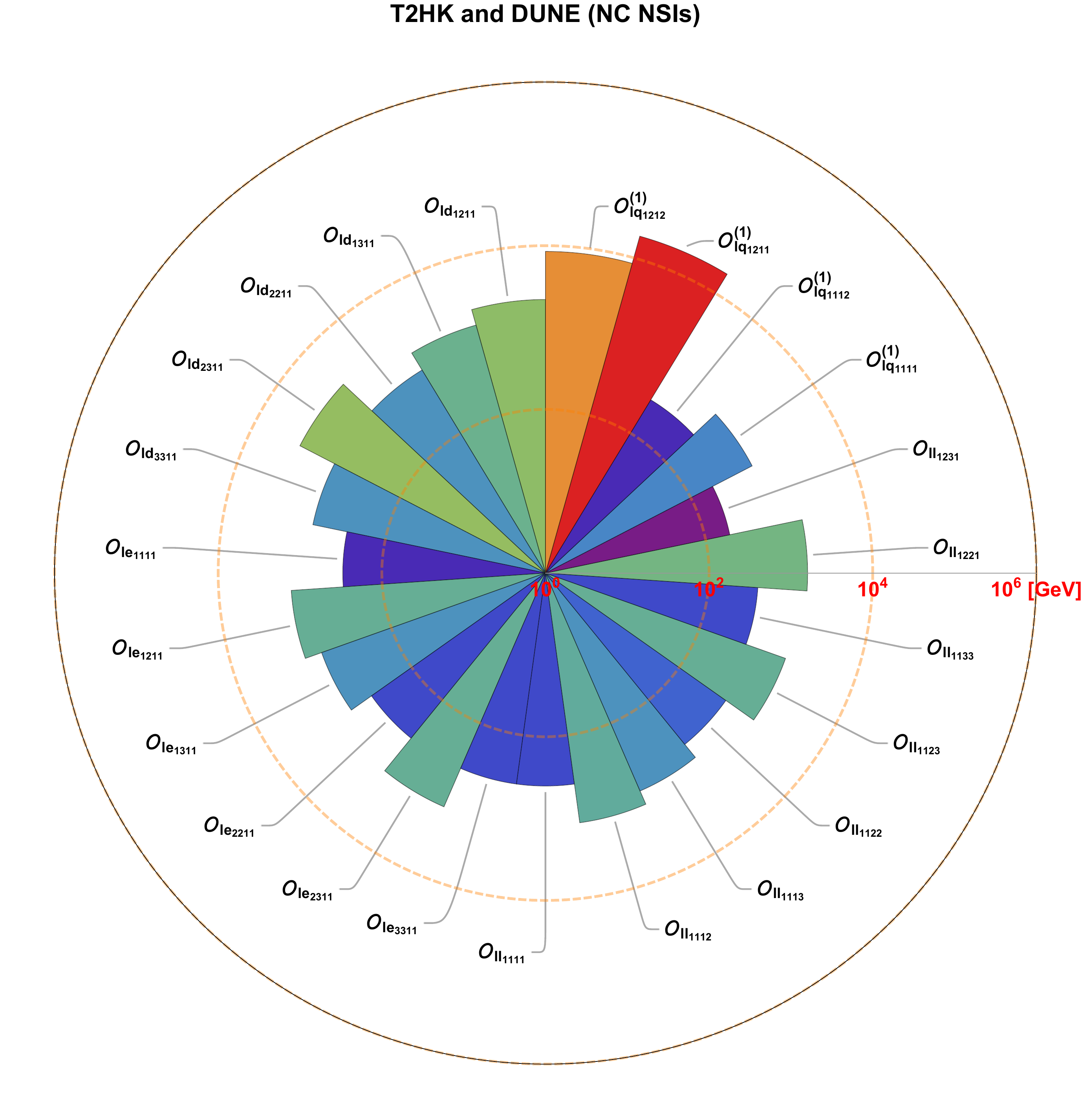} ~& ~ \includegraphics[scale=0.5]{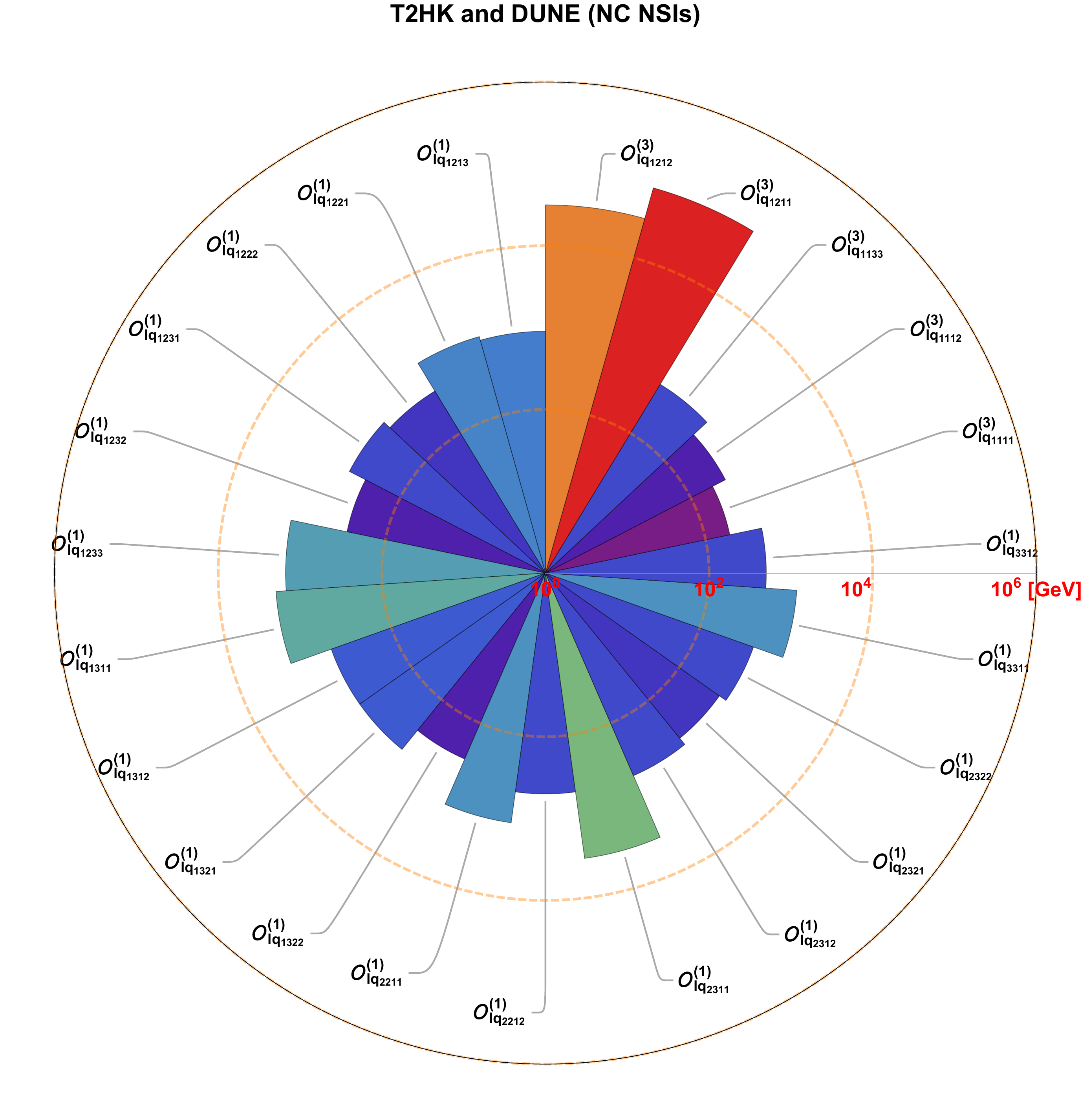}\\
\includegraphics[scale=0.5]{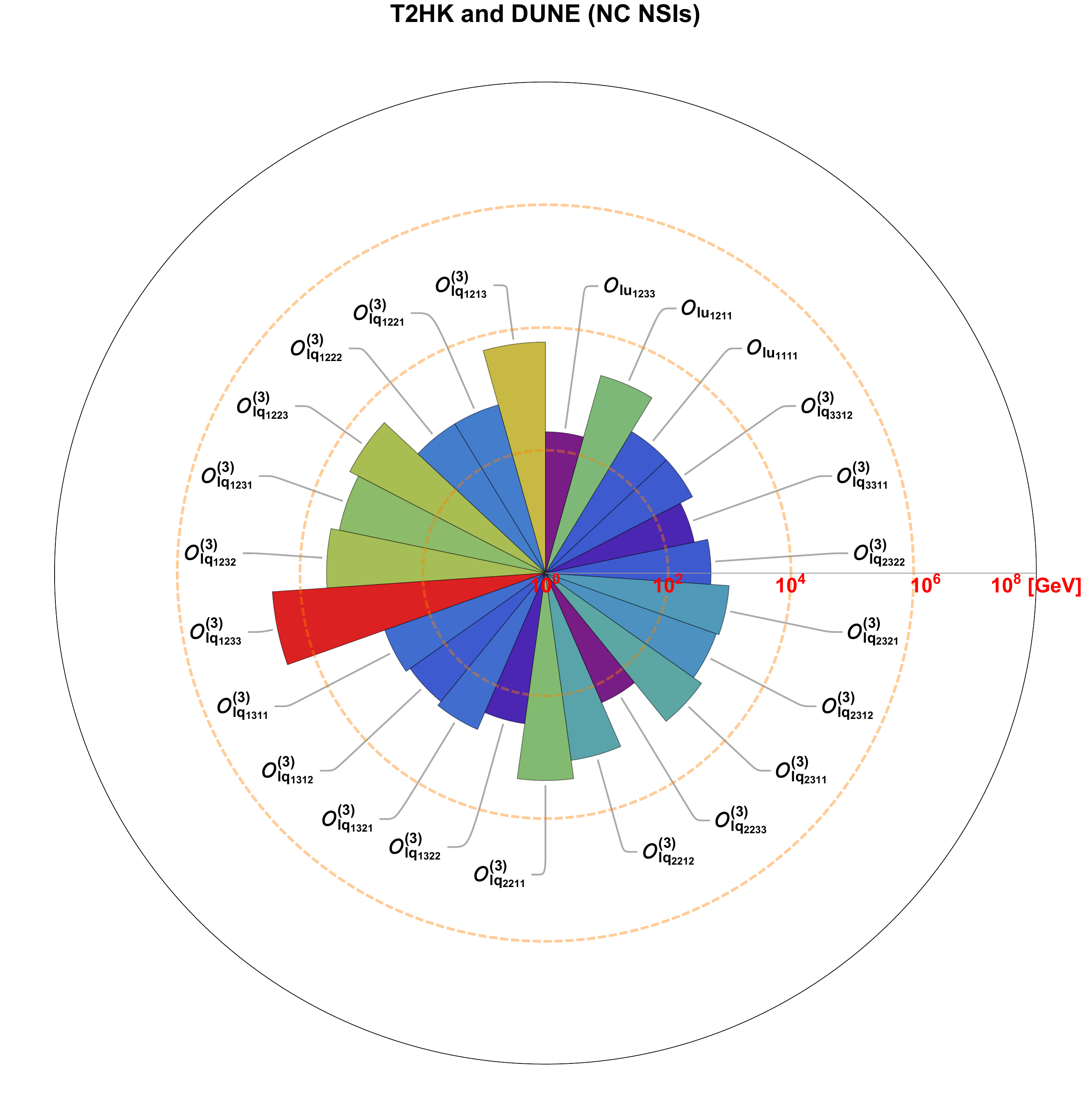} ~& ~ \includegraphics[scale=0.5]{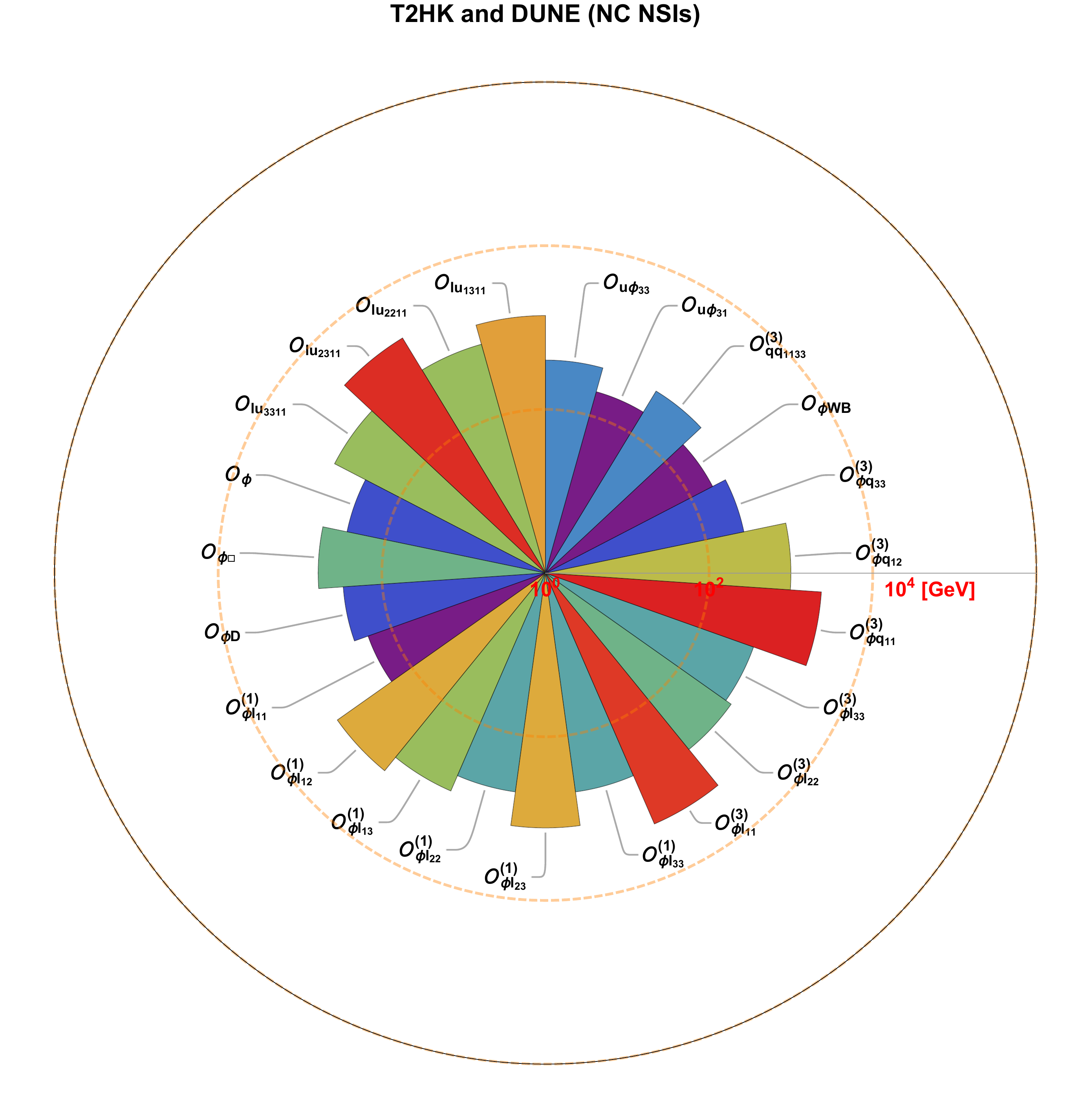}
\end{tabular}
  \end{adjustbox}}
\caption{Figure\,\ref{NC1} continued.}\label{NC2}
\end{figure}

\begin{figure}[t]
\centering{
  \begin{adjustbox}{max width = \textwidth}
\begin{tabular}{cc}
\includegraphics[scale=0.5]{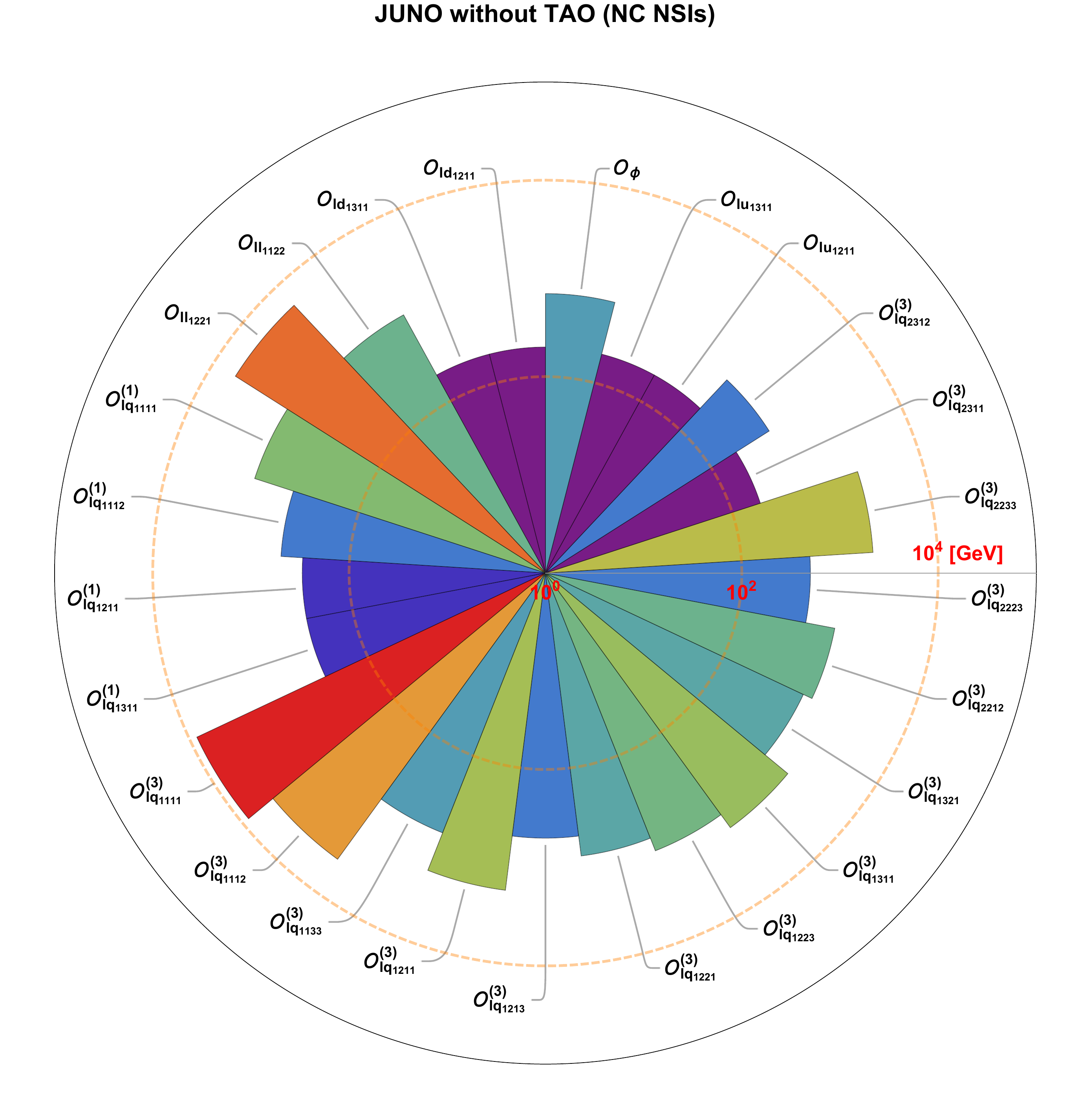} ~& ~ \includegraphics[scale=0.5]{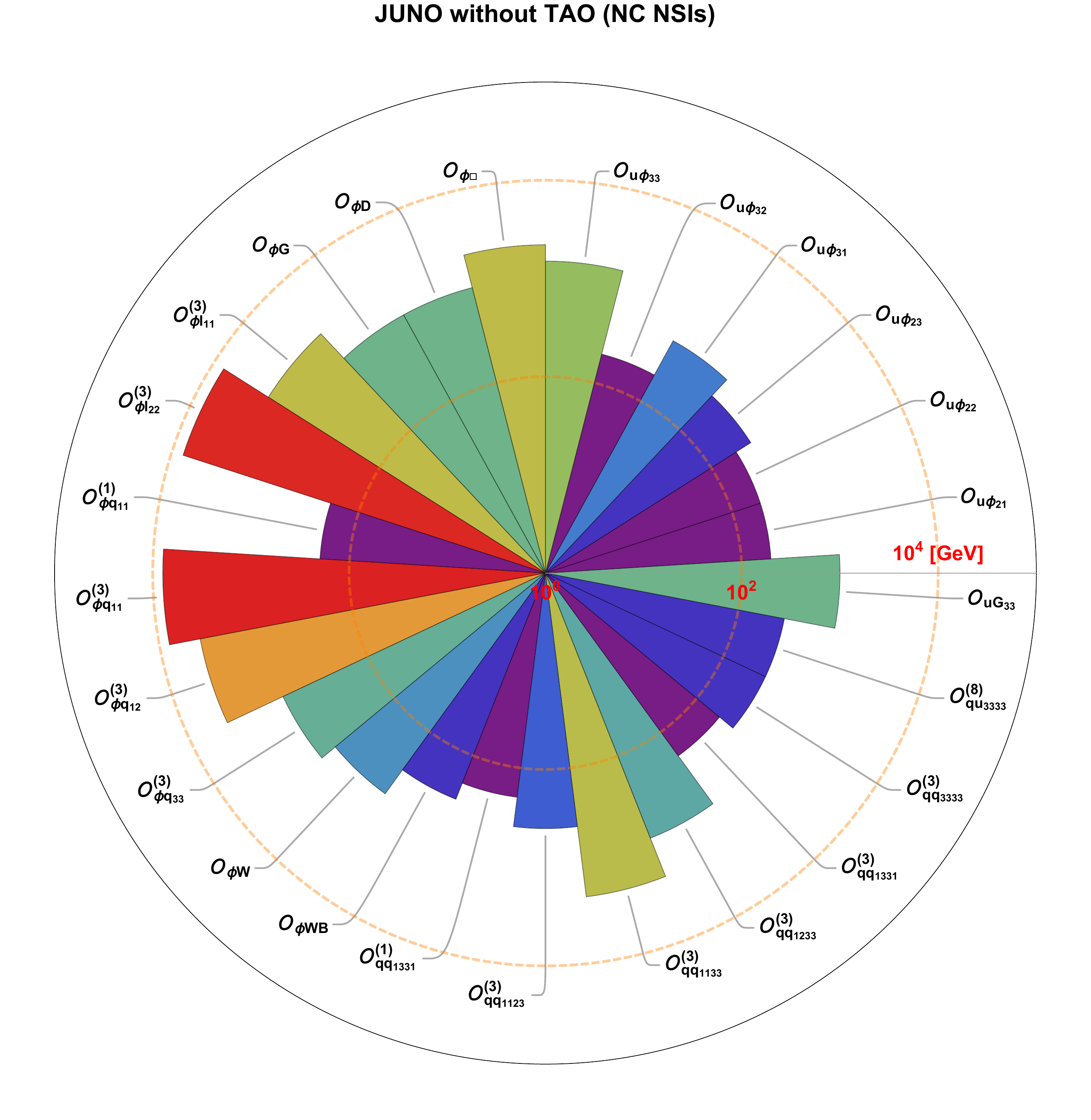}\\
\includegraphics[scale=0.5]{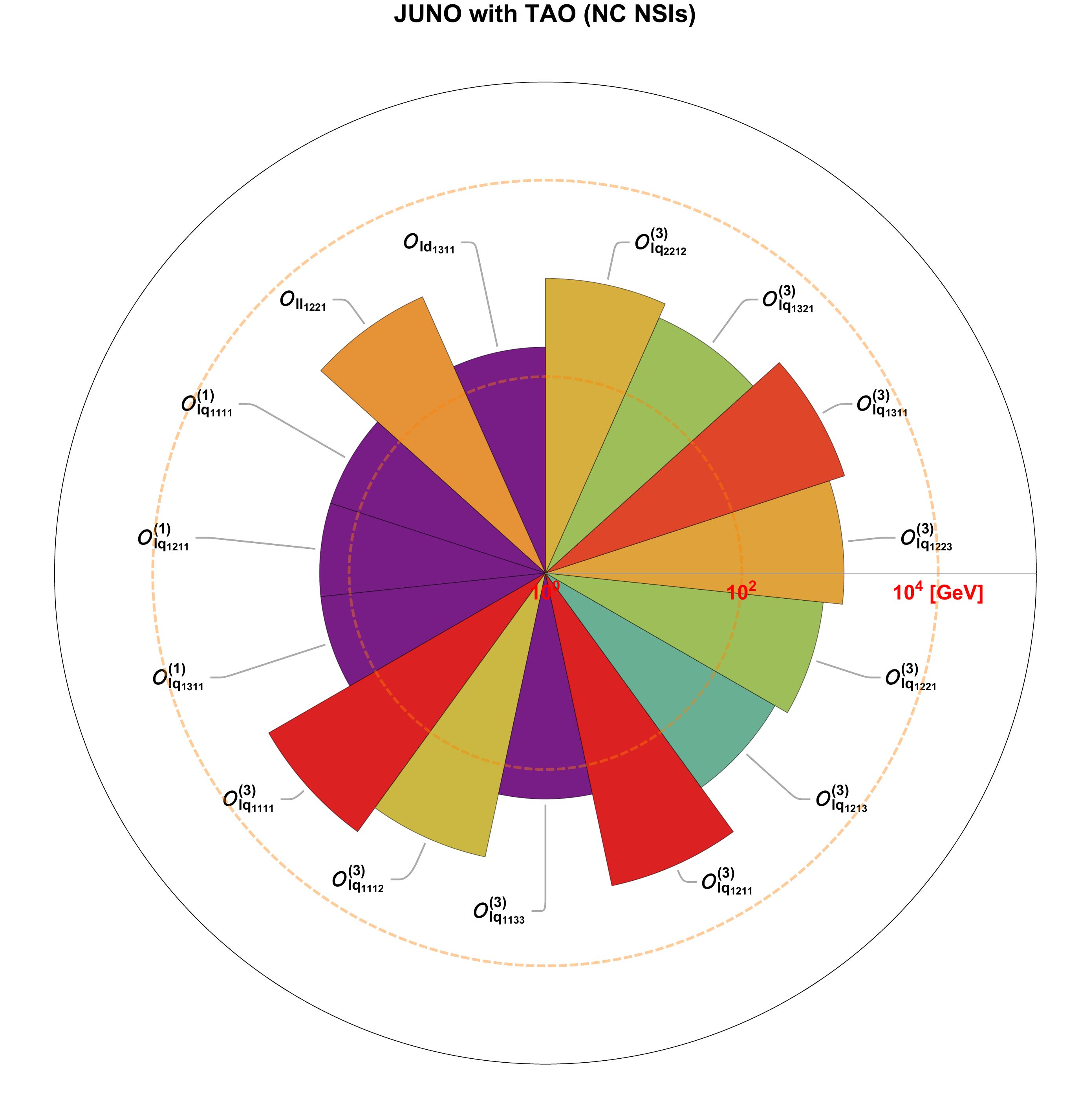} ~& ~ \includegraphics[scale=0.5]{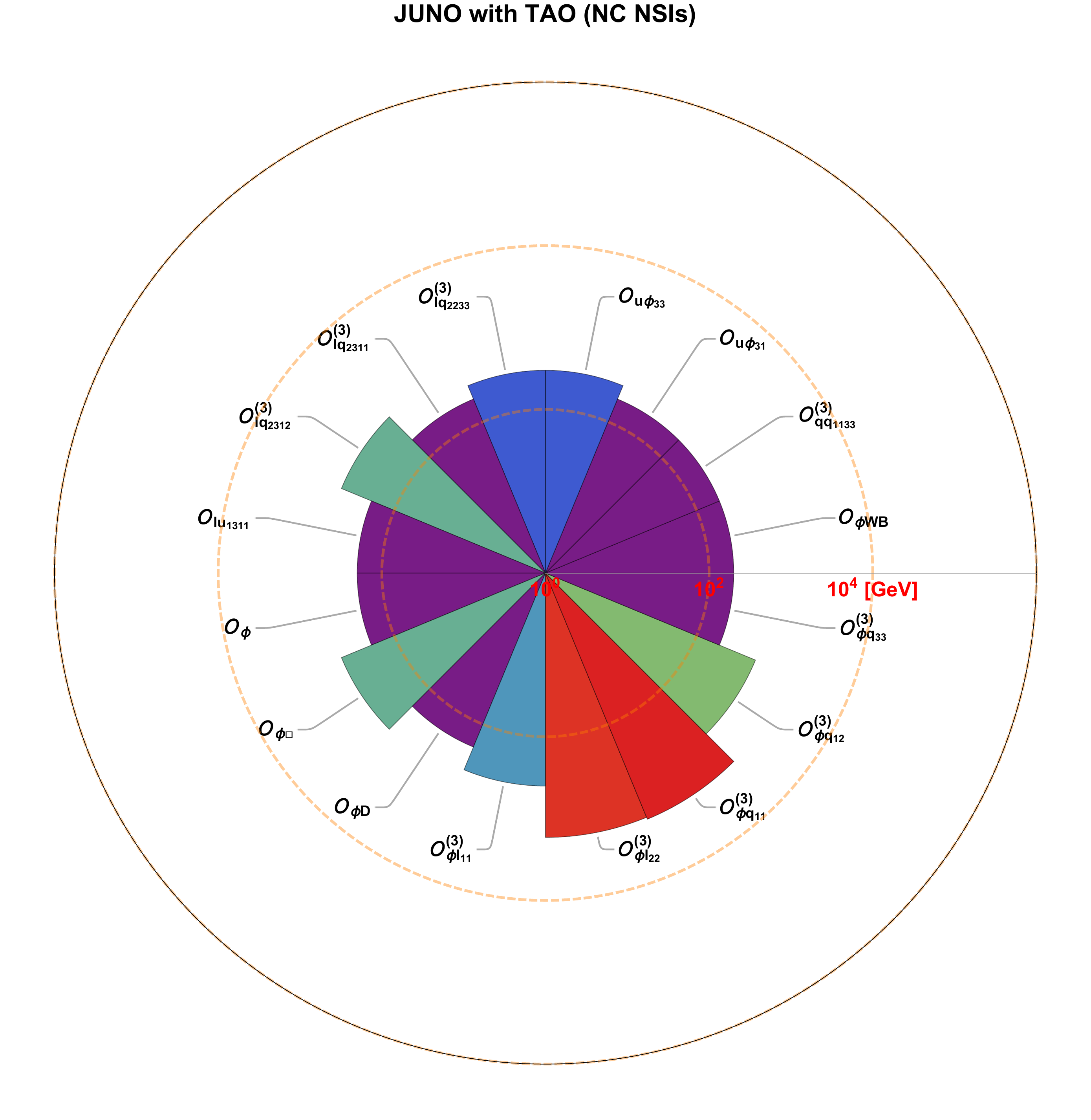}
\end{tabular}
  \end{adjustbox}}
\caption{Figure\,\ref{NC2} continued.}\label{NC3}
\end{figure}

\clearpage

\begin{table}[!thb]
\caption{\label{tab:operator-constraints-2} Sensitivity to the dimension-6 SMEFT operators at 95\% CL in the long-baseline experiments T2HK and DUNE and the reactor experiment JUNO. This table provides constraints for operators from ${\cal O}_{{H}}$ to ${\cal O}_{{Hl}_{33}}$.}
\begin{center}
\resizebox{\linewidth}{!}{%
\begin{tabular}{cccccc}\hline\hline
Operator & T2HK and DUNE limit (TeV) & T2HK limit (TeV) & DUNE limit (TeV) & JUNO and TAO limit (TeV) & JUNO limit (TeV) \\ \hline
\rule{0pt}{3ex}${\cal O}_{H}$	&	0.3	&	0.2	&	0.2	&	0.2	&	0.7	 \\ 
\rule{0pt}{3ex}${\cal O}_{HBox}$	&	0.6	&	0.5	&	0.6	&	0.5	&	2.2	 \\ 
\rule{0pt}{3ex}${\cal O}_{HD}$	&	0.3	&	0.2	&	0.3	&	0.2	&	1.0	 \\ 
\rule{0pt}{3ex}${\cal O}_{HG}$	&	0.1	&	0.1	&	0.1	&	0.1	&	1.0	 \\ 
\rule{0pt}{3ex}${\cal O}_{HB}$	&	0	&	0	&	0	&	0	&	0.1	 \\ 
\rule{0pt}{3ex}${\cal O}_{HW}$	&	0.1	&	0.1	&	0.1	&	0.1	&	0.6	 \\ 
\rule{0pt}{3ex}${\cal O}_{HWB}$	&	0.2	&	0.2	&	0.2	&	0.2	&	0.3	 \\ 
\rule{0pt}{3ex}${\cal O}_{{uH}_{11}}$	&	0	&	0	&	0	&	0	&	0.1	 \\ 
\rule{0pt}{3ex}${\cal O}_{{uH}_{12}}$	&	0.1	&	0.1	&	0.1	&	0.1	&	0.1	 \\ 
\rule{0pt}{3ex}${\cal O}_{{uH}_{13}}$	&	0	&	0	&	0	&	0	&	0.1	 \\ 
\rule{0pt}{3ex}${\cal O}_{{uH}_{21}}$	&	0	&	0	&	0	&	0	&	0.2	 \\ 
\rule{0pt}{3ex}${\cal O}_{{uH}_{22}}$	&	0	&	0	&	0	&	0	&	0.2	 \\ 
\rule{0pt}{3ex}${\cal O}_{{uH}_{23}}$	&	0.1	&	0.1	&	0.1	&	0.1	&	0.3	 \\ 
\rule{0pt}{3ex}${\cal O}_{{uH}_{31}}$	&	0.2	&	0.2 &	0.2	&	0.2	&	0.5	 \\ 
\rule{0pt}{3ex}${\cal O}_{{uH}_{32}}$	&	0	&	0	&	0	&	0	&	0.2	 \\ 
\rule{0pt}{3ex}${\cal O}_{{uH}_{33}}$	&	0.4	&	0.3	&	0.4	&	0.3	&	1.5	 \\ 
\rule{0pt}{3ex}${\cal O}_{{dH}_{13}}$	&	0.1	&	0	&	0.1	&	0	&	0.1	 \\ 
\rule{0pt}{3ex}${\cal O}_{{dH}_{23}}$	&	0	&	0	&	0	&	0	&	0.1	 \\ 
\rule{0pt}{3ex}${\cal O}_{{dH}_{31}}$	&	0	&	0	&	0	&	0	&	0.1	 \\ 
\rule{0pt}{3ex}${\cal O}_{{dH}_{32}}$	&	0	&	0	&	0	&	0	&	0.1	 \\ 
\rule{0pt}{3ex}${\cal O}_{{dH}_{33}}$	&	0	&	0	&	0	&	0	&	0.1	 \\ 
\rule{0pt}{3ex}${\cal O}_{{eH}_{13}}$	&	0	&	0	&	0	&	0	&	0.1	 \\ 
\rule{0pt}{3ex}${\cal O}_{{eH}_{23}}$	&	0	&	0	&	0	&	0	&	0.1	 \\ 
\rule{0pt}{3ex}${\cal O}_{{eH}_{31}}$	&	0	&	0	&	0	&	0	&	0.1	 \\ 
\rule{0pt}{3ex}${\cal O}_{{eH}_{32}}$	&	0	&	0	&	0	&	0	&	0.1	 \\ 
\rule{0pt}{3ex}${\cal O}_{{eH}_{33}}$	&	0	&	0	&	0	&	0	&	0.1	 \\ 
\rule{0pt}{3ex}${\cal O}_{{uG}_{33}}$	&	0	&	0	&	0	&	0	&	1	 \\ 
\rule{0pt}{3ex}${\cal O}_{{Hl}_{11}}^{(1)}$	&	0.2	&	0	&	0.1	&	0	&	0.1	 \\ 
\rule{0pt}{3ex}${\cal O}_{{Hl}_{12}}^{(1)}$	&	1.3	&	1.2	&	0.9	&	0.1	&	0.1	 \\ 
\rule{0pt}{3ex}${\cal O}_{{Hl}_{13}}^{(1)}$	&	0.8	&	0.3	&	0.7	&	0.1	&	0.1	 \\ 
\rule{0pt}{3ex}${\cal O}_{{Hl}_{22}}^{(1)}$	&	0.5	&	0.2	&	0.4	&	0	&	0	 \\ 
\rule{0pt}{3ex}${\cal O}_{{Hl}_{23}}^{(1)}$	&	1.3	&	0.7	&	1.3	&	0	&	0	 \\ 
\rule{0pt}{3ex}${\cal O}_{{Hl}_{33}}^{(1)}$	&	0.5	&	0.2	&	0.4	&	0	&	0	 \\ 
\rule{0pt}{3ex}${\cal O}_{{Hl}_{11}}^{(3)}$	&	2.2	&	1.7	&	2.1	&	0.4	&	2.2	 \\ 
\rule{0pt}{3ex}${\cal O}_{{Hl}_{12}}^{(3)}$	&	84.9	&	84.9	&	3.7	&	1.7	&	1.7	 \\ 
\rule{0pt}{3ex}${\cal O}_{{Hl}_{13}}^{(3)}$	&	0.7	&	0.2	&	0.7	&	1.5	&	1.6	 \\ 
\rule{0pt}{3ex}${\cal O}_{{Hl}_{22}}^{(3)}$	&	0.6	&	0.4	&	0.6	&	1.7	&	7.6	 \\ 
\rule{0pt}{3ex}${\cal O}_{{Hl}_{23}}^{(3)}$	&	1.5	&	1.0	&	1.4	&	0.1	&	0.3	 \\ 
\rule{0pt}{3ex}${\cal O}_{{Hl}_{33}}^{(3)}$	&	0.5	&	0.2	&	0.5	&	0	&	0	 \\ 
\hline\hline
\end{tabular}}
\end{center}
\end{table}

\begin{table}[!ht]
\caption{\label{tab:operator-constraints-3} Sensitivity to the dimension-6 SMEFT operators at 95\% CL in the long-baseline experiments T2HK and DUNE and the reactor experiment JUNO. This table provides constraints for operators from ${\cal O}_{{Hq}_{11}^{(1)}}$ to ${\cal O}_{{lq}_{1311}^{(1)}}$.}
\begin{center}
\resizebox{\linewidth}{!}{%
\begin{tabular}{cccccc}\hline\hline
Operator & T2HK and DUNE limit (TeV) & T2HK limit (TeV) & DUNE limit (TeV) & JUNO and TAO limit (TeV) & JUNO limit (TeV) \\ \hline
\rule{0pt}{3ex}${\cal O}_{{Hq}_{11}}^{(1)}$	&	0	&	0	&	0	&	0	&	0.2	 \\ 
\rule{0pt}{3ex}${\cal O}_{{Hq}_{12}}^{(1)}$	&	0	&	0	&	0	&	0	&	0.1	 \\ 
\rule{0pt}{3ex}${\cal O}_{{Hq}_{11}}^{(3)}$	&	2.4	&	1.7	&	2.2	&	1.8	&	7.9	 \\ 
\rule{0pt}{3ex}${\cal O}_{{Hq}_{12}}^{(3)}$	&	1.0	&	0.6	&	0.8	&	0.6	&	3.8	 \\ 
\rule{0pt}{3ex}${\cal O}_{{Hq}_{33}}^{(3)}$	&	0.3	&	0.2	&	0.3	&	0.2	&	0.9	 \\ 
\rule{0pt}{3ex}${\cal O}_{{Hud}_{11}}$	&	1.6	&	1.2	&	1.5	&	1.1	&	5.0	 \\ 
\rule{0pt}{3ex}${\cal O}_{{ll}_{1111}}$	&	0.4	&	0.2	&	0.4	&	0	&	0	 \\ 
\rule{0pt}{3ex}${\cal O}_{{ll}_{1112}}$	&	1.2	&	1.1	&	0.9	&	0.1	&	0.1	 \\ 
\rule{0pt}{3ex}${\cal O}_{{ll}_{1113}}$	&	0.8	&	0.3	&	0.7	&	0.1	&	0.1	 \\ 
\rule{0pt}{3ex}${\cal O}_{{ll}_{1122}}$	&	0.5	&	0.2	&	0.4	&	0	&	1.0	 \\ 
\rule{0pt}{3ex}${\cal O}_{{ll}_{1123}}$	&	1.3	&	0.7	&	1.3	&	0	&	0	 \\ 
\rule{0pt}{3ex}${\cal O}_{{ll}_{1133}}$	&	0.4	&	0.2	&	0.4	&	0	&	0	 \\ 
\rule{0pt}{3ex}${\cal O}_{{ll}_{1221}}$	&	1.6	&	1.2	&	1.5	&	1.2	&	5.5	 \\ 
\rule{0pt}{3ex}${\cal O}_{{ll}_{1231}}$	&	0.2	&	0	&	0.2	&	0	&	0	 \\ 
\rule{0pt}{3ex}${\cal O}_{{qq}_{1231}}^{(1)}$	&	0	&	0	&	0	&	0	&	0.1	 \\ 
\rule{0pt}{3ex}${\cal O}_{{qq}_{1331}}^{(1)}$	&	0.1	&	0.1	&	0.1	&	0.1	&	0.2	 \\ 
\rule{0pt}{3ex}${\cal O}_{{qq}_{1332}}^{(1)}$	&	0	&	0	&	0	&	0	&	0.1	 \\ 
\rule{0pt}{3ex}${\cal O}_{{qq}_{1123}}^{(3)}$	&	0	&	0	&	0	&	0	&	0.4	 \\ 
\rule{0pt}{3ex}${\cal O}_{{qq}_{1133}}^{(3)}$	&	0.4	&	0.1	&	0.3	&	0.2	&	2.1	 \\ 
\rule{0pt}{3ex}${\cal O}_{{qq}_{1231}}^{(3)}$	&	0	&	0	&	0	&	0	&	0.1	 \\ 
\rule{0pt}{3ex}${\cal O}_{{qq}_{1233}}^{(3)}$	&	0.1	&	0	&	0.1	&	0	&	0.8	 \\ 
\rule{0pt}{3ex}${\cal O}_{{qq}_{1331}}^{(3)}$	&	0.1	&	0.1	&	0.1	&	0.1	&	0.2	 \\ 
\rule{0pt}{3ex}${\cal O}_{{qq}_{1332}}^{(3)}$	&	0	&	0	&	0	&	0	&	0.1	 \\ 
\rule{0pt}{3ex}${\cal O}_{{qq}_{3333}}^{(3)}$	&	0	&	0	&	0	&	0	&	0.3	 \\ 
\rule{0pt}{3ex}${\cal O}_{{lq}_{1111}}^{(1)}$	&	0.7	&	0.3	&	0.7	&	0.2	&	1.3	 \\ 
\rule{0pt}{3ex}${\cal O}_{{lq}_{1112}}^{(1)}$	&	0.3	&	0.1	&	0.3	&	0	&	0.5	 \\ 
\rule{0pt}{3ex}${\cal O}_{{lq}_{1122}}^{(1)}$	&	0.1	&	0	&	0.1	&	0	&	0	 \\ 
\rule{0pt}{3ex}${\cal O}_{{lq}_{1211}}^{(1)}$	&	18.8	&	18.6	&	2.3	&	0.2 &   0.3	 \\ 
\rule{0pt}{3ex}${\cal O}_{{lq}_{1212}}^{(1)}$	&	8.5	&	8.5	&	0.7	&	0	&	0	 \\ 
\rule{0pt}{3ex}${\cal O}_{{lq}_{1213}}^{(1)}$	&	0.9	&	0.9	&	0	&	0	&	0	 \\ 
\rule{0pt}{3ex}${\cal O}_{{lq}_{1221}}^{(1)}$	&	1.0	&	0.8	&	0.7	&	0	&	0.1	 \\ 
\rule{0pt}{3ex}${\cal O}_{{lq}_{1222}}^{(1)}$	&	0.4	&	0.4	&	0.3	&	0	&	0	 \\ 
\rule{0pt}{3ex}${\cal O}_{{lq}_{1231}}^{(1)}$	&	0.5	&	0.5	&	0.1	&	0	&	0	 \\ 
\rule{0pt}{3ex}${\cal O}_{{lq}_{1232}}^{(1)}$	&	0.3	&	0.3	&	0	&	0	&	0	 \\ 
\rule{0pt}{3ex}${\cal O}_{{lq}_{1233}}^{(1)}$	&	1.5	&	1.5	&	0.1	&	0	&	0	 \\ 
\rule{0pt}{3ex}${\cal O}_{{lq}_{1311}}^{(1)}$	&	2.0	&	0.7	&	1.8	&	0.2	&	0.3	 \\ 
\hline\hline
\end{tabular}}
\end{center}
\end{table}

\begin{table}[!ht]
\caption{\label{tab:operator-constraints-4} Sensitivity to the dimension-6 SMEFT operators at 95\% CL in the long-baseline experiments T2HK and DUNE and the reactor experiment JUNO. This table provides constraints for operators from ${\cal O}_{{lq}_{1312}^{(1)}}$ to ${\cal O}_{{lq}_{2212}}^{(3)}$.}
\begin{center}
\resizebox{\linewidth}{!}{%
\begin{tabular}{cccccc}\hline\hline
Operator & T2HK and DUNE limit (TeV) & T2HK limit (TeV) & DUNE limit (TeV) & JUNO and TAO limit (TeV) & JUNO limit (TeV) \\ \hline
\rule{0pt}{3ex}${\cal O}_{{lq}_{1312}}^{(1)}$	&	0.6	&	0.2	&	0.6	&	0	&	0.1	 \\ 
\rule{0pt}{3ex}${\cal O}_{{lq}_{1321}}^{(1)}$	&	0.6	&	0.2	&	0.6	&	0.1	&	0.1	 \\ 
\rule{0pt}{3ex}${\cal O}_{{lq}_{1322}}^{(1)}$	&	0.3	&	0.1	&	0.3	&	0	&	0	 \\ 
\rule{0pt}{3ex}${\cal O}_{{lq}_{1333}}^{(1)}$	&	0.1	&	0	&	0.1	&	0	&	0	 \\ 
\rule{0pt}{3ex}${\cal O}_{{lq}_{2211}}^{(1)}$	&	1.2	&	0.6	&	1.1	&	0.1	&	0.1	 \\ 
\rule{0pt}{3ex}${\cal O}_{{lq}_{2212}}^{(1)}$	&	0.5	&	0.3	&	0.5	&	0	&	0	 \\ 
\rule{0pt}{3ex}${\cal O}_{{lq}_{2222}}^{(1)}$	&	0.1	&	0.1	&	0.1	&	0	&	0	 \\ 
\rule{0pt}{3ex}${\cal O}_{{lq}_{2311}}^{(1)}$	&	3.3	&	1.7	&	3.1	&	0.1	&	0.1	 \\ 
\rule{0pt}{3ex}${\cal O}_{{lq}_{2312}}^{(1)}$	&	0.5	&	0.2	&	0.5	&	0	&	0	 \\ 
\rule{0pt}{3ex}${\cal O}_{{lq}_{2313}}^{(1)}$	&	0.1	&	0	&	0.1	&	0	&	0	 \\ 
\rule{0pt}{3ex}${\cal O}_{{lq}_{2321}}^{(1)}$	&	0.4	&	0.4	&	0.4	&	0	&	0	 \\ 
\rule{0pt}{3ex}${\cal O}_{{lq}_{2322}}^{(1)}$	&	0.5	&	0.2	&	0.5	&	0	&	0	 \\ 
\rule{0pt}{3ex}${\cal O}_{{lq}_{2333}}^{(1)}$	&	0.1	&	0.1	&	0.1	&	0	&	0	 \\ 
\rule{0pt}{3ex}${\cal O}_{{lq}_{3311}}^{(1)}$	&	1.2	&	0.6	&	1.1	&	0.1	&	0.1	 \\ 
\rule{0pt}{3ex}${\cal O}_{{lq}_{3312}}^{(1)}$	&	0.5	&	0.3	&	0.5	&	0	&	0	 \\ 
\rule{0pt}{3ex}${\cal O}_{{lq}_{3322}}^{(1)}$	&	0.1	&	0.1	&	0.1	&	0	&	0	 \\ 
\rule{0pt}{3ex}${\cal O}_{{lq}_{1111}}^{(3)}$	&	0.2	&	0	&	0.2	&	1.8	&	8.4	 \\ 
\rule{0pt}{3ex}${\cal O}_{{lq}_{1112}}^{(3)}$	&	0.3	&	0.1	&	0.3	&	0.9	&	4	 \\ 
\rule{0pt}{3ex}${\cal O}_{{lq}_{1122}}^{(3)}$	&	0.1	&	0	&	0.1	&	0	&	0	 \\ 
\rule{0pt}{3ex}${\cal O}_{{lq}_{1133}}^{(3)}$	&	0.5	&	0.3	&	0.4	&	0.2	&	0.7	 \\ 
\rule{0pt}{3ex}${\cal O}_{{lq}_{1211}}^{(3)}$	&	77.1	&	77.1	&	3.9	&	1.8	&	1.8	 \\ 
\rule{0pt}{3ex}${\cal O}_{{lq}_{1212}}^{(3)}$	&	31.5	&	31.5	&	1.8	&	0.1	&	0.1	 \\ 
\rule{0pt}{3ex}${\cal O}_{{lq}_{1213}}^{(3)}$	&	5.8	&	5.8	&	0.5	&	0.5	&	0.5	 \\ 
\rule{0pt}{3ex}${\cal O}_{{lq}_{1221}}^{(3)}$	&	0.7	&	0.7	&	0.6	&	0.7	&	0.8	 \\ 
\rule{0pt}{3ex}${\cal O}_{{lq}_{1222}}^{(3)}$	&	0.7	&	0.7	&	0.3	&	0.1	&	0.1	 \\ 
\rule{0pt}{3ex}${\cal O}_{{lq}_{1223}}^{(3)}$	&	3.9	&	3.9	&	1.1	&	1.1	&	1.1	 \\ 
\rule{0pt}{3ex}${\cal O}_{{lq}_{1231}}^{(3)}$	&	2.7	&	2.7	&	0	&	0.1	&	0.1	 \\ 
\rule{0pt}{3ex}${\cal O}_{{lq}_{1232}}^{(3)}$	&	3.7	&	3.7	&	0	&	0	&	0	 \\ 
\rule{0pt}{3ex}${\cal O}_{{lq}_{1233}}^{(3)}$	&	28.9	&	28.9	&	0.8	&	0.1	&	0.1	 \\ 
\rule{0pt}{3ex}${\cal O}_{{lq}_{1311}}^{(3)}$	&	0.6	&	0.2	&	0.6	&	1.6	&	1.6	 \\ 
\rule{0pt}{3ex}${\cal O}_{{lq}_{1312}}^{(3)}$	&	0.5	&	0.2	&	0.5	&	0	&	0.1	 \\ 
\rule{0pt}{3ex}${\cal O}_{{lq}_{1321}}^{(3)}$	&	0.6	&	0.2	&	0.6	&	0.7	&	0.8	 \\ 
\rule{0pt}{3ex}${\cal O}_{{lq}_{1322}}^{(3)}$	&	0.3	&	0.1	&	0.3	&	0	&	0	 \\ 
\rule{0pt}{3ex}${\cal O}_{{lq}_{1333}}^{(3)}$	&	0	&	0	&	0	&	0.1	&	0.1	 \\ 
\rule{0pt}{3ex}${\cal O}_{{lq}_{2211}}^{(3)}$	&	2.4	&	1.8	&	2.2	&	0.1	&	0.1	 \\ 
\rule{0pt}{3ex}${\cal O}_{{lq}_{2212}}^{(3)}$	&	1.2	&	1.0	&	1.1	&	1.0	&	1.0	 \\ 
\hline\hline
\end{tabular}}
\end{center}
\end{table}

\begin{table}[!ht]
\caption{\label{tab:operator-constraints-5} Sensitivity to the dimension-6 SMEFT operators at 95\% CL in the long-baseline experiments T2HK and DUNE and the reactor experiment JUNO. This table provides constraints for operators from ${\cal O}_{{lq}_{2223}^{(3)}}$ to ${\cal O}_{{ledq}_{1213}}$.}
\begin{center}
\resizebox{\linewidth}{!}{%
\begin{tabular}{cccccc}\hline\hline
Operator & T2HK and DUNE limit (TeV) & T2HK limit (TeV) & DUNE limit (TeV) & JUNO and TAO limit (TeV) & JUNO limit (TeV) \\ \hline
\rule{0pt}{3ex}${\cal O}_{{lq}_{2223}}^{(3)}$	&	0	&	0	&	0	&	0	&	0.5	 \\ 
\rule{0pt}{3ex}${\cal O}_{{lq}_{2233}}^{(3)}$	&	0.2	&	0.1	&	0.2	&	0.3	&	2.2	 \\ 
\rule{0pt}{3ex}${\cal O}_{{lq}_{2311}}^{(3)}$	&	1.3	&	1.0	&	1.3	&	0.2	&	0.2	 \\ 
\rule{0pt}{3ex}${\cal O}_{{lq}_{2312}}^{(3)}$	&	0.9	&	0.5	&	0.9	&	0.5	&	0.5	 \\ 
\rule{0pt}{3ex}${\cal O}_{{lq}_{2321}}^{(3)}$	&	1.0	&	0.9	&	0.5	&	0.1	&	0.1	 \\ 
\rule{0pt}{3ex}${\cal O}_{{lq}_{2322}}^{(3)}$	&	0.5	&	0.1	&	0.4	&	0.1	&	0.1	 \\ 
\rule{0pt}{3ex}${\cal O}_{{lq}_{3311}}^{(3)}$	&	0.3	&	0.1	&	0.3	&	0	&	0	 \\ 
\rule{0pt}{3ex}${\cal O}_{{lq}_{3312}}^{(3)}$	&	0.5	&	0.3	&	0.5	&	0	&	0	 \\ 
\rule{0pt}{3ex}${\cal O}_{{lq}_{3322}}^{(3)}$	&	0.1	&	0.1	&	0.1	&	0	&	0	 \\ 
\rule{0pt}{3ex}${\cal O}_{{le}_{1111}}$	&	0.3	&	0.1	&	0.3	&	0	&	0	 \\ 
\rule{0pt}{3ex}${\cal O}_{{le}_{1211}}$	&	1.3	&	1.2	&	0.9	&	0.1	&	0.1	 \\ 
\rule{0pt}{3ex}${\cal O}_{{le}_{1311}}$	&	0.8	&	0.3	&	0.7	&	0.1	&	0.1	 \\ 
\rule{0pt}{3ex}${\cal O}_{{le}_{2211}}$	&	0.4	&	0.2	&	0.4	&	0	&	0	 \\ 
\rule{0pt}{3ex}${\cal O}_{{le}_{2311}}$	&	1.3	&	0.7	&	1.3	&	0	&	0	 \\ 
\rule{0pt}{3ex}${\cal O}_{{le}_{3311}}$	&	0.4	&	0.2	&	0.4	&	0	&	0	 \\ 
\rule{0pt}{3ex}${\cal O}_{{lu}_{1111}}$	&	0.5	&	0.2	&	0.5	&	0	&	0	 \\ 
\rule{0pt}{3ex}${\cal O}_{{lu}_{1211}}$	&	2.2	&	2.0	&	1.6	&	0.1	&	0.2	 \\ 
\rule{0pt}{3ex}${\cal O}_{{lu}_{1233}}$	&	0.2	&	0.2	&	0.1	&	0	&	0	 \\ 
\rule{0pt}{3ex}${\cal O}_{{lu}_{1311}}$	&	1.4	&	0.5	&	1.2	&	0.2	&	0.2	 \\ 
\rule{0pt}{3ex}${\cal O}_{{lu}_{1333}}$	&	0.1	&	0	&	0.1	&	0	&	0	 \\ 
\rule{0pt}{3ex}${\cal O}_{{lu}_{2211}}$	&	0.8	&	0.4	&	0.8	&	0	&	0	 \\ 
\rule{0pt}{3ex}${\cal O}_{{lu}_{2311}}$	&	2.3	&	1.2	&	2.2	&	0	&	0.1	 \\ 
\rule{0pt}{3ex}${\cal O}_{{lu}_{2333}}$	&	0.1	&	0.1	&	0.1	&	0	&	0	 \\ 
\rule{0pt}{3ex}${\cal O}_{{lu}_{3311}}$	&	0.8	&	0.4	&	0.8	&	0	&	0	 \\ 
\rule{0pt}{3ex}${\cal O}_{{ld}_{1211}}$	&	2.2	&	2.0	&	1.6	&	0.1	&	0.2	 \\ 
\rule{0pt}{3ex}${\cal O}_{{ld}_{1311}}$	&	1.4	&	0.5	&	1.3	&	0.2	&	0.2	 \\ 
\rule{0pt}{3ex}${\cal O}_{{ld}_{2211}}$	&	0.8	&	0.4	&	0.8	&	0	&	0	 \\ 
\rule{0pt}{3ex}${\cal O}_{{ld}_{2311}}$	&	2.4	&	1.2	&	2.3	&	0	&	0.1	 \\ 
\rule{0pt}{3ex}${\cal O}_{{ld}_{3311}}$	&	0.8	&	0.4	&	0.8	&	0	&	0	 \\ 
\rule{0pt}{3ex}${\cal O}_{{qu}_{3333}}^{(8)}$	&	0	&	0	&	0	&	0	&	0.3	 \\ 
\rule{0pt}{3ex}${\cal O}_{{ledq}_{1111}}$	&	0	&	0	&	0	&	0.2	&	1.2	 \\ 
\rule{0pt}{3ex}${\cal O}_{{ledq}_{1112}}$	&	0	&	0	&	0	&	0.1	&	0.5	 \\ 
\rule{0pt}{3ex}${\cal O}_{{ledq}_{1211}}$	&	454.2	&	454.2	&	19.3	&	1.2	&	1.2	 \\ 
\rule{0pt}{3ex}${\cal O}_{{ledq}_{1212}}$	&	155.6	&	155.6	&	9.1	&	0.6	&	0.6	 \\ 
\rule{0pt}{3ex}${\cal O}_{{ledq}_{1213}}$	&	29.2	&	29.2	&	1.2	&	0	&	0	 \\ 
\hline\hline
\end{tabular}}
\end{center}
\end{table}

\begin{table}[!t]
\caption{\label{tab:operator-constraints-6} Sensitivity to the dimension-6 SMEFT operators at 95\% CL in the long-baseline experiments T2HK and DUNE and the reactor experiment JUNO. This table provides constraints for operators from ${\cal O}_{{ledq}_{2111}}$ to ${\cal O}_{{lequ}_{3231}}^{(3)}$.}
\begin{center}
\resizebox{\linewidth}{!}{%
\begin{tabular}{cccccc}\hline\hline
Operator & T2HK and DUNE limit (TeV) & T2HK limit (TeV) & DUNE limit (TeV) & JUNO and TAO limit (TeV) & JUNO limit (TeV) \\ \hline
\rule{0pt}{3ex}${\cal O}_{{ledq}_{2111}}$	&	0	&	0	&	0	&	0.2	&	0.2	 \\ 
\rule{0pt}{3ex}${\cal O}_{{ledq}_{2112}}$	&	0	&	0	&	0	&	0.1	&	0.1	 \\ 
\rule{0pt}{3ex}${\cal O}_{{ledq}_{2211}}$	&	12.3	&	9.1	&	11.2	&	0.7	&	0.7	 \\ 
\rule{0pt}{3ex}${\cal O}_{{ledq}_{2212}}$	&	5.8	&	4.3	&	5.3	&	0.7	&	0.7	 \\ 
\rule{0pt}{3ex}${\cal O}_{{ledq}_{2213}}$	&	0.4	&	0.3	&	0.4	&	0	&	0	 \\ 
\rule{0pt}{3ex}${\cal O}_{{ledq}_{3111}}$	&	0	&	0	&	0	&	0.2	&	0.2	 \\ 
\rule{0pt}{3ex}${\cal O}_{{ledq}_{3112}}$	&	0	&	0	&	0	&	0.1	&	0.1	 \\ 
\rule{0pt}{3ex}${\cal O}_{{ledq}_{3211}}$	&	5.7	&	5.1	&	5.4	&	1.2	&	1.2	 \\ 
\rule{0pt}{3ex}${\cal O}_{{ledq}_{3212}}$	&	3.2	&	2.6	&	2.6	&	0.6	&	0.6	 \\ 
\rule{0pt}{3ex}${\cal O}_{{ledq}_{3213}}$	&	0.8	&	0.7	&	0.5	&	0	&	0	 \\ 
\rule{0pt}{3ex}${\cal O}_{{lequ}_{1111}}^{(1)}$	&	0	&	0	&	0	&	0.2	&	1.2	 \\ 
\rule{0pt}{3ex}${\cal O}_{{lequ}_{1211}}^{(1)}$	&	393.2	&	393.2	&	20.0	&	1.2	&	1.2	 \\ 
\rule{0pt}{3ex}${\cal O}_{{lequ}_{1231}}^{(1)}$	&	7.4	&	7.4	&	0	&	0	&	0	 \\ 
\rule{0pt}{3ex}${\cal O}_{{lequ}_{2111}}^{(1)}$	&	0	&	0	&	0	&	0.2	&	0.2	 \\ 
\rule{0pt}{3ex}${\cal O}_{{lequ}_{2211}}^{(1)}$	&	12.5	&	9.3	&	11.5	&	2.0	&	2.0	 \\ 
\rule{0pt}{3ex}${\cal O}_{{lequ}_{3111}}^{(1)}$	&	0	&	0	&	0	&	0.2	&	0.2	 \\ 
\rule{0pt}{3ex}${\cal O}_{{lequ}_{3211}}^{(1)}$	&	5.8	&	5.2	&	5.5	&	1.2	&	1.2	 \\ 
\rule{0pt}{3ex}${\cal O}_{{lequ}_{1111}}^{(3)}$	&	0	&	0	&	0	&	1.0	&	4.6	 \\ 
\rule{0pt}{3ex}${\cal O}_{{lequ}_{1211}}^{(3)}$	&	256.2	&	256.1	&	10.5	&	0	&	0	 \\ 
\rule{0pt}{3ex}${\cal O}_{{lequ}_{1231}}^{(3)}$	&	7.2	&	7.2	&	0.2	&	0	&	0	 \\ 
\rule{0pt}{3ex}${\cal O}_{{lequ}_{2111}}^{(3)}$	&	0	&	0	&	0	&	0.9	&	1	 \\ 
\rule{0pt}{3ex}${\cal O}_{{lequ}_{2211}}^{(3)}$	&	6.2	&	4.5	&	5.6	&	0.1	&	0.1	 \\ 
\rule{0pt}{3ex}${\cal O}_{{lequ}_{2231}}^{(3)}$	&	0.1	&	0.1	&	0.1	&	0	&	0	 \\ 
\rule{0pt}{3ex}${\cal O}_{{lequ}_{3111}}^{(3)}$	&	0	&	0	&	0	&	0.9	&	0.9	 \\ 
\rule{0pt}{3ex}${\cal O}_{{lequ}_{3211}}^{(3)}$	&	2.6	&	2.3	&	2.5	&	0.4	&	0.4	 \\ 
\rule{0pt}{3ex}${\cal O}_{{lequ}_{3231}}^{(3)}$	&	0.1	&	0.1	&	0	&	0	&	0	 \\ 
\hline\hline
\end{tabular}}
\end{center}
\end{table}

\clearpage

\section{\label{app:B}Effects of electron appearance in T2HK and DUNE near detectors}

\begin{figure}[!h]
\center{\includegraphics[width=\textwidth]{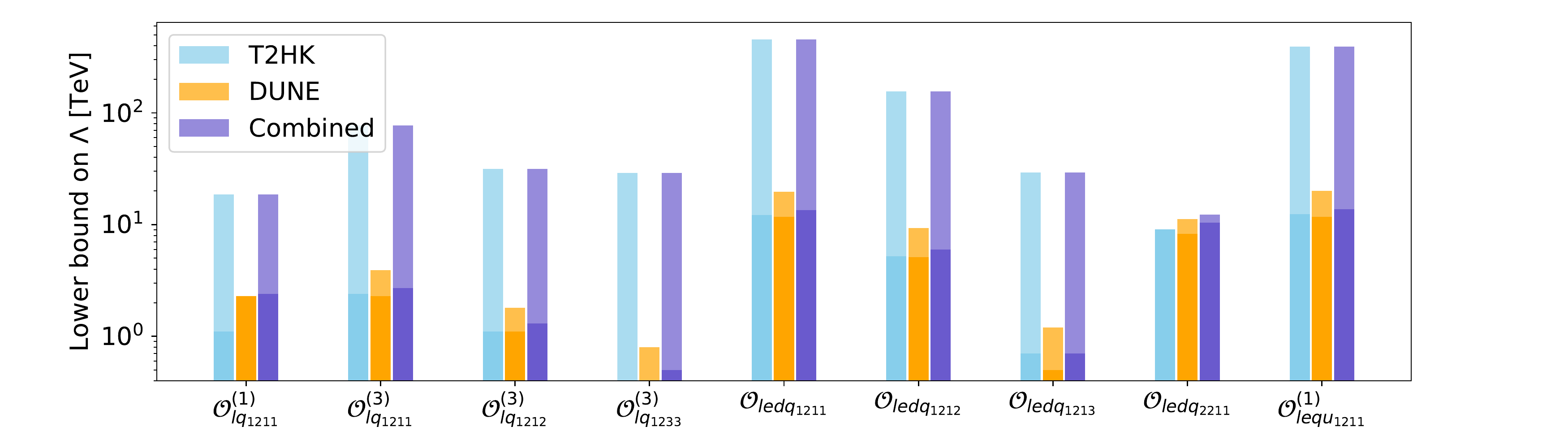}}
\caption{\label{fig:nd280_without_eapp} {  Total contribution of electron appearance channels $\nu_\mu \rightarrow \nu_e$ and $\bar{\nu}_\mu \rightarrow \bar{\nu}_e$ in T2HK and DUNE near detectors. The shaded parts correspond to the sensitivities to operators in figure\,\ref{fig:epsmemu2} without electron appearance data in near detectors, whereas the light-coloured parts indicate the effect of other channels T2HK and DUNE statistics.}
}
\end{figure}

{ 
In this section, we briefly discuss the significance of electron appearance channels $\nu_\mu \rightarrow \nu_e$ and $\bar{\nu}_\mu \rightarrow \bar{\nu}_e$ in T2HK and DUNE near detectors. In the main text, these channels were taken into account in T2HK, DUNE and T2HK+DUNE discovery potentials. Owing to the zero-distance oscillation effect~\cite{Ohlsson:2012kf}, the electron appearance channels play a significant role in probing the source NSI parameter $\epsilon^s_{\mu e}$ through pion decay. Here, we show how much the electron appearance channels contribute to the total sensitivities.

The operators where $\epsilon^s_{\mu e}$ makes the most significant contribution were displayed in figure\,\ref{fig:epsmemu2} in section\,\ref{sec:results_NC_NSI}. In figure\,\ref{fig:nd280_without_eapp}, we show the effect on the operator sensitivities that arise from source NSI parameter $\epsilon^s_{\mu e}$ in T2HK and DUNE. The shaded regions correspond to the contributions from source, detection and matter NSI parameters other than $\epsilon^s_{\mu e}$. The light-coloured regions on the other hand illustrate the sensitivities arising from $\epsilon^s_{\mu e}$. The most significant contribution is seen in T2HK near detector ND280, which is able to yield the most stringent constraint on $\epsilon^s_{\mu e}$ thanks to its formidable statistics and detector position. The absence of $\nu_e$ and $\bar{\nu}_e$ samples in T2HK ND280 statistics leads to a significant decrease in sensitivity to $\epsilon^s_{\mu e}$, shifting the 95\% CL constraint from $|\epsilon^s_{\mu e}| \lesssim$ 10$^{-6}$ to 10$^{-3}$, when no other NSI parameter is taken into account.
}

\clearpage
\bibliographystyle{JHEP}
\bibliography{ref}

\end{document}